\documentclass[aps,prd,twocolumn,superscriptaddress,floatfix,showpacs,showkeys]{revtex4-1}

\usepackage[utf8]{inputenc}
\usepackage[T1]{fontenc}
\usepackage{slashed}
\usepackage{times}
\usepackage{mathrsfs}

\usepackage{amssymb,amsfonts,amsmath,amsthm,wasysym}
\usepackage{dsfont,bbm}
\usepackage{dcolumn}
\usepackage{placeins}

\usepackage{graphicx}
\usepackage[caption=false]{subfig}
\usepackage{color}

\newcommand{\MS}{{\overline{\mathrm{MS}}}}
\newcommand{\cO}{\mathcal {O}}
\newcommand{\Dl}[1]{\overset{\leftarrow}{D}_{#1}}
\newcommand{\Dr}[1]{\overset{\rightarrow}{D}_{#1}}
\newcommand{\Dd}[1]{\overset{\leftrightarrow}{D}_{#1}}
\newcommand{\RI}{RI$^\prime$}
\newcommand{\gc}{\bar{g}}
\newcommand{\third}{\mbox{\small $\frac{1}{3}$}}

\begin{document}

\title{Nonperturbative Renormalization in Lattice QCD with three Flavors
of Clover Fermions: Using Periodic and Open Boundary Conditions}

\author{G.S.~Bali}
\author{S.~B\"urger}
\author{S.~Collins}
\author{M.~{G\"ockeler}}
\author{M.~Gruber}
\author{S.~Piemonte}
\author{A.~{Sch\"afer}}
    \affiliation{Institut f\"ur Theoretische Physik, Universit\"at Regensburg, 93040 Regensburg, Germany}
\author{A.~{Sternbeck}}
    \affiliation{Theoretisch-Physikalisches Institut,
     Friedrich-Schiller-Universit\"at Jena, 07743 Jena, Germany}
\author{P.~Wein}
   \affiliation{Institut f\"ur Theoretische Physik, Universit\"at Regensburg, 93040 Regensburg, Germany}

\collaboration{RQCD Collaboration}
   

\begin{abstract}
We present the nonperturbative computation of renormalization
factors in the \RI-(S)MOM (regularization independent (symmetric) momentum
subtraction) schemes for the QCD gauge field ensembles
generated by the CLS (coordinated lattice simulations) effort with
three flavors of nonperturbatively improved Wilson (clover) quarks.
We use ensembles with
the standard (anti-)periodic boundary conditions in the time direction
as well as gauge field configurations with open boundary conditions.
Besides flavor-nonsinglet quark-antiquark operators with up to two
derivatives we also consider three-quark operators with up to one
derivative.
For the \RI-SMOM scheme results we make use of the recently
calculated three-loop conversion factors to the $\MS$ scheme.
The present version of the paper contains an Addendum with additional
analytical expressions and updated results.

\end{abstract}

\maketitle

\section{Introduction}

Matrix elements of local operators between hadron states contain valuable
information on hadron structure. For example, decay constants, moments of
parton distributions and light-cone distribution amplitudes can be
written in this form. In order to compute such quantities from
QCD, nonperturbative methods are required. One possibility is lattice QCD:
After a Wick rotation from Minkowski to Euclidean space 
the ensuing Euclidean functional integral of QCD is regularized
with the help of a lattice discretization such that a numerical evaluation
through Monte Carlo simulations becomes possible.

Hadronic matrix elements of local operators in general are
ultraviolet (UV) divergent quantities. These are regularized on the
lattice and their dependence on the cutoff is then treated by the
renormalization procedure. In a perturbative continuum calculation, such as
in dimensional regularization, the UV divergences are removed order
by order in the expansion in the coupling constant by imposing a given
condition on how the divergent part is subtracted (MS schemes).
In lattice Monte Carlo simulations, expectation values and matrix elements
of bare operators are determined nonperturbatively. Such quantities are
however not interesting \emph{per se}, but only as input for further
computations in particle phenomenology. A calculation of the renormalization
constants, e.g., in lattice perturbation theory, is therefore required to
absorb the divergences of bare operators that appear as the lattice
spacing $a$ is sent to zero, and to convert the lattice results to the
continuum schemes commonly adopted in phenomenology. 

Unfortunately, the convergence of lattice perturbation theory is rather slow.
Many strategies have been developed to address this limitation in the last
decades. An alternative possibility is to employ an intermediate scheme
which is applicable both in the continuum and on the lattice. The \RI-MOM
scheme~\cite{Martinelli:1994ty} is a Regularization Independent scheme
where the renormalization
condition is imposed on amputated vertex functions in momentum space
at a renormalization scale determined by the external momenta.
The renormalization constants in the \RI-MOM scheme can be calculated
nonperturbatively on the lattice just by computing the expectation value
of the vertex function on an ensemble of gauge-fixed configurations. The
conversion (or matching) from the \RI-MOM to the MS schemes is then
performed straightforwardly by a perturbative calculation in the
continuum with dimensional regularization.

The continuum limit $a \rightarrow 0$ is achieved by extrapolating
results obtained at many nonvanishing lattice spacings $a$.
As the lattice spacing is reduced,
standard QCD simulation algorithms suffer from critical slowing down, meaning
that an increasing Hybrid Monte Carlo (HMC) simulation time is required to
fully sample the configuration space. If gauge (and fermion) fields fulfill
(anti)periodic boundary conditions in all directions, the simulation will
eventually get stuck in a fixed topological sector and ergodicity is lost.
Open boundary conditions in the time direction have been proposed to solve
this topological freezing problem at small lattice
spacings~\cite{Luscher:2011kk}. The modification of the
space-time manifold allows the topological charge $Q_{\textrm{top}}$
to differ from an integer value, even in the continuum limit,
thereby allowing for small continuous fluctuations of $Q_{\textrm{top}}$
during the HMC trajectory. This approach has been adopted within the
CLS (coordinated lattice simulations)
effort~\cite{Bruno:2014jqa,Mohler:2017wnb,Bali:2016umi}.

The breaking of translational invariance in the time direction is a
drawback of open boundary conditions. Space-time correlators corresponding
to physical particles can still be safely measured in the bulk of the
lattice, far away from the boundaries. However, it is not a priori clear
how the same procedure can be applied to vertex functions and
correlators in momentum space. Here we show that this is indeed possible and 
present our approach to the computation of the renormalization constants
in the \RI-(S)MOM schemes on $n_f=2+1$ CLS ensembles~\cite{Bruno:2014jqa}
with degenerate light and strange quark masses.

We consider a large variety of nonsinglet quark-antiquark operators
with up to two derivatives as well as three-quark operators with up to
one derivative employing lattices with lattice spacings down to about
$0.04 \, \mbox{fm}$. Treating the quark-antiquark operators in the
\RI-SMOM scheme, we make use of the recently calculated three-loop
conversion factors~\cite{Kniehl:2020sgo,Bednyakov:2020ugu,Kniehl:2020nhw}
in order to reduce the systematic error due to the truncation of the
perturbative expansion.

With the help of Ward identities, the renormalization and improvement
of the vector current is studied in Ref.~\cite{Gerardin:2018kpy} within
the CLS setup. On a subset of the CLS ensembles with
periodic boundary conditions some renormalization factors have been
evaluated for a restricted range of lattice spacings in
Ref.~\cite{Harris:2019bih} using the \RI-MOM scheme. Utilizing
Schr\"odinger functional techniques, the ALPHA collaboration has computed
several (ratios of) renormalization factors within the CLS setup,
see, e.g.,
Refs.~\cite{Heitger:2017njs,DallaBrida:2018tpn,Campos:2018ahf,deDivitiis:2019xla,Heitger:2020mkp,Heitger:2020zaq}.

The paper is organized as follows. In the next section we describe the
quark-antiquark operators studied, the employed renormalization schemes
as well as their numerical implementation on lattices with periodic and
open boundary conditions. Section~\ref{sec.3quark} discusses the
renormalization of three-quark operators. In Sec.~\ref{sec.subtraction}
we explain briefly the perturbative subtraction of lattice artifacts.
Section~\ref{sec.chiex} is devoted to the chiral extrapolation required to
obtain a mass-independent renormalization procedure. Our conventions
regarding continuum perturbation theory are collected in the following
section. In Sec.~\ref{sec.window} we discuss the dependence of our
results on the renormalization scale. The next section details the two
basic methods that we apply to extract our final numbers. Some of these
results are discussed in Sec.~\ref{sec.discussion}. Section~\ref{sec.summary}
contains a short summary. Technical details and results from perturbation
theory that are used in our computations are given in the
Appendices~\ref{sec.opmulti} -- \ref{sec.confac3}. Tables of our results
can be found in Appendix~\ref{sec.tables}.
The mixing matrices of the quark-antiquark and the three-quark
operators are given in ancillary files.

\section{Quark-antiquark operators}

\subsection{Multiplets of quark-antiquark operators}

The main focus of our calculations is the renormalization of quark-antiquark
operators involving up to two covariant derivatives acting on the quark
fields. Three-quark operators, relevant for the computation of moments of
baryon distribution amplitudes, can be treated analogously and will be
discussed in Sec.~\ref{sec.3quark}.

The elementary building blocks of local quark-antiquark operators are of
the form
\begin{align}
\bar{\psi}_\alpha^f(x) &\Gamma_{\alpha\beta} \psi_\beta^{f'}(x)\,, \\
\bar{\psi}_\alpha^f(x) &\Gamma_{\alpha\beta} \Dd{\mu} \psi_\beta^{f'}(x)\,, \\
\bar{\psi}_\alpha^f(x) &\Gamma_{\alpha\beta} \Dd{\mu} \Dd{\nu}
                           \psi_\beta^{f'}(x)\,, 
\end{align}
where $f$, $f'$, {\ldots} are flavor indices, $\alpha$, $\beta$, {\ldots} are
spinor indices and $\Gamma$ denotes a Dirac matrix. Color indices are
suppressed and we define $\Dd{\mu} = \Dr{\mu} - \Dl{\mu}$. On the lattice
the covariant derivatives are replaced by the standard (symmetric)
discretized versions. We restrict ourselves to the flavor-nonsinglet
case choosing, e.g., $f \neq f'$. Flavor-singlet operators could be
constructed by summing over $f=f'$.

In order to retain as much of the continuum symmetry as possible on the
lattice we consider multiplets of operators which transform irreducibly
under the hypercubic group H(4) and enjoy a definite charge conjugation
parity. Since the constraints imposed by space-time symmetry are less
stringent on the lattice than in the continuum the possibilities for
mixing increase. The choice of the operator multiplets is hence guided
by the desire to avoid mixing as far as possible, especially mixing
with operators of lower dimension. 

When nonforward matrix elements of operators with derivatives are considered,
either in the renormalization procedure or between hadronic states, 
we must be prepared to find mixing with so-called total-derivative
operators, i.e., operators of the generic form
$\partial_\mu \partial_\nu \cdots \bar{\psi} (\cdots) \psi$. This type of
mixing occurs already in the continuum and is therefore unavoidable on
the lattice. However, due to charge conjugation invariance, which is exact
also on the lattice, we encounter this phenomenon only in the case of
operators with two derivatives. The operator multiplets that we consider
are compiled in Appendix~\ref{sec.opmulti}.

\subsection{Renormalization schemes} \label{sec.schemes}

In this section we describe our implementation of the \RI-MOM
scheme~\cite{Martinelli:1994ty} and the \RI-SMOM scheme~\cite{Sturm:2009kb}.

Let $\cO^{(i)}_m (x)$ ($i=1,2,\ldots,d$, $m=1,2,\ldots,M$) denote $M$ 
multiplets of local quark-antiquark operators which transform 
identically according to an irreducible, unitary, $d$-dimensional 
representation of H(4). We call the unrenormalized, but
(lattice-)regularized vertex functions (in Landau gauge)
$V^{(i)}_m (p,q)$, where $p$ and $q$ are the external quark momenta.
The corresponding renormalized (in the $\MS$ scheme) vertex functions
are denoted by $\bar{V}^{(i)}_m (p,q)$. The dependence of $\bar{V}^{(i)}_m$
on the renormalization scale $\mu$ is suppressed for brevity. Note
that $V^{(i)}_m$ as well as $\bar{V}^{(i)}_m$
carry spinor indices and are therefore to be considered as
$4 \times 4$-matrices. (The color indices have been averaged over.)
When all mixing multiplets are taken into account, we should have
(up to power corrections in the lattice spacing)
\begin{equation} 
\bar{V}^{(i)}_m (p,q) = Z_q^{-1} 
   \sum_{m'=1}^M Z_{m m'} V^{(i)}_{m'} (p,q) \,,
\end{equation}
where $Z$ is the matrix of renormalization and mixing coefficients and
$Z_q$ is the wave function renormalization constant of the quark 
fields.

The renormalization factors depend on the renormalization scale $\mu$
as well as on the cutoff, i.e., the lattice spacing $a$. A third
dimensionful quantity appearing in this connection is the asymptotic
scale parameter $\Lambda$. Being dimensionless, the renormalization
factors in a mass-independent renormalization scheme such as the 
\RI-(S)MOM scheme can only depend on two dimensionless combinations
of these three quantities, e.g., on $a \mu$ and $a \Lambda$, or
functions thereof, such as the bare lattice or the renormalized
coupling constant. In the following, we shall suppress
all arguments that are not needed in the respective context and write,
e.g., $Z(\mu,a)$, $Z(\mu)$ or simply $Z$.

In the \RI-(S)MOM scheme we denote the matrix of renormalization
and mixing coefficients by $\hat{Z}$ and the quark field renormalization
factor by $\hat{Z}_q$. The definition of $\hat{Z}_q$ will be given
below in Eq.~(\ref{eq.defzq}). The renormalized vertex function is then
written as 
\begin{equation} 
V^{(i)}_m (p,q)^{\mathrm R} = \hat{Z}_q^{-1} 
   \sum_{m'=1}^M \hat{Z}_{m m'} V^{(i)}_{m'} (p,q) \,.
\end{equation}
With the lattice Born term $\hat{B}^{(i)}_m (p,q)$ corresponding to
$V^{(i)}_m (p,q)$, we obtain
\begin{equation} 
\begin{split}
\sum_{i=1}^d &\mathrm {tr} 
\left( V^{(i)}_m (p,q)^{\mathrm R}
                    \, \hat{B}^{(i) \dagger}_{m'} (p,q) \right) \\
 =& \hat{Z}_q^{-1}  \sum_{m''=1}^M \hat{Z}_{m m''} \sum_{i=1}^d 
   \mathrm {tr} \left( V^{(i)}_{m''} (p,q)
                    \, \hat{B}^{(i) \dagger}_{m'} (p,q) \right)\,.
\end{split}
\end{equation}
The \RI-(S)MOM scheme is now defined by requiring that for a given
momentum geometry $p = \hat{p}$, $q = \hat{q}$ the left-hand side
of this equation coincides with the corresponding tree-level expression:
\begin{equation} 
\begin{split}
\sum_{i=1}^d &\mathrm {tr} 
\left( V^{(i)}_m (\hat{p},\hat{q})^{\mathrm R}
                    \, \hat{B}^{(i) \dagger}_{m'} (\hat{p},\hat{q}) \right) \\
 =& \sum_{i=1}^d 
   \mathrm {tr} \left( \hat{B}^{(i)}_{m} (\hat{p},\hat{q})
                    \, \hat{B}^{(i) \dagger}_{m'} (\hat{p},\hat{q}) \right)\,.
\end{split}
\end{equation}
In this way we arrive at our renormalization condition
\begin{equation} \label{eq.renco}     
\begin{split}
\sum_{i=1}^d &\mathrm {tr} 
   \left( \hat{B}^{(i)}_m (\hat{p},\hat{q})
          \hat{B}^{(i) \dagger}_{m'} (\hat{p},\hat{q}) \right) \\
 =& \hat{Z}_q^{-1}  \sum_{m''=1}^M \hat{Z}_{m m''} \sum_{i=1}^d 
   \mathrm {tr} \left( V^{(i)}_{m''} (\hat{p},\hat{q})
          \hat{B}^{(i) \dagger}_{m'} (\hat{p},\hat{q}) \right)\,.
\end{split}
\end{equation}
As the \RI-(S)MOM scheme should be mass independent, we have to impose
this condition in the chiral limit. The sum over all members of the
operator multiplets ensures that all lattice symmetries are preserved,
if the individual operators are normalized such that they transform
according to a unitary representation of H(4).

In the \RI-MOM scheme we choose 
\begin{equation} \label{eq.rimom}
\hat{p} = \hat{q} = \frac{\mu}{2}(1,1,1,1)
\end{equation}
while in the \RI-SMOM scheme we require
$\hat{p}^2 = \hat{q}^2 = (\hat{p}-\hat{q})^2 = \mu^2$,
which may be achieved by taking 
\begin{equation}  \label{eq.rismom}
\hat{p} = \frac {\mu}{\sqrt{2}} (1,1,0,0) \; , \; 
\hat{q} = \frac {\mu}{\sqrt{2}} (0,1,1,0) \;. 
\end{equation}
Note that in our conventions the time component is the last
component of the momenta. The choice of the momentum directions is to
be considered as belonging to the definition of the renormalization
scheme. It should be mentioned that the mixing with total-derivative
operators can be taken into account only within the \RI-SMOM scheme,
because these operators do not contribute in forward matrix elements.

If there is no mixing ($M=1$), the formulas simplify. Omitting the
superfluous multiplet indices $m$ and $m'$ in this case, we can write
\begin{equation} 
\hat{Z} \hat{Z}_q^{-1} = \frac
  {\sum_{i=1}^d \mathrm {tr} \left( \hat{B}^{(i)} \hat{B}^{(i) \dagger} \right)}
  {\sum_{i=1}^d \mathrm {tr} \left( V^{(i)} \hat{B}^{(i) \dagger} \right)} \,.
\end{equation}

In the case of the (flavor-nonsinglet) local vector and axialvector
currents one may want to modify the above renormalization conditions such
that they are consistent with the respective Ward identities. Calling the
vertex function of the local vector current $\mathcal V_\mu$ (with the
flavor indices suppressed) $V_\mu (p,q)$, we use the renormalization condition
\begin{equation} \label{eq.vecwi}
\begin{split}
\hat{Z}_q^{-1} \hat{Z} &\sum_\mu \mathrm {tr} \left( V_\mu (\hat{p},\hat{p})
    (\gamma_\mu - \hat{\slashed{p}} \hat{p}_\mu/\hat{p}^2) \right) \\
=& \sum_\mu \mathrm {tr}
  \left( \gamma_\mu (\gamma_\mu
    - \hat{\slashed{p}} \hat{p}_\mu/\hat{p}^2) \right) = 12
\end{split}
\end{equation}
in the case of the \RI-MOM scheme. In the \RI-SMOM scheme we employ
the renormalization condition~\cite{Sturm:2009kb}
\begin{equation} \label{eq.svecwi}
\hat{Z}_q^{-1} \hat{Z} \sum_\mu \mathrm {tr}
  \left( V_\mu (\hat{p},\hat{q}) (\hat{p}_\mu - \hat{q}_\mu)
  (\hat{\slashed{p}} - \hat{\slashed{q}}) \right) = 4 (\hat{p}-\hat{q})^2 \,.
\end{equation}
Similarly, we have for the axialvector current $\mathcal A_\mu$ with
vertex function $A_\mu (p,q)$ the renormalization condition 
\begin{equation}  \label{eq.axvecwi}
\hat{Z}_q^{-1} \hat{Z} \sum_\mu \mathrm {tr}
  \left( A_\mu (\hat{p},\hat{p}) \gamma_5
    (\gamma_\mu - \hat{\slashed{p}} \hat{p}_\mu/\hat{p}^2) \right) = 12
\end{equation}
in the \RI-MOM scheme and 
\begin{equation} \label{eq.saxvecwi}
\hat{Z}_q^{-1} \hat{Z} \sum_\mu \mathrm {tr} \left( A_\mu (\hat{p},\hat{q})
  (\hat{p}_\mu - \hat{q}_\mu)\gamma_5
  (\hat{\slashed{p}} - \hat{\slashed{q}}) \right) = 4 (\hat{p}-\hat{q})^2 
\end{equation}
in the \RI-SMOM scheme.

In the \RI-MOM scheme as well as in the \RI-SMOM scheme the wave
function renormalization constant of the quark fields $\hat{Z}_q$ is
determined from the quark propagator $S(\hat{p})$ according to
\begin{equation} \label{eq.defzq}
\hat{Z}_q =
  \frac{ {\rm tr} \left( - {\mathrm i} \sum_\lambda \gamma_\lambda 
           \sin (a \hat{p}_\lambda) a  S^{-1} (\hat{p}) \right) }
           {4 \sum_\lambda \sin^2 (a \hat{p}_\lambda) } 
\end{equation}
with $\hat{p}^2 = \mu^2$.
Other definitions of $\hat{Z}_q$ have been proposed in the literature,
mainly with the aim of reducing lattice artifacts, see, e.g.,
Refs.~\cite{Sturm:2009kb,Capitani:2000xi,Maillart:2008pv,Oliveira:2018lln}.

Using the lattice Born term instead of the continuum
Born term in the renormalization condition (\ref{eq.renco}) and proceeding
analogously in the calculation of $\hat{Z}_q$ ensures that $\hat{Z}$
is the unit matrix in the free case. Note, however, that in many cases
there is no difference between employing the lattice or the continuum
Born terms for our choice of the momentum directions. 

The renormalization matrix $\hat{Z}$ leads from the bare lattice
operators to renormalized operators in our regularization independent
\RI-(S)MOM scheme. The matrix $Z$ transforming the bare operators into
renormalized operators in the $\MS$ scheme of dimensional regularization
is then given by $Z = C \hat{Z}$, where the matrix $C$ is calculated from
\begin{equation} \label{eq.match}
\begin{split}
\sum_{m''=1}^M \sum_{i=1}^d &C_{m m''} \mathrm {tr} 
   \left( B^{(i)}_{m''}(\hat{p},\hat{q})
            B^{(i) \dagger}_{m'}(\hat{p},\hat{q}) \right) \\
 &= C_q \sum_{i=1}^d \mathrm {tr} 
   \left( \bar{V}^{(i)}_m (\hat{p},\hat{q})
            B^{(i) \dagger}_{m'} (\hat{p},\hat{q}) \right) \,.
\end{split}
\end{equation}
Here $B^{(i)}_m$ is the continuum Born term, and the factor $C_q$
is the analogue of $C$ for the quark wave function renormalization constant,
i.e., $Z_q = C_q \hat{Z}_q$. Note that $C_q$ and the conversion matrix
$C$ are completely determined from a calculation in continuum
perturbation theory. If the renormalization of the vector or
axialvector current is performed consistently with the Ward identities,
cf.\ Eqs.~(\ref{eq.vecwi}) - (\ref{eq.saxvecwi}), the conversion factor
$C$ is equal to one in these cases.

One can avoid the use of the quark wave function renormalization constant
by computing ratios $Z/Z_V$ with the help of the \RI-(S)MOM scheme and
determining the renormalization constant $Z_V$ of the local vector current
$\mathcal V_\mu$ by other methods, e.g., from matrix elements of
$\mathcal V_\mu$ between hadronic states of given electric charge.

As already mentioned, all of the above calculations should be performed for
massless quarks so that our renormalization schemes are mass independent.
In practice, Monte Carlo simulations with our boundary conditions
require nonvanishing quark masses. Therefore we compute $\hat{Z}$
first at nonzero masses and perform an extrapolation to the chiral limit in
the end, see Sec.~\ref{sec.chiex}. In order to avoid unnecessary
complications with this extrapolation we take the quark masses in the
simulations to be flavor independent.

\vspace*{0.5cm}
\subsection{Numerical implementation} \label{sec.implement}

The CLS ensembles that we use are listed in Table~\ref{tab.ensembles}
along with their most relevant properties. On the coarser lattices
($\beta = 3.34, \, 3.40, \, 3.46, \, 3.55$) we have ensembles with the
standard boundary conditions at our disposal, i.e., periodic boundary
conditions in all four directions for the gluons
and periodic (antiperiodic) boundary conditions in space (time) for the
quarks. In this case we use the term periodic boundary conditions as shorthand.
The corresponding ensembles are labeled `p' in Table~\ref{tab.ensembles}.
On the finer lattices ($\beta = 3.70, \, 3.85$) only ensembles with open
boundary conditions in time, labeled `o', are
available~\cite{Bruno:2014jqa}. It should be noted that the ensembles
H101, U103, H200, and N202 are only used for the assessment of
systematic uncertainties and not for the evaluation of our final results.

\begin{table}[h]
\caption{\label{tab.ensembles} List of ensembles. The inverse gauge
coupling~$\beta$ determines the lattice spacing, while the spatial and
temporal extents fix the lattice geometry~$N_s^3\times N_t$.
Boundary conditions in the time direction
are either periodic~(p) or open~(o). The hopping parameter $\kappa$
determines the corresponding quark mass; the resulting approximate pion
mass~$m_\pi$ is given in units of MeV, followed by the spatial lattice
size in pion mass units.} 
\begin{ruledtabular}
\begin{tabular}{rccrcllccc}
\multicolumn{1}{c}{Ens.} & $\beta$ & $N_s$ & \multicolumn{1}{c}{$N_t$} & bc
  & \multicolumn{1}{c}{$\kappa$} & $m_\pi$ & $m_\pi L$ \\
\hline
A650    & $3.34$ & $24$ & $ 48$ & p  & $0.1366      $  & $368$ & $4.4$ \\
A652    & $3.34$ & $24$ & $ 48$ & p  & $0.1365695   $  & $429$ & $5.1$ \\
A651    & $3.34$ & $24$ & $ 48$ & p  & $0.1365      $  & $552$ & $6.6$ \\
\hline
rqcd017 & $3.40$ & $32$ & $ 32$ & p  & $0.136865    $  & $235$ & $3.3$ \\
rqcd021 & $3.40$ & $32$ & $ 32$ & p  & $0.136813    $  & $338$ & $4.7$ \\
rqcd016 & $3.40$ & $32$ & $ 32$ & p  & $0.13675962  $  & $420$ & $5.9$ \\
H101    & $3.40$ & $32$ & $ 96$ & o  & $0.13675962  $  & $420$ & $5.9$ \\
U103    & $3.40$ & $24$ & $128$ & o  & $0.13675962  $  & $420$ & $4.4$ \\
rqcd019 & $3.40$ & $32$ & $ 32$ & p  & $0.1366      $  & $603$ & $8.4$ \\
\hline
X450    & $3.46$ & $48$ & $ 64$ & p  & $0.136994    $  & $263$ & $4.9$ \\
rqcd030 & $3.46$ & $32$ & $ 64$ & p  & $0.1369587   $  & $317$ & $3.9$ \\
B450    & $3.46$ & $32$ & $ 64$ & p  & $0.13689     $  & $418$ & $5.2$ \\
rqcd029 & $3.46$ & $32$ & $ 64$ & p  & $0.1366      $  & $707$ & $8.7$ \\
\hline
X251    & $3.55$ & $48$ & $ 64$ & p  & $0.1371      $  & $266$ & $4.2$ \\
X250    & $3.55$ & $48$ & $ 64$ & p  & $0.13705     $  & $348$ & $5.4$ \\
rqcd025 & $3.55$ & $32$ & $ 64$ & p  & $0.137       $  & $412$ & $4.3$ \\
H200    & $3.55$ & $32$ & $ 96$ & o  & $0.137       $  & $412$ & $4.3$ \\
N202    & $3.55$ & $48$ & $128$ & o  & $0.137       $  & $412$ & $6.4$ \\
B250    & $3.55$ & $32$ & $ 64$ & p  & $0.1367      $  & $707$ & $7.4$ \\
\hline
N300    & $3.70$ & $48$ & $128$ & o  & $0.137       $  & $422$ & $5.1$ \\
N303    & $3.70$ & $48$ & $128$ & o  & $0.1368      $  & $641$ & $7.8$ \\
\hline
J500    & $3.85$ & $64$ & $192$ & o  & $0.136852    $  & $410$ & $5.2$ \\
N500    & $3.85$ & $48$ & $128$ & o  & $0.13672514  $  & $599$ & $5.7$ \\
\end{tabular}
\end{ruledtabular}

\end{table}

The lattice spacings have been determined from the Wilson flow time at
the SU(3) symmetric point in lattice units $t_0^*/a^2$, where the SU(3)
symmetric point is defined by $12 t_0^* m_\pi^2 = 1.11$.
Equating $t_0^*$ with the result
$\mu_{\mathrm{ref}}^*=(8t_0^*)^{-1/2}\approx 478 \, \mathrm{MeV}$
of Ref.~\cite{Bruno:2017gxd} we arrive at the values given in
Table~\ref{tab.spacings}.

\begin{table}[h]
\caption{\label{tab.spacings} Lattice spacings.}
\begin{ruledtabular}
\begin{tabular}{crrrrrr}
$\beta$ & 3.34 & 3.40 & 3.46 & 3.55 & 3.70 & 3.85 \\
$1/a^2 \, [\mbox{GeV}^2]$ &
  4.05  & 5.31  & 6.77  & 9.46  & 15.77  & 25.54 \\
$a \, [\mbox{fm}]$ &
  0.098 & 0.086 & 0.076 & 0.064 & 0.050  & 0.039
\end{tabular}
\end{ruledtabular}
\end{table}

Our calculations are based on correlation functions with external quark
lines. Therefore we are forced to work in a fixed gauge. As usual we choose
the Landau gauge, because this gauge can be implemented on the lattice as
well as in continuum perturbation theory. The required correlation
functions are evaluated with the help of momentum sources. On the ensembles
with periodic boundary conditions this procedure is
straightforward to implement~\cite{Gockeler:1998ye,Gockeler:2010yr},
and the relevant formulas are given
in Sec.~\ref{sec.periodicbc}. In the case of open boundary conditions
some modifications are necessary, which will be discussed in
Sec.~\ref{sec.openbc}.

\subsubsection{Periodic boundary conditions} \label{sec.periodicbc}

For a given operator $\cO (x)$ we start from the three-point function
\begin{equation} \label{eq.3ptf}
\begin{split}
&G_{\alpha\beta}(p,q) \\ & {}
  = \frac{a^{12}}{V} \sum_{x,y,z} 
  \mathrm e^{- \mathrm i p \cdot x - \mathrm i (q-p) \cdot z 
                                       + \mathrm i q \cdot y } 
\langle \psi_\alpha (x) \cO (z) \bar{\psi}_\beta (y) \rangle
\end{split}
\end{equation}
and the quark propagator
\begin{equation}  \label{eq.prop}
 S_{\alpha\beta} (p) = 
 \frac{a^8}{V} \sum_{x,y} {\mathrm e}^{- {\mathrm i} 
    p \cdot (x-y) } \langle \psi_\alpha (x) \bar{\psi}_\beta (y) \rangle \,,
\end{equation}
where $V$ denotes the (dimensionful) volume of the lattice.
Since we use flavor-independent quark masses, flavor indices have been
omitted. In the case of a flavor-nonsinglet operator $\cO$ we have 
only quark-line connected contributions in the three-point function $G$,
while for a flavor-singlet operator there would be an additional quark-line
disconnected contribution. The quark propagator $S$ always refers to a
single flavor.

Note that due to translation invariance one of the sums in the above
expression~(\ref{eq.3ptf}) is redundant. For example, one could restrict
the sum over $z$ to a single lattice point omitting at the same time
a factor $a^4/V$. However, the volume averaging connected with this sum
suppresses statistical fluctuations very efficiently and is therefore
highly advantageous.

Due to the invariance under global color transformations, which survives
the Landau gauge fixing, both of these correlation functions are
proportional to the unit matrix in color space. We assume that the color
indices have been averaged over, as we already did in
Sec.~\ref{sec.schemes}, so that $G$ and $S$ are $4 \times 4$ matrices. 

The vertex function of the operator $\cO$ is then constructed as
\begin{equation} 
V(p,q) = S^{-1} (p) G(p,q) S^{-1} (q) \,.
\end{equation}

The calculation of the correlation functions with the help of momentum 
sources proceeds as follows. If $M(x,y)$ represents the fermion matrix
on a given gauge field configuration, we compute the quark propagator
$\hat{S}(x,y)$ with a momentum source by solving the lattice Dirac equation
\begin{equation} \label{eq.dirac}
a^4 \sum_z M(y,z) 
\left( a^4 \sum_x  \hat{S}(z,x) {\mathrm e}^{\mathrm i p \cdot x} 
      \right) = {\mathrm e}^{\mathrm i p \cdot y} \,.
\end{equation}
The quark propagator $S(p)$ in momentum space is then evaluated as the
gauge field average of 
\begin{align}
\frac{a^8}{V} &\sum_{x,y} \mathrm e^{- \mathrm i p \cdot (x-y)} \hat{S}(x,y)
\nonumber \\
 &= \frac{a^8}{V} \sum_x \mathrm e^{- \mathrm i p \cdot x}
  \left( \sum_y  \hat{S}(x,y) \mathrm e^{\mathrm i p \cdot y} \right) \,.
\end{align}

Representing $\sum_z \mathrm e^{- \mathrm i (q-p) \cdot z} \cO (z)$ as 
\begin{equation} 
\sum_{z,z'} \mathrm e^{- \mathrm i (q-p) \cdot (z+z')/2}
  \bar{\psi} (z) J(z,z') \psi (z') \,,
\end{equation}
the quark-line connected part of $G(p,q)$ is obtained as the gauge field
average of 
\begin{widetext}
\begin{equation} 
\begin{split}
\hat{G}(p,q) & = \frac{a^{12}}{V} \sum_{x,y,z,z'} 
 \mathrm e^{- \mathrm i p \cdot x - \mathrm i (q-p) \cdot (z+z')/2 
                                       + \mathrm i q \cdot y } 
  \hat{S}(x,z) J(z,z')\hat{S}(z',y) \\
& = \frac{a^4}{V} \sum_{z,z'} 
 \mathrm e^{ - \mathrm i (q-p) \cdot (z+z')/2} \, \gamma_5
 \left( a^4 \sum_x \hat{S}(z,x) \mathrm e^{\mathrm i p \cdot x} \right)^\dagger
 \gamma_5 J(z,z') 
 \left( a^4 \sum_y \hat{S}(z',y) \mathrm e^{\mathrm i q \cdot y} \right) \,.
\end{split} \label{eq.gconn}
\end{equation}
\end{widetext}

The statistical errors are computed with the help of the (single
elimination) jackknife procedure. As a relatively small number of
configurations is sufficient for our purposes, we can choose them such
that they are statistically independent to a good accuracy. Hence we
refrain from binning.

If one imposes the standard boundary conditions on the quark fields,
the possible spatial components of the quark momenta $p$ on an
$N_s^3 \times N_t$ lattice are integer multiples of $2 \pi/(a N_s)$
while the time components are of the form $(n+1/2) 2 \pi/(a N_t)$
with $n \in \mathbb Z$. Such momenta do not allow us to satisfy
the condition (\ref{eq.rimom}) (or the condition (\ref{eq.3qsmom})
for three-quark operators below) exactly. Therefore
we employ twisted boundary conditions~\cite{Sachrajda:2004mi}
when solving the lattice Dirac equation (\ref{eq.dirac}). For twist $\tau$
in the spatial directions we get spatial momentum components
$p_1 = p_2 = p_3 = (n+\tau/2) 2 \pi/(a N_s)$. In the \RI-MOM scheme
the corresponding temporal twist is then chosen such that
$p_4 = p_1 = p_2 = p_3$. We employ the five values
$\tau = 0.0, \, 0.4, \, 0.8, \, 1.2, \, 1.6$.
In the \RI-SMOM scheme with the momenta (\ref{eq.rismom}) twisted
boundary conditions are not required. Nevertheless, we use the same
twist values as in the \RI-MOM scheme, because a larger number of
momenta appears to be beneficial in the analysis.

\begin{figure*}
\includegraphics[width=.95\textwidth]{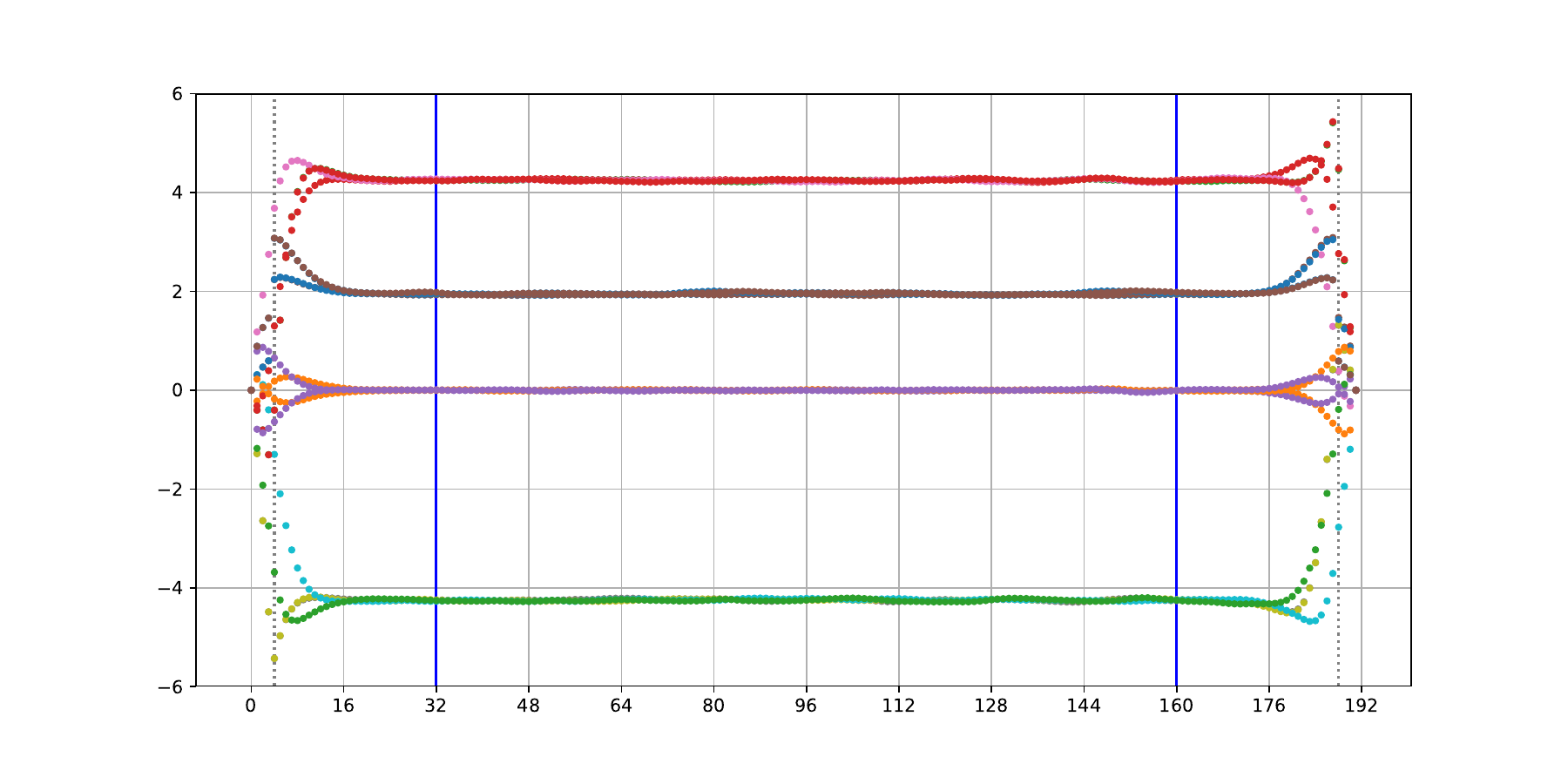}
\caption{\label{fig.J500prop} Components of the quark propagator on
ensemble J500 summed over space as a function of time
(in lattice units). The magnitude of the momentum is given by
$\mu \approx 2 \, \mathrm {GeV}$ and $a^{-1} \approx 5 \, \mathrm {GeV}$.
The volume for the momentum source is limited by the dotted grey lines.
The volume for the final summation is indicated by the blue lines.}
\end{figure*}

\begin{figure*}
\includegraphics[width=.95\textwidth]{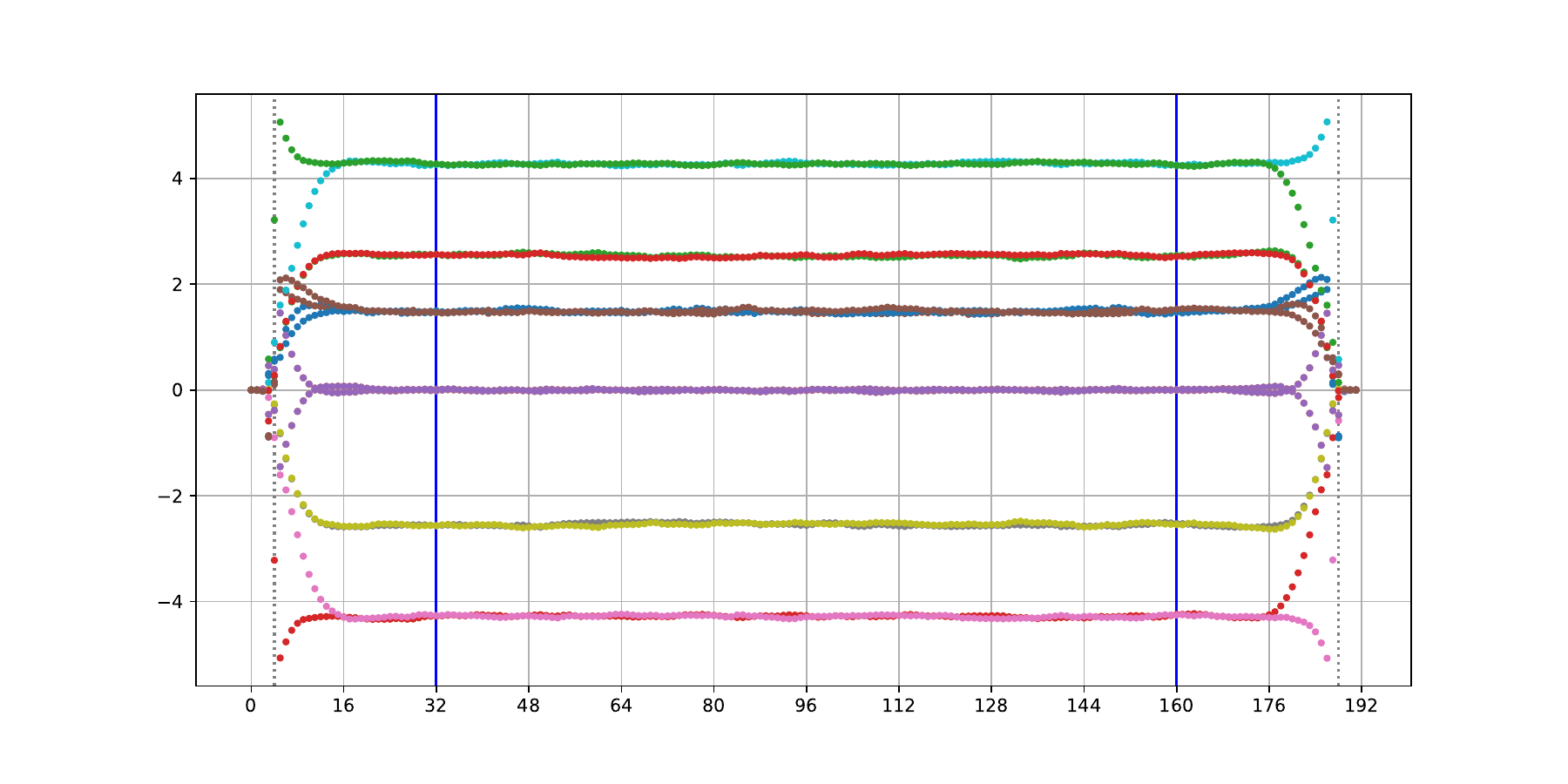}
\caption{\label{fig.J500bilinear} Components of the three-point function
of the operator $\bar{\psi} \gamma_1 \Dd{4} \psi$ on ensemble J500
summed over space as a function of  time (in lattice units).
The magnitude of the momentum is given by $\mu \approx 2 \, \mathrm {GeV}$
and $a^{-1} \approx 5 \, \mathrm {GeV}$. The blue and dotted grey
lines are as in Fig.~\ref{fig.J500prop}.}
\end{figure*}

Besides the renormalization factors themselves one might also be interested
in ratios of these factors, either in order to avoid the appearance of the
quark wave function renormalization constant or because the physical
quantity to be studied is a ratio of matrix elements of different operators
and is hence renormalized by the ratio of the corresponding renormalization
factors. The latter case appears for example when moments of hadron
distribution amplitudes are
computed~\cite{Bali:2015ykx,Bali:2019ecy,Braun:2016wnx,Bali:2019dqc,addendum}.
Such ratios of renormalization coefficients can be evaluated ``directly''
by computing them on the single jackknife ensembles and then analyzing
the results as for individual renormalization factors.

\subsubsection{Open boundary conditions} \label{sec.openbc}

On the ensembles with open boundary conditions we have to modify our
computational strategy. In order to avoid instabilities in the inversion
of the fermion matrix, i.e., in the evaluation of the quark propagators,
the support of the momentum sources is no longer taken to be the whole
$N_s^3 \times N_t$ lattice, but replaced with a subvolume of size
$N_s^3 \times (N_t - 2 \Delta)$ situated symmetrically about the center
of the lattice, thus keeping a distance of $\Delta \cdot a$ from the
temporal boundaries. Moreover, we have to take into account that the
invariance under translations in the time
direction is broken. Still, in the bulk of the lattice, i.e., at
sufficiently large distances from the temporal boundaries, the effects of
this symmetry loss should be negligible. Therefore we restrict the
sum over the operator position $z$ in the three-point function
(\ref{eq.3ptf}) to an even smaller volume of size
$N_s^3 \times (N_t - 2 \tilde{\Delta})$, again
centered at the midpoint of the lattice.

In our calculations we choose $\Delta = 4$ and $\tilde{\Delta} = 32$.
To motivate these choices we plot spatial sums of correlation functions 
against time on the ensemble J500 in Figs.~\ref{fig.J500prop} and
\ref{fig.J500bilinear}. In Fig.~\ref{fig.J500prop} we do this for the
quark propagator (\ref{eq.prop}) and in Fig.~\ref{fig.J500bilinear}
for the three-point function (\ref{eq.3ptf}) of the operator
$\bar{\psi} \gamma_1 \Dd{4} \psi$. In both cases we show the real parts
of all 16 components for the momentum geometry (\ref{eq.rimom})
of the \RI-MOM scheme with $\mu \approx 2 \, \mathrm {GeV}$. The dotted
vertical lines indicate the source volume, while the blue vertical lines 
limit the subvolume that will be used for the final summation. It should
be noted that the ``flatness'' of the data in the bulk improves with
increasing $\mu$ and that $\mu = 2 \, \mathrm {GeV}$ is the lowest scale
entering our final analysis. We remark that J500 corresponds to
our smallest lattice spacing. Therefore, at the other $\beta$ values,
$\tilde{\Delta} \cdot a = 32a$ is bigger if expressed in physical units.

Also the implementation of the Landau gauge requires some modifications
on lattices with open boundary conditions. On a given lattice gauge field
configuration the Landau gauge is imposed by maximizing the functional
\begin{equation}\label{eq.gauge_functional}
F[U] = \frac{a^4}{V} \sum_{x,\mu}
       \textrm{Re} \left[ \textrm{tr} \, U_\mu(x) \right] 
\end{equation}
with respect to the gauge transformation field $\Omega(x)$ acting on
the link variables $U_\mu(x)$ as
\begin{equation}
U_\mu(x) \rightarrow \Omega(x) U_\mu(x) \Omega^\dag(x + \hat{\mu})\,.
\end{equation}
The corresponding Lie-algebra valued field $A_\mu(x)$ is given by the
traceless antihermitian part of the gauge-fixed link field
$U_\mu(x)$, i.e.,
\begin{equation}
A_\mu(x) = \left. \frac{1}{2a} \left( U_\mu(x) - U^\dag_\mu(x) \right)
  \right |_{\textrm{traceless}} \,.
\end{equation} 

The maximization can be performed, for instance, by a local
overrelaxation algorithm. We continue the minimization iterations until
\begin{equation}
\frac{a^2}{12 V} \sum_{x, \mu}
  \textrm{tr} \left[ \left( A_\mu(x)-A_\mu(x-\hat{\mu}) \right)^2 \right] \,,
\end{equation} 
the deviation from the Landau gauge condition, is smaller than $10^{-10}$.
In the case of open boundary conditions we somewhat modify this 
procedure: In the sum in Eq.~(\ref{eq.gauge_functional}) an additional 
factor $1/2$ is attached to the spatial links living on the boundaries 
at $x_4=0$ and $x_4=a N_t$, in analogy to the modification of the standard 
plaquette gauge action~\cite{Luscher:2011kk}.

\subsubsection{Finite size effects}

The finite volume of our lattices may distort our results.
On the one hand, we should therefore consider lattices of different size
with otherwise identical parameters and perform an infinite volume limit
in the end. This would amount to a rather demanding procedure. On the other
hand, renormalization is a short-distance phenomenon. Hence one may expect
that the evaluation of renormalization factors is not severely affected
by finite size effects, at least for not too small renormalization scales.

\begin{figure}
\includegraphics[width=.5\textwidth]{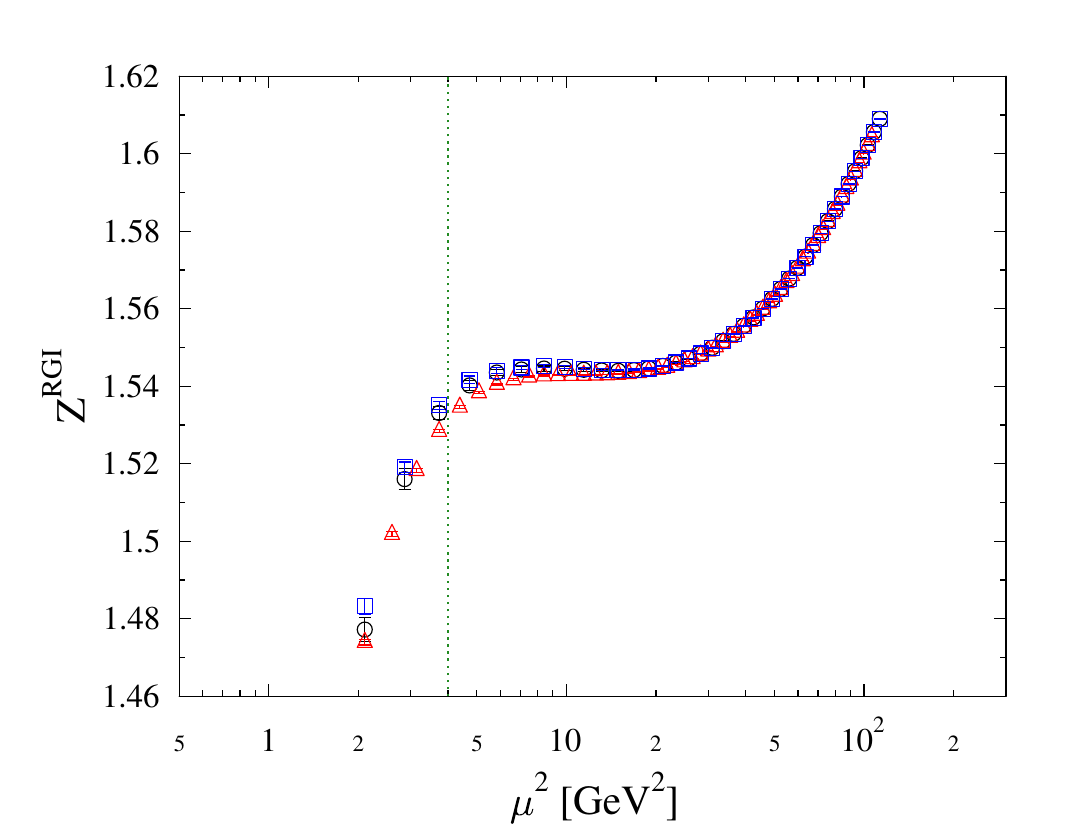}
\caption{\label{fig.fsize} Results for $Z^{\mathrm {RGI}}$ for
  the $v_{2b}$ operators obtained at $\beta=3.55$, $\kappa=0.137$ on the
  ensembles rqcd025 (volume $32^3 \times 64$, periodic boundary conditions,
  blue squares), H200 (volume $32^3 \times 96$, open boundary conditions,
  black circles), and N202 (volume $48^3 \times 128$, open boundary conditions,
  red triangles). Data to the left of the vertical green dotted line do
  not enter our analysis.}
\end{figure}

In the end we are restricted to the ensembles in Table~\ref{tab.ensembles},
which do not allow us to perform a systematic infinite volume limit.
However, for two simulation points ($\beta=3.40$, $\kappa=0.13675962$ and
$\beta=3.55$, $\kappa=0.137$) we do have ensembles with different spatial
volumes at our disposal so that we can get at least some hints at the
size of finite volume effects. This is illustrated in Fig.~\ref{fig.fsize}
for the $v_{2b}$ operators given in Eq.~(\ref{eq.v2b}). We show
$Z^{\mathrm {RGI}}$, as defined in Eq.~(\ref{eq.defRGI}) below, obtained on a
spatial volume $N_s^3 = 32^3$ at $\beta=3.55$, $\kappa=0.137$
with periodic boundary conditions (blue squares)
as well as on two lattices with open boundary conditions, $32^3$
(black circles) and $48^3$ (red triangles). The amount of agreement between
the blue squares and the black circles indicates the consistency between
the results from open and periodic boundary conditions, while the
comparison of the black circles with the red triangles gives an impression
of the finite size effects. Note that only results with
$\mu^2 \geq 4 \, \mathrm{GeV}^2$ will enter our final analysis.

\section{Three-quark operators} \label{sec.3quark}

In this section we describe briefly the renormalization of three-quark
operators, which appear in the description of baryon distribution
amplitudes~\cite{Bali:2015ykx,Bali:2019ecy}.

Every local three-quark operator can be represented as a linear
combination of the operators
\begin{equation} 
\begin{split}
&\Psi^{f_1 f_2 f_3}_{\alpha_1 \alpha_2 \alpha_3}
(\bar{l}_1,\bar{l}_2,\bar{l}_3;x) \\
& {} = \epsilon^{ijk} (D_{\bar{l}_1} \psi^{f_1} (x))^i_{\alpha_1}
(D_{\bar{l}_2} \psi^{f_2} (x))^j_{\alpha_2}
(D_{\bar{l}_3} \psi^{f_3} (x))^k_{\alpha_3} \,.
\end{split}
\end{equation} 
Here we use a multi-index notation for the covariant derivatives,
$D_{\bar{l}} \equiv D_{\lambda_1} \cdots  D_{\lambda_l}$.
Aiming again at a mass-independent renormalization scheme we assign the
same mass to all flavors and eventually consider the chiral limit.
In an abbreviated notation we write the above operators as
\begin{equation} 
\Psi^f_\alpha (\bar{l};x) \,.
\end{equation} 
Then we have
\begin{equation} \label{eq.symm}
\Psi^{f_\pi}_{\alpha_\pi} (\bar{l}_\pi;x) = \Psi^f_\alpha (\bar{l};x) 
\end{equation} 
for all permutations $\pi$ in the symmetric group $\mathcal S_3$ of
three elements, where
\begin{equation} 
\Psi^{f_\pi}_{\alpha_\pi} (\bar{l}_\pi;x) =  
\Psi^{f_{\pi(1)} f_{\pi(2)} f_{\pi(3)}}
_{\alpha_{\pi(1)} \alpha_{\pi(2)} \alpha_{\pi(3)}}
(\bar{l}_{\pi(1)},\bar{l}_{\pi(2)},\bar{l}_{\pi(3)};x) \,.
\end{equation} 
From these ``elementary'' operators we construct the operators of interest
with the help of flavor structures $F$ and spinor structures $S$ according to  
\begin{equation} \label{eq.genop}
F^f S^{\bar{l}}_\alpha \Psi^f_\alpha (\bar{l};x) \,,
\end{equation} 
where a sum over all (multi-)indices that appear twice is implied.

In the case of the quark-antiquark operators one has to distinguish
between flavor-singlet and flavor-nonsinglet operators corresponding to
the decomposition $3 \otimes \bar{3} = 1 \oplus 8$ under flavor SU(3).
For our three-quark operators we have the decomposition
$3 \otimes 3 \otimes 3 = 1 \oplus 8 \oplus 8 \oplus 10$.
The flavor-singlet (flavor-decuplet) representation corresponds to the
totally antisymmetric (totally symmetric or trivial) representation of
$\mathcal S_3$. The two flavor octets, called mixed symmetric (MS) and
mixed antisymmetric (MA), form a basis for the two-dimensional
representation of $\mathcal S_3$. More explicitly, we have the singlet
flavor structure $F^{B,f_1 f_2 f_3}_s$ with
\begin{equation} 
F^{B,f_\pi}_s 
= {\mathrm {sgn}}(\pi) F^{B,f}_s \,,
\end{equation} 
decuplet flavor structures $F^{B,f_1 f_2 f_3}_d$ with
\begin{equation} 
F^{B,f_\pi}_d = F^{B,f}_d \,,
\end{equation} 
and the octet flavor structures $F^{B,f_1 f_2 f_3}_{ot}$, where
the second subscript $t$ takes the value $t=1$ for MS and $t=2$ for MA.

The spinor structures should be chosen to yield a flavor-spinor structure
that is totally symmetric under simultaneous permutations of the flavor,
spinor and derivative indices $f_a$, $\alpha_a$, $\bar{l}_a$ ($a=1,2,3$).
Furthermore, the operator multiplets should transform irreducibly
under the spinorial hypercubic group $\overline{\mathrm {H(4)}}$, which
replaces the hypercubic group H(4) in the case of fermionic
operators~\cite{Dai:2001zu}. The group $\overline{\mathrm {H(4)}}$ has
five irreducible spinorial representations:
$\tau^{\underbar{$\scriptstyle 4$}}_1$, $\tau^{\underbar{$\scriptstyle 4$}}_2$,
$\tau^{\underbar{$\scriptstyle 8$}}_{\phantom{1}}$,
$\tau^{\underbar{$\scriptstyle 12$}}_1$ and
$\tau^{\underbar{$\scriptstyle 12$}}_2$.
(The superscripts indicate the dimension of these representations.)
Multiplets transforming according to these representations have been given in
Ref.~\cite{Kaltenbrunner:2008pb}. Starting from these operators we
construct multiplets of spinor structures 
\begin{equation} 
S^{(m,i),\bar{l}}_{s,\alpha} \,, \;
S^{(m,i),\bar{l}}_{d,\alpha} \,, \;
S^{(m,i),\bar{l}}_{ot,\alpha} \,,
\end{equation} 
which transform under $\mathcal S_3$ identically to their flavor counterparts:
\begin{equation}
S^{(m,i),\bar{l}_\pi} _{s,\alpha_\pi} = 
  {\mathrm {sgn}}(\pi) S^{(m,i),\bar{l}}_{s,\alpha}  \; \mbox{etc.}
\end{equation} 
Here $m$ labels the different $\overline{\mathrm {H(4)}}$ multiplets and
$i$ labels the different members of the multiplets. Then 
\begin{equation} 
\sum_{t=1}^2 F^{B,f}_{ot} S^{(m,i),\bar{l}}_{ot,\alpha}
\end{equation} 
is indeed totally symmetric under simultaneous permutations of the
flavor, spinor and derivative indices. An analogous statement holds for
the singlet and the decuplet. The corresponding operators
$S^{\bar{l}}_\alpha \Psi^f_\alpha (\bar{l};x)$ with zero and one derivative
have been given in Ref.~\cite{Bali:2015ykx} for generic flavors.
For the reader's convenience they are collected in Appendix~\ref{sec.3qops}.
They are chosen such that they do not mix with operators of lower dimension.
Moreover, there is no mixing between operators transforming according
to nonequivalent representations of $\mathcal S_3$.

In the case of the flavor-octet operators we find
\begin{equation} 
F^{B,f}_{ot} S^{(m,i),\bar{l}}_{ot,\alpha} \Psi^f_\alpha (\bar{l};x) 
= \frac{1}{2} \sum_{t'=1}^2 F^{B,f}_{ot'} S^{(m,i),\bar{l}}_{ot',\alpha}
  \Psi^f_\alpha (\bar{l};x) 
\end{equation} 
for $t \in \{ 1,2 \}$. Therefore we can always work with the MA flavor
structure ($t=2$) and assume that the flavor-spinor structure factorizes
into a flavor structure and a spinor structure as in Eq.~(\ref{eq.genop}).
For the singlet and decuplet operators this factorization is trivially
satisfied.

Our renormalization procedure for the three-quark operators is similar to
the RI$^\prime$-SMOM scheme used in the case of the quark-antiquark
operators. In particular, we compute the quark field renormalization
factor $Z_q$ from the quark propagator as above, see Eq.~(\ref{eq.defzq}).

\begin{figure}[tb]
\includegraphics[width=.3\textwidth]{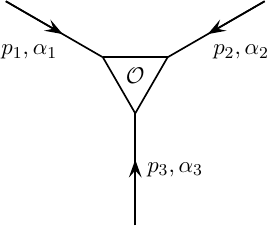}
\caption{\label{fig.4pfunc}Four-point function of a three-quark operator.}
\end{figure}

For an operator of the form (\ref{eq.genop}) we consider (in Landau gauge)
the vertex function 
\begin{equation} 
\begin{split}
& \Lambda (p_1,p_2,p_3)^{f_1 f_2 f_3}_{\alpha_1 \alpha_2 \alpha_3} \equiv
\Lambda (p)^f_\alpha \\
& {} = \sum_{\pi \in \mathcal S_3}  F^{f_\pi} 
    S^{\bar{l}}_\beta H^\beta_{\alpha_\pi} ({\bar{l}};p_\pi) 
= \sum_{\pi \in \mathcal S_3} F^{f_\pi} 
  S^{\bar{l}_\pi}_{\beta_\pi} H^{\beta_\pi}_{\alpha_\pi} (\bar{l}_\pi;p_\pi) \,.
\end{split}
\end{equation} 
Here $H^\beta_\alpha (\bar{l};p) \equiv H^{\beta_1 \beta_2 \beta_3}_{\alpha_1 \alpha_2 \alpha_3} (\bar{l}_1,\bar{l}_2,\bar{l}_3;p_1,p_2,p_3)$
denotes the ``flavorless'' amputated four-point function with open spinor
indices $\alpha_1$, $\alpha_2$, $\alpha_3$ ($\beta_1$, $\beta_2$, $\beta_3$) 
at the external quark lines (at the operator), pictorially represented in
Fig.~\ref{fig.4pfunc}. More explicitly, we have
\begin{widetext}
\begin{equation} 
\begin{split}
H^{\beta_1 \beta_2 \beta_3}_{\alpha_1 \alpha_2 \alpha_3} (\bar{l}_1, \bar{l}_2, \bar{l}_3;p_1, p_2, p_3) = \sum_{x_1,x_2,x_3} 
 \mathrm e^{\mathrm i (p_1 \cdot x_1 + p_2 \cdot x_2 + p_3 \cdot x_3)}
 \epsilon^{i_1 i_2 i_3} \epsilon^{j_1 j_2 j_3}  
 & \left \langle \hat{S}^{i_1 j_1}_{\beta_1 \alpha'_1} (\bar{l}_1;0,x_1) \, 
        \hat{S}^{i_2 j_2}_{\beta_2 \alpha'_2} (\bar{l}_2;0,x_2) \,  
        \hat{S}^{i_3 j_3}_{\beta_3 \alpha'_3} (\bar{l}_3;0,x_3) \right \rangle
\\ & \times
  S^{-1} (p_1)_{\alpha'_1 \alpha_1} S^{-1} (p_2)_{\alpha'_2 \alpha_2} 
  S^{-1} (p_3)_{\alpha'_3 \alpha_3} \,.
\end{split}
\end{equation} 
\end{widetext}
As in the case of the quark-antiquark operators, $S(p)$ denotes the
(color averaged) quark propagator. Propagators with covariant derivatives
acting at $x$ are denoted by $\hat{S}^{ij}_{\alpha \beta}(\bar{l};x,y)$, and 
$\langle \cdots \rangle$ indicates the average over the gauge fields
fixed to Landau gauge.

Since in the present context all quark masses are equal, the propagators
do not need a flavor label. In our setup, the external momenta are
chosen such that 
$p_1^2 = p_2^2 = p_3^2 = (p_1+p_2+p_3)^2 = (p_1+p_2)^2 = (p_1+p_3)^2 = \mu^2$
with the renormalization scale $\mu$.

We write the mixing operator multiplets for a fixed flavor structure
$F^{B,f}_{o2}$ in the form
\begin{equation} 
\cO^{(i)}_m (x) =
F^{B,f}_{o2} S^{(m,i),\bar{l}}_{o2,\alpha} \Psi^f_\alpha (\bar{l};x) 
\end{equation} 
and analogously with $o2$ replaced by $o1$, $s$ or $d$. 
The corresponding vertex functions are given by
\begin{equation} 
\Lambda (\cO^{(i)}_m | p)^f_\alpha = 
\sum_{\pi \in \mathcal S_3} F^{B,f_\pi}_{o2} 
    S^{(m,i),\bar{l}}_{o2,\beta} H^\beta_{\alpha_\pi} (\bar{l};p_\pi) \,.
\end{equation}
The renormalized vertex functions take the form
\begin{equation} 
\Lambda^{\mathrm R} (\cO^{(i)}_m | p)^f_\alpha = 
\sum_{\pi \in \mathcal S_3} F^{B,f_\pi}_{o2}
  \left[ S^{(m,i),\bar{l}}_{o2,\beta} H^\beta_{\alpha_\pi} (\bar{l};p_\pi) 
  \right]^{\mathrm R} \,,
\end{equation}
where 
\begin{equation} 
  \left[ S^{(m,i),\bar{l}}_{o2,\beta} H^\beta_\alpha (\bar{l};p) 
  \right]^{\mathrm R} = \sum_{m'} \hat{Z}_q^{-3/2} \hat{Z}_{m m'} 
  S^{(m',i),\bar{l}}_{o2,\beta} H^\beta_\alpha (\bar{l};p) \,.
\end{equation}

The renormalization and mixing coefficients $\hat{Z}_{m m'}$ are fixed by the
renormalization condition
\begin{equation} 
\begin{split}
&\sum_i \Lambda^{\mathrm R} ( \cO^{(i)}_m | p)^f_\alpha 
\left( \Lambda^{\mathrm {Born}} ( \cO^{(i)}_{m'} | p)^f_\alpha \right)^* \\
& {} = \sum_i \Lambda^{\mathrm {Born}} ( \cO^{(i)}_m | p)^f_\alpha 
\left( \Lambda^{\mathrm {Born}} ( \cO^{(i)}_{m'} | p)^f_\alpha \right)^*
\,,
\end{split}
\end{equation} 
which is analogous to Eq.~(\ref{eq.renco}) for quark-antiquark operators.
The superscript ``Born'' indicates the corresponding tree level expression
(Born term). This is again taken with all lattice artifacts included.

More explicitly our renormalization condition can be written as
\begin{equation} 
\sum_{m''} \hat{Z}_q^{-3/2} \hat{Z}_{m m''} L_{m'' m'} = R_{m m'}
\end{equation} 
with 
\begin{equation} 
\begin{split}
L_{m m'} = \sum_i \sum_{t=1}^2 S^{(m,i),\bar{l}}_{ot,\beta}
  & \left( S^{(m',i),\bar{l}'}_{ot,\beta'} \right)^*
                         H^\beta_\alpha (\bar{l};p) \\
  & \times \left( H^{\beta'}_\alpha (\bar{l}';p)^{\mathrm {Born}} \right)^*
\end{split}
\end{equation} 
and 
\begin{equation} 
\begin{split}
R_{m m'} = \sum_i \sum_{t=1}^2 S^{(m,i),\bar{l}}_{ot,\beta}
  & \left( S^{(m',i),\bar{l}'}_{ot,\beta'} \right)^* 
               H^\beta_\alpha (\bar{l};p)^{\mathrm {Born}} \\
  & \times \left( H^{\beta'}_\alpha (\bar{l}';p)^{\mathrm {Born}} \right)^* \,.
\end{split}
\end{equation} 
For singlet and decuplet one gets analogous equations where, of course,
no sums over $t$ appear. So we have
\begin{equation} 
\hat{Z}_{m m'} = \hat{Z}_q^{3/2} \left( R L^{-1} \right)_{m m'} \,.
\end{equation}

The corresponding (matrices of) renormalization factors leading to
operators renormalized in the $\MS$ scheme are constructed
with the help of (matrices of) conversion factors calculated
in continuum perturbation theory, where we use the particular version of 
the $\MS$ scheme introduced in Ref.~\cite{Kraenkl:2011qb}. Due to the 
complexity of higher-loop calculations we had to limit ourselves to
one-loop accuracy~\cite{Gruber:2017ozo}. Also the anomalous dimensions of our
operators are in general only known to one loop, with the exception of
the operators without derivatives, for which the anomalous dimensions
have been calculated to three loops~\cite{Gracey:2012gx}.

The evaluation of the correlation functions on lattices with periodic
and open boundary conditions as well as the chiral extrapolation proceed
in complete analogy to the case of quark-antiquark operators.
For the external momenta we have taken
\begin{equation} \label{eq.3qsmom}
\begin{split}
p_1 & {} = \frac{\mu}{2} (+1,+1,+1,+1) \,, \\
p_2 & {} = \frac{\mu}{2} (-1,-1,-1,+1) \,, \\
p_3 & {} = \frac{\mu}{2} (+1,-1,-1,-1) \,,
\end{split}
\end{equation}
employing twisted boundary conditions. 

\section{Perturbative subtraction of lattice artifacts} \label{sec.subtraction}

For larger values of the renormalization scale $\mu$ lattice artifacts
will show up. Given the fact that for most operators $Z$ has to 
diverge as $a \to 0$, it is not immediately obvious what one should call 
lattice artifacts in the present context. In order to clarify this point 
it is useful to have a look at the calculation of $Z$ in lattice 
perturbation theory. Evaluating the required correlation functions to 
one-loop order at vanishing quark mass yields results of the form
\begin{equation} \label{eq.Zmua}
Z(\mu,a) = 1 - \frac{g^2}{16 \pi^2}
  \left( (\gamma_0/2) \ln (a^2 \mu^2) + \Delta + F(a^2 \mu^2) \right) \,,
\end{equation}
where $g$ is the bare coupling constant and $F(0)=0$.
What is usually quoted as the one-loop result
from lattice perturbation theory is the above expression with all
contributions that go to zero for $a \to 0$ omitted, i.e.,
\begin{equation}
1 - \frac{g^2}{16 \pi^2}
  \left( (\gamma_0/2) \ln (a^2 \mu^2) + \Delta \right) \,.
\end{equation}
From this point of view the quantity $F(a^2 \mu^2)$ would be considered
as a lattice artifact, which vanishes for fixed $\mu$ like $a^2$ (up to
logarithms) as $a \to 0$~\cite{Constantinou:2009tr}. However, keeping
such contributions (or part of them) should not change the continuum
limit of the renormalized quantities. Nevertheless, it seems to be
generally expected that suppressing the above sort of lattice artifacts
in the renormalization factors would also reduce the lattice artifacts
in the renormalized quantities.

One can realize such a suppression by calculating expressions for the
lattice artifacts in (one-loop) lattice perturbation theory and subtracting
these from the data~\cite{Becirevic:2004ny,Gockeler:2010yr}. Using the above
notation this means subtracting
\begin{equation}
- \frac{g^2}{16 \pi^2} F(a^2 \mu^2) \,.
\end{equation}
We have computed $F(a^2 \mu^2)$ for the quark-antiquark operators
with less than two derivatives in the RI$^\prime$-MOM and
the RI$^\prime$-SMOM schemes (see Appendix~\ref{sec.LPT} for details).

Of course, the bare coupling $g^2$ in this expression could be replaced by
other sensible definitions of the coupling constant such as the boosted
coupling
\begin{equation}
g^2_\Box = \frac{g^2}{\third \mbox{tr} \, U_\Box} \,,
\end{equation}
where $\third \mbox{tr} \, U_\Box$ denotes the average plaquette.
In most cases, both choices of the coupling constant seem to reduce
the discretization effects by about the same amount, though with opposite
signs of the remaining lattice artifacts.

In the following we shall restrict ourselves to the straightforward
case of the bare coupling. A scale-dependent choice of the coupling
has been proposed in Ref.~\cite{Harris:2019bih}.

\section{Chiral extrapolation} \label{sec.chiex}

Aiming at a mass-independent renormalization scheme we should use massless
quarks. However, the boundary conditions that we employ in
our Monte Carlo simulations require massive quarks. Hence we must
extrapolate the results obtained in our simulations with three
mass-degenerate quarks to the chiral limit, where all quark masses vanish. 

At $\beta = 3.34$ there are ensembles with periodic boundary conditions
for three different quark masses available, while for each of the next
three $\beta$ values ($\beta = 3.40, \, 3.46, \, 3.55$) we have ensembles
with periodic boundary conditions for four different masses, see
Table~\ref{tab.ensembles}. On the two finest lattices
($\beta = 3.70, \, 3.85$) we have only two different masses each
at our disposal. All of these ensembles will be used for the chiral
extrapolation.

\begin{figure}[tb]
\includegraphics[width=.45\textwidth]{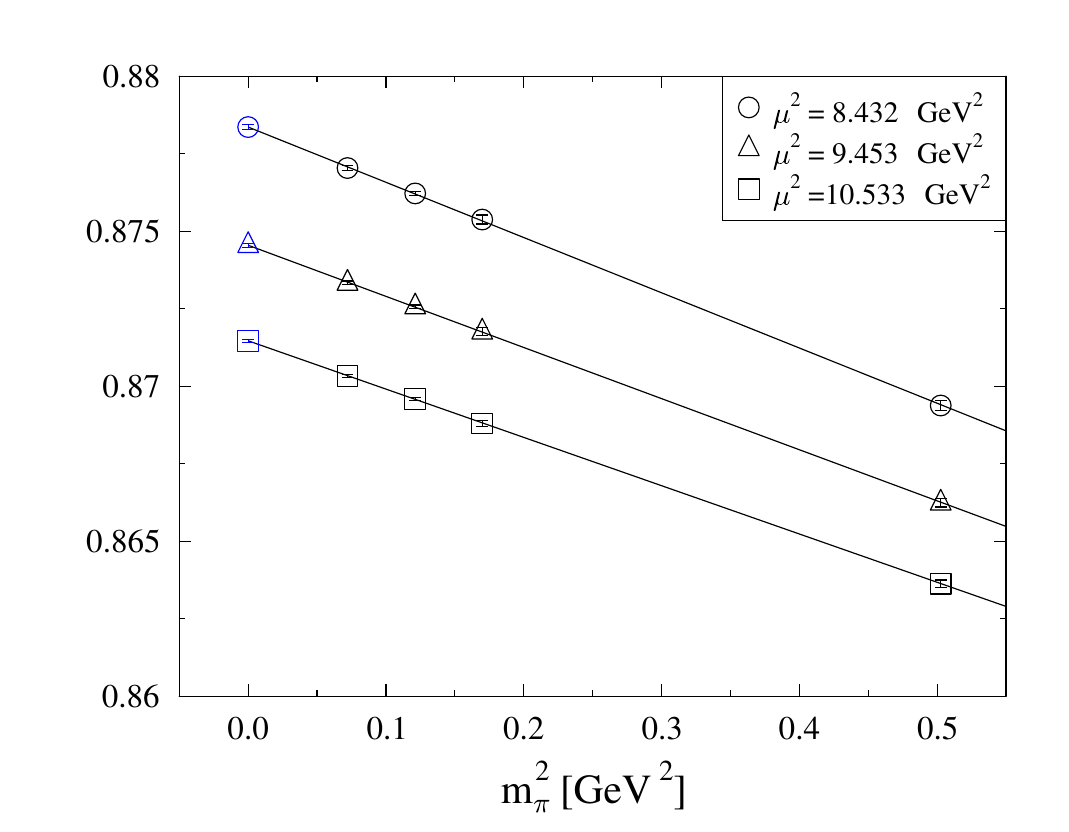}
\caption{\label{fig.chiexloc} Local chiral extrapolation for the tensor
density in the \RI-SMOM scheme at $\beta = 3.55$.}
\end{figure}

One-loop lattice perturbation theory~\cite{Capitani:2000xi} suggests that 
the leading mass dependence is linear in the quark mass. Therefore,
in a first approach, we extrapolate linearly in $m_\pi^2$. On the coarser
lattices we observe a rather mild mass dependence of the vertex functions,
especially for larger scales, so that we are confident that the linear
extrapolations yield reliable results also on the two finest lattices.
The chiral extrapolations are performed at fixed external momenta.
As these depend on the lattice size (see Sec.~\ref{sec.periodicbc}) it 
is in some cases necessary to interpolate between the simulated momenta.
This is done linearly in $\mu^2$. As an example we show the chiral
extrapolation of the tensor density at $\beta = 3.55$ for three scales
in Fig.~\ref{fig.chiexloc}.

On the coarser lattices we could perform additional quadratic
(in $m_\pi^2$)
extrapolations in order to estimate the uncertainties connected with
taking the chiral limit. However, this is not possible on the two finest
lattices. In order to get nevertheless an impression of the importance
of $m_\pi^4$ terms in the full range of $\beta$ we have used the following
procedure as a second approach.

Since the renormalization factors computed in the massless theory suffice to
renormalize also the vertex functions evaluated with nonvanishing quark
masses, it follows that
\begin{equation} 
\hat{Z}(\mu,a,m_\pi) \hat{Z}(\mu,a,0)^{-1}
\end{equation}
has a finite continuum limit $a \to 0$. Here we have indicated the dependence
of the renormalization matrix $\hat{Z}$ on the renormalization scale $\mu$,
the lattice spacing $a$ and the pion mass $m_\pi$ explicitly. Note that 
$\mu$ has to be kept fixed in physical units as $a \to 0$. The existence 
of the continuum limit of $\hat{Z}(\mu,a,m_\pi) \hat{Z}(\mu,a,0)^{-1}$
motivates the following ansatz for the mass dependence:
\begin{equation} 
\begin{split}
\hat{Z}(\mu,a,m_\pi) = \big[1 &+ (b_0(\mu) + b_1(\mu) a) m_\pi^2 \\ 
&+ (c_0(\mu) + c_1(\mu) a) m_\pi^4 \big] \hat{Z}(\mu,a,0) \,.
\end{split}
\end{equation}
Note that the coefficients $b_0$, $b_1$, $c_0$, and $c_1$ are of
mass dimension $-2$, $-1$, $-4$, and $-3$, respectively. 
One-loop lattice perturbation theory~\cite{Capitani:2000xi} reveals
contributions proportional to $a m_q$. Therefore we include terms linear
in $a$ to describe the lattice spacing dependence of
$\hat{Z}(\mu,a,m_\pi) \hat{Z}(\mu,a,0)^{-1}$. Since we have to work at
the same value of $\mu$ for all lattice spacings, it is necessary to
interpolate the scale dependence of the $\hat{Z}$ data. For this
purpose we use cubic splines in $\ln (a^2 \mu^2)$.

If we consider $M$ mixing operator multiplets, the quantities 
$b_0(\mu)$, $b_1(\mu)$, $c_0(\mu)$, $c_1(\mu)$ and $\hat{Z}(\mu,a,0)$
are $M \times M$ matrices. For a given value of $\mu$ we have therefore 
$(4+n_a)M^2$ parameters, if we use data for $n_a$ values of $a$.
In the case of the quark-antiquark operators, where $n_a=6$, these
$10 M^2$ parameters are fitted to $19 M^2$ data points corresponding
to the 19 available combinations of $a$ and $m_\pi$. In the case of the
three-quark operators we do not have data for $\beta = 3.34$ at our
disposal, hence $n_a=5$, and $9 M^2$ parameters must be fitted to
$16 M^2$ data points. For $M>2$ the number of free parameters becomes
prohibitively large, and we have to restrict ourselves to separate
fits for each $\beta$.

\begin{figure}[tb]
\includegraphics[width=.45\textwidth]{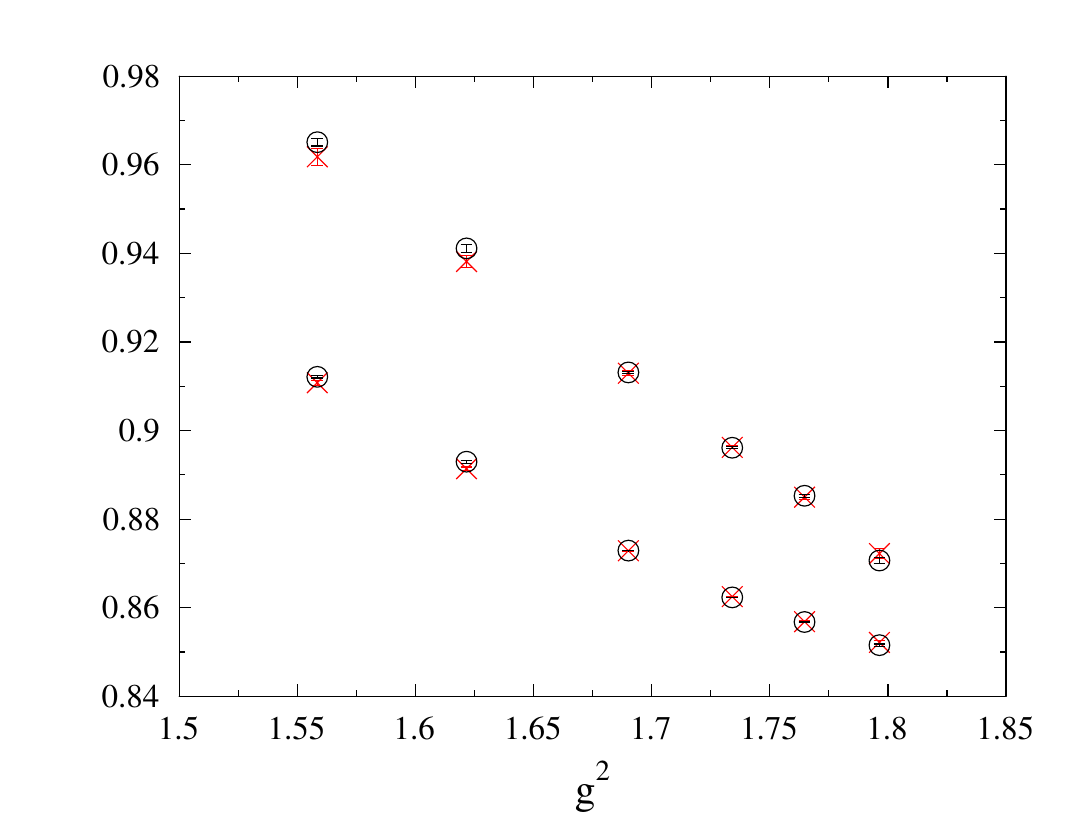}
\caption{\label{fig.chiex} Comparison between the local chiral
extrapolation (black circles) and the global chiral extrapolation (red
crosses) for the tensor density in the \RI-SMOM scheme at the scales
$\mu^2 = 4 \, \mathrm{GeV}^2$ (upper set of data points) and
$\mu^2 = 10 \, \mathrm{GeV}^2$ (lower set of data points).
The extrapolated $\hat{Z}$ values are plotted
against the bare coupling $g^2$.}
\end{figure}

In the following we shall call the first approach
``local chiral extrapolation''
because the extrapolation is performed independently for each $\beta$.
The second approach will be referred to as a
``global chiral extrapolation''.
The latter method addresses the limitation that we do not have the
same coverage of pion masses at all the lattice spacings, while the former
approach has been used in our previous publications. One finds rather
good agreement between the two procedures, which improves for larger scales.
An example is shown in Fig.~\ref{fig.chiex}. Nevertheless
we consider the global chiral extrapolation to be more reliable, because
in this case we expect the extrapolation on the two finest lattices to
benefit from the information on the behavior at small masses, which is
available only on the coarser lattices.

\section{Input from continuum perturbation theory}

Continuum perturbation theory is used for the calculation of the
conversion factors. Moreover, in order to control the scale dependence
of the renormalization factors we need perturbative expressions for
$W(\mu, \mu_0)= Z(\mu) Z^{-1} (\mu_0)$. In both cases the running
coupling $\gc(\mu)$ is required.

The scale dependence of $\gc(\mu)$ is controlled by the $\beta$ function
\begin{equation}
\label{eq.betadef}  
\beta(\gc) = \mu \, d \gc /d\mu \,,
\end{equation}
where the derivative is to be taken at fixed bare parameters and fixed cutoff.
The perturbative expansion of $\beta(\gc)$ can be written as
\begin{equation} \label{eq.betaexpansion}
\beta(\gc) = - \beta_0 \frac{\gc (\mu)^3}{16 \pi^2}
             - \beta_1 \frac{\gc (\mu)^5}{(16 \pi^2)^2} 
             - \beta_2 \frac{\gc (\mu)^7}{(16 \pi^2)^3}  + \cdots \,.
\end{equation}
The values of the coefficients $\beta_0$ - $\beta_4$ in the $\MS$ scheme
are listed in Eqs.~(\ref{eq.beta0}) - (\ref{eq.beta4}).
Integrating Eq.~(\ref{eq.betadef}) one obtains
\begin{equation} \label{eq.grunning}
\begin{split}
\frac{\mu}{\Lambda} =& 
 \left( \frac{\beta_0}{16 \pi^2} {\gc (\mu)^2} \right)
                             ^{\frac{\beta_1}{2 \beta_0^2}}
 \exp \left(\frac{1}{2 \beta_0} \cdot 
              \frac{16 \pi^2}{\gc (\mu)^2} \right) \\
 & \times \exp \left \{ \int_0^{\gc (\mu)} \! d g'
  \left( \frac{1}{\beta (g')} 
     + \frac{1}{\beta_0} \frac{16 \pi^2}{g'^3}
       - \frac{\beta_1}{\beta_0^2} \frac{1}{g'} \right) \right \} 
\end{split}
\end{equation}
with the $\Lambda$ parameter appearing as an integration constant.

The running of the $Z$ matrices is governed by the anomalous dimension matrix
\begin{equation} \label{eq.gammadef}
\gamma (\gc) = - \left( \mu \frac{d Z}{d \mu} \right) Z^{-1} \,,
\end{equation}
whose perturbative expansion reads
\begin{equation} \label{eq.gammaexpansion}
\gamma(\gc) = \gamma_0 \frac{\gc (\mu)^2}{16 \pi^2}
         + \gamma_1 \left( \frac{\gc (\mu)^2}{16 \pi^2} \right)^2
         + \gamma_2 \left( \frac{\gc (\mu)^2}{16 \pi^2} \right)^3 + \cdots \,.
\end{equation}
Again, the bare parameters and the cutoff are kept constant when taking
the derivative. 

We define the quark field renormalization constant $Z_q$ so that the
renormalized quark propagator is obtained from the bare
propagator by multiplication with $Z_q$, i.e., we use the continuum analogue
of Eq.~(\ref{eq.defzq}). For the anomalous dimension of the quark field
we adopt the definition
\begin{equation} \label{eq.qfanodim}
\gamma_q (\gc) = - \mu \frac{d Z_q}{d \mu} Z_q^{-1} \,.
\end{equation}
Collections of coefficients $\gamma_i$ in the $\MS$ scheme can be found
in Appendix~\ref{sec.rgfunc} for our quark-antiquark operators as well
as for the quark field and in Appendix~\ref{sec.anodim3q} for our
three-quark operators.

Taking the derivative with respect to the running coupling $\gc(\mu)$, we get
\begin{equation} \label{eq.rundiff}
\frac{d Z}{d \gc} = - \frac{\gamma (\gc)}{\beta (\gc)} Z \,.
\end{equation}
This system of differential equations can formally be solved in the form
\begin{widetext}
\begin{equation} \label{eq.zmixrunning}
Z(\mu) Z^{-1} (\mu_0) = 
\sum_{n=0}^\infty (-1)^n \int_{\gc(\mu_0)}^{\gc(\mu)} dg_n 
  \int_{\gc(\mu_0)}^{g_n} dg_{n-1} \cdots 
  \int_{\gc(\mu_0)}^{g_2} dg_1 \frac{\gamma(g_n)}{\beta(g_n)} \cdots 
   \frac{\gamma(g_2)}{\beta(g_2)} \frac{\gamma(g_1)}{\beta(g_1)} \,.
\end{equation}
\end{widetext}
In the general case of $M$ mixing multiplets of operators this expression
may be difficult to evaluate, because the $M \times M$ matrices $\gamma (g)$
do not necessarily commute for different values of the coupling $g$.
If there is no mixing ($M=1$) the formula simplifies to
\begin{equation} \label{eq.zrunning}
Z(\mu) Z^{-1} (\mu_0) = \exp \left\{ - \int_{\gc(\mu_0)}^{\gc(\mu)} dg'
\frac{\gamma(g')}{\beta(g')} \right\} \,.
\end{equation}
In analogy to the definition of the $\Lambda$ parameter we can define
the so-called RGI (renormalization group invariant) $Z$, which is
independent of scale and scheme:
\begin{equation} \label{eq.defRGI}
\begin{split}
Z^{\mathrm {RGI}} =&
  \left( 2 \beta_0 \frac {\gc (\mu)^2}{16 \pi^2}\right)
                             ^{-\frac{\gamma_0}{2 \beta_0}} \\
 &\times \exp \left \{ \int_0^{\gc (\mu)} \! d g'
  \left( \frac{\gamma (g')}{\beta (g')} 
   + \frac{\gamma_0}{\beta_0 g'} \right) \right \} Z(\mu) \,.
\end{split}
\end{equation}

Of course, the independence of scale and scheme is strictly realized only
if the $\beta$ function and the anomalous dimension
$\gamma$ are exactly known (including nonperturbative contributions).
In practical applications this is usually not the case, except for the
vector current and the nonsinglet axialvector current, whose anomalous
dimensions are known to vanish exactly. In all the other cases
one must be prepared to encounter violations of the scale and scheme
independence due to truncation errors in the perturbative approximations.

With the help of the methods described in Secs.~\ref{sec.schemes} and
\ref{sec.implement} we can compute $\hat{Z}$, the renormalization matrix
leading from the bare lattice operators to operators renormalized in
the \RI-(S)MOM scheme, for some range of renormalization scales $\mu$. 
In order to evaluate the matrix $Z$ transforming the bare operators into
$\MS$ operators we need in addition the conversion matrix (or conversion
factor if there is no mixing) $C$, cf.\ Eq.~(\ref{eq.match}) in the case
of quark-antiquark operators. This is calculated in continuum
perturbation theory leading to results of the form
\begin{equation} \label{eq.matchexpand}
C(\mu) = 1 + c_1 \frac{\gc (\mu)^2}{16 \pi^2}
           + c_2 \left( \frac{\gc (\mu)^2}{16 \pi^2} \right)^2
           + c_3 \left( \frac{\gc (\mu)^2}{16 \pi^2} \right)^3 + \cdots \,.
\end{equation}
For selected quark-antiquark operators, explicit expressions for the
coefficients $c_1$, $c_2$, {\ldots} are compiled in Appendix~\ref{sec.confac}. 
For three-quark operators, see Appendix~\ref{sec.confac3}.

With the $\beta$ function and the anomalous dimension function $\gamma$
given to some order in perturbation theory, we calculate the integrals in
(\ref{eq.grunning}) and (\ref{eq.zrunning}) analytically if possible or else
by numerical integration. In the case of two mixing multiplets with a
nondiagonal matrix of anomalous dimensions, the integral representation
(\ref{eq.zmixrunning}) of the scale dependence of $Z$ is not very helpful.
Instead of trying to evaluate the expression (\ref{eq.zmixrunning})
we solve the system of differential equations (\ref{eq.rundiff})
with the help of a power series expansion in $\gc$ following
Ref.~\cite{Blumlein:1997em}.

In principle, one has a lot of freedom in choosing the renormalization 
scheme for the anomalous dimension, the coupling constant in the conversion
factor etc., cf.~Ref.~\cite{Gockeler:2010yr}. If there were no truncation
errors, $Z^{\mathrm {RGI}}$ would be independent of all these choices.
In practice, we consider only the $\MS$ scheme. In order to obtain an
estimate of the truncation errors we vary the number of loops taken into
account.

\section{Looking for a window} \label{sec.window}

At this point, the results for the $Z$ factors (extrapolated to the chiral
limit and converted to the $\MS$ scheme) suffer from two problems:
the truncation errors of the perturbative
expressions for the $\gamma$ and $\beta$ functions and the conversion
matrix (factor) $C$ on the one side and the lattice artifacts on the other side.
The former grow as $\mu$ becomes smaller, while the latter increase as
$\mu$ gets larger. Ideally, one would like to find a window, i.e., a
$\mu$ interval where both errors are negligible. This would require
\begin{equation} 
  1/L^2 \ll \Lambda^2_{\mathrm {QCD}} \ll \mu^2 \ll \pi^2/a^2 
\end{equation}
for a lattice of linear extent $L$. Then lattice artifacts would be
negligible and the scale dependence could be described by low-order
continuum perturbation theory. In such a window the
results for $Z^{\mathrm {RGI}}$ would be independent of the scale $\mu$,
i.e., when plotted against $\mu$ they would show a plateau, and the
final value for $Z^{\mathrm {RGI}}$ could be read off at a value $\mu$
within this window.

Unfortunately, such an ideal situation is hard to achieve 
at the present lattice spacings. Although in recent years there has
been considerable progress in the perturbative evaluation of anomalous
dimensions and conversion factors, the convergence of the perturbative
expansions, in particular for the conversion factors, is not always
satisfactory. Therefore the results for $Z^{\mathrm {RGI}}$ will
usually not be independent of $\mu$ even for values
$\mu \approx 2 \, \mathrm {GeV}$.

For increasing values of $\mu$ lattice artifacts will become larger.
Nevertheless, the continuum limit of renormalized quantities remains
in principle well-defined also for higher renormalization scales. Therefore
one can simply use $Z$ evaluated at some convenient, fixed scale $\mu$
and hope that the continuum limit is under control. However, one may
expect that suppressing the lattice artifacts in the $Z$ factors would also
make the behavior of the renormalized quantities for $a \to 0$ more benign.

One possibility to realize such a suppression consists in fitting the
$Z^{\mathrm {RGI}}$ data with a suitable ansatz for the lattice artifacts.
Since the values of $p^2 = a^2 \mu^2$ actually appearing in the data
cover only a finite interval which does not extend down to 0, a
polynomial in $p^2$ would be a reasonable choice.

Alternatively one can calculate expressions for the lattice artifacts
in lattice perturbation theory and subtract these from the
data as explained in Sec.~\ref{sec.subtraction}. While this procedure
reduces the scale dependence of the $Z^{\mathrm {RGI}}$ data already
substantially, some lattice artifacts will in general remain, which
can still be fitted.

Lattice artifacts can be studied in a much clearer way for quantities
that have a decent continuum limit. This is not the case for the $Z$
factors themselves, but for ratios $Z(\mu) Z^{-1} (\mu_0)$ of
renormalization factors (matrices) at different
renormalization scales~\cite{Arthur:2010ht}.
Therefore these ratios offer a possibility to investigate the impact of
discretization effects in our calculations. After performing a
continuum limit we can see how well the scale dependence is described by
continuum perturbation theory.
Such investigations have already been performed for various operators
in a number of different settings, see, e.g.,
Refs.~\cite{Durr:2010aw,Aoki:2010pe,Arthur:2012opa}.
In this way we should be able, at least for
these quantities, to disentangle lattice artifacts and truncation errors,
which otherwise are hard to separate unambiguously.

\begin{figure}
\includegraphics[width=.45\textwidth]{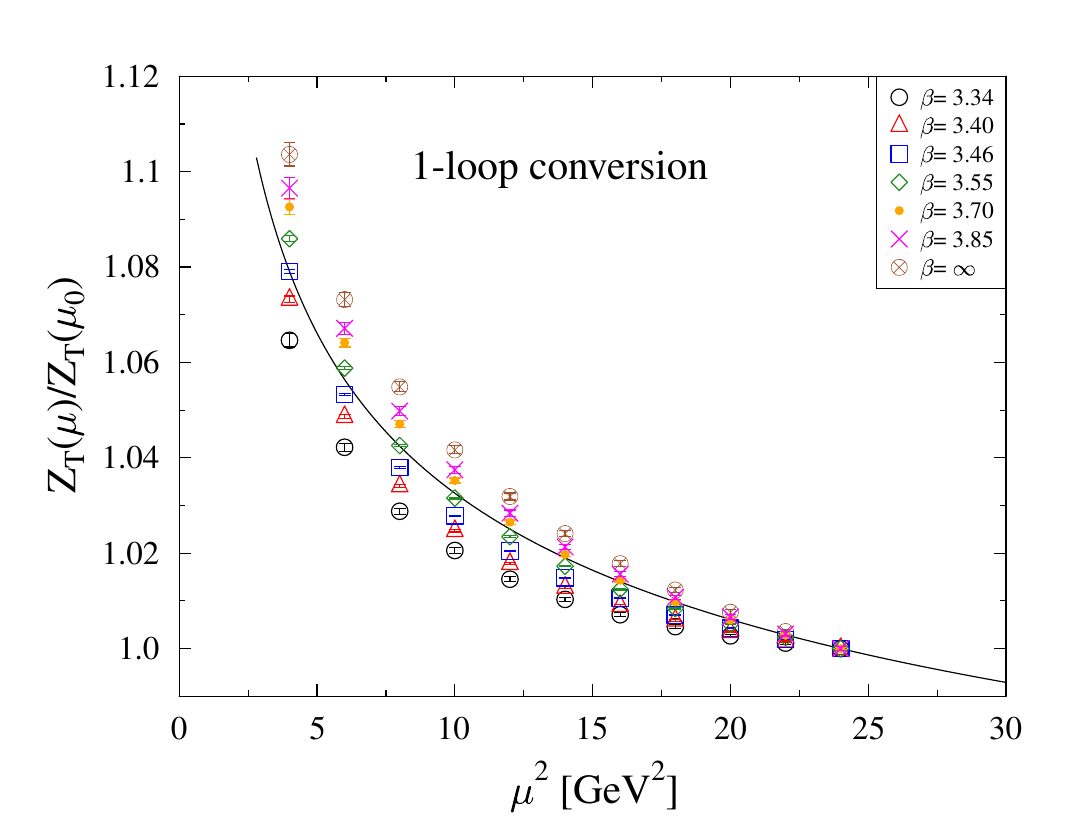}
\includegraphics[width=.45\textwidth]{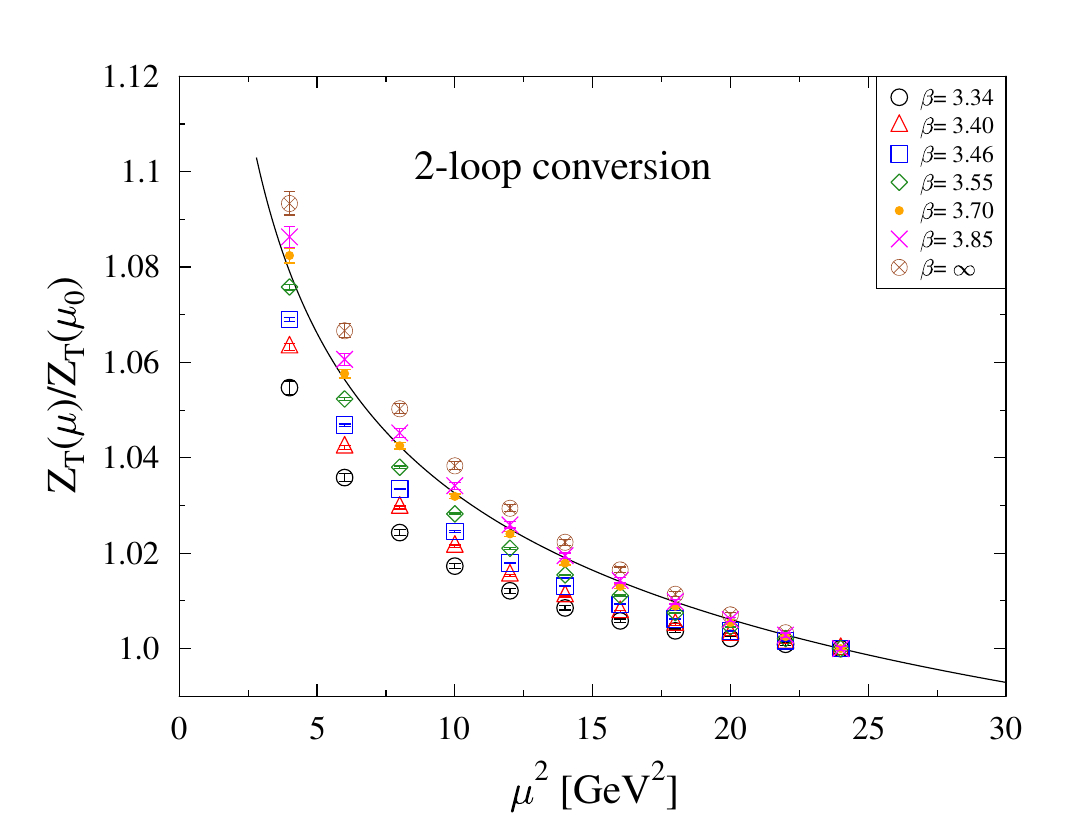}
\includegraphics[width=.45\textwidth]{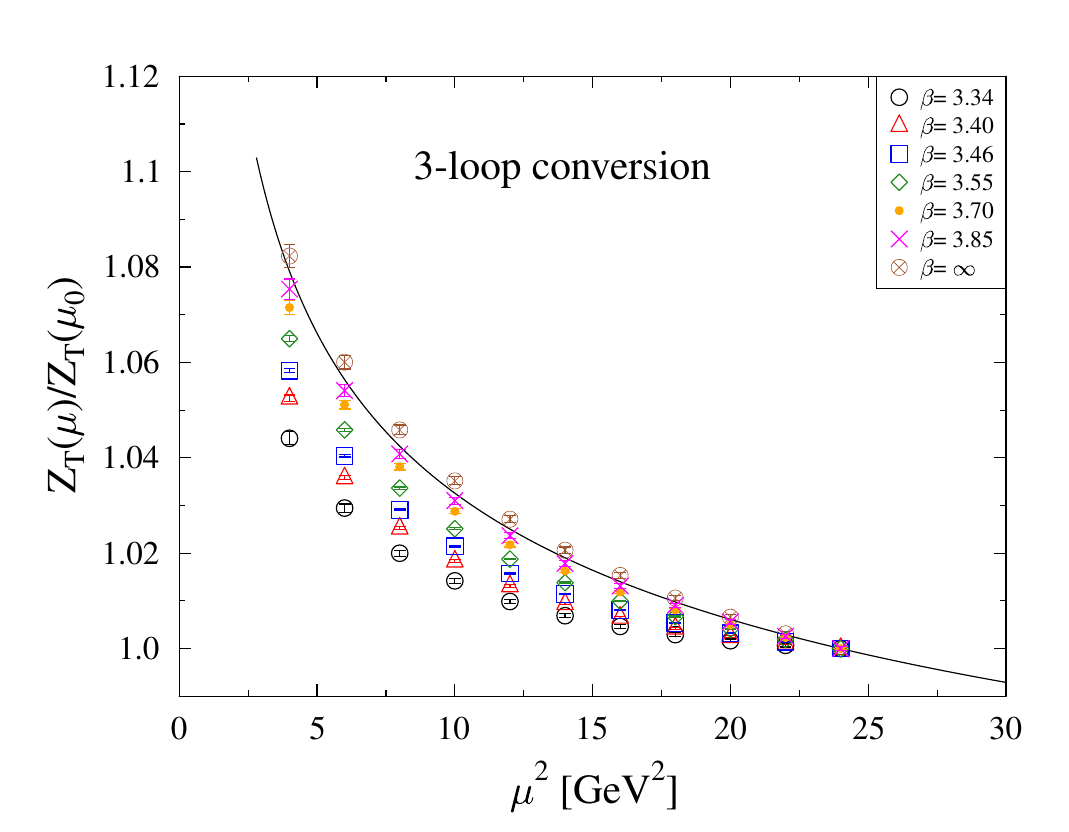}
\caption{\label{fig.tensumrun} Ratios of the renormalization factor of
the tensor density evaluated in the \RI-SMOM scheme at different scales
with the scale in the denominator fixed at $\mu_0^2 = 24 \, \mathrm {GeV}^2$.
The nonperturbative results at finite $\beta$ have been computed in the
\RI-SMOM scheme, lattice artifacts have been subtracted perturbatively, and
a global chiral extrapolation has been performed. Finally, the values have
been converted to the $\MS$ scheme using one-loop, two-loop and three-loop (from
top to bottom) perturbation theory and extrapolated to the continuum
limit ($\beta = \infty$). The curves show the behavior calculated
in three-loop perturbation theory in the $\MS$ scheme.}
\end{figure}

We show $Z(\mu) Z^{-1} (\mu_0)$ for the tensor density in
Fig.~\ref{fig.tensumrun} comparing results obtained with
conversion factors at different orders in perturbation theory.
The statistical errors of the ratios have been calculated from the
errors of numerator and denominator by means of error propagation.
For the continuum extrapolation we have employed a second-order
polynomial in $a^2$. The agreement with the perturbative scale
dependence shown by the curves improves significantly as we go from
one-loop to three-loop conversion factors.
In general the perturbative series for the conversion factors is less well
behaved than the expansions of the anomalous dimensions, which tend to
converge quite fast. When plotting $Z^{\mathrm {RGI}}$ against $\mu$ we
must therefore be prepared to find visible deviations from a plateau
also for moderate values of $\mu$, where lattice artifacts should be small.

\begin{figure}
\includegraphics[width=.45\textwidth]{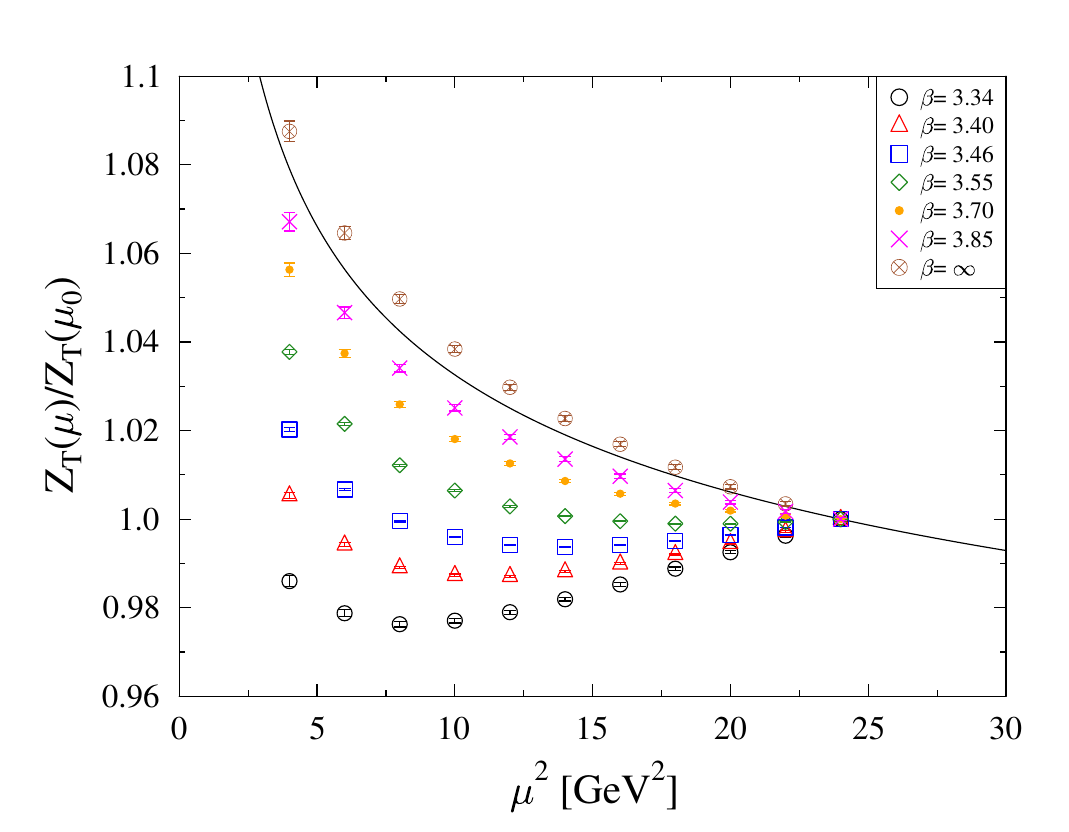}
\caption{\label{fig.tensumrun2} The same as the lowest panel in 
Fig.~\ref{fig.tensumrun}, but without the perturbative subtraction of
lattice artifacts.}
\end{figure}

If the one-loop lattice artifacts are not subtracted, the agreement with the
perturbative running is less satisfactory, even if the three-loop conversion
factor is used. For the tensor density this is demonstrated in
Fig.~\ref{fig.tensumrun2}, which should be compared with the lowest panel
in Fig.~\ref{fig.tensumrun}. 

Particularly good agreement with the perturbative running is observed
for the pseudoscalar density as Fig.~\ref{fig.g5run} shows. In this case
we can even use the five-loop expression for the anomalous dimension (see
Appendix~\ref{sec.rgfunc}), but the difference with the three-loop results
would be hardly visible in the figure.

\begin{figure}
\includegraphics[width=.45\textwidth]{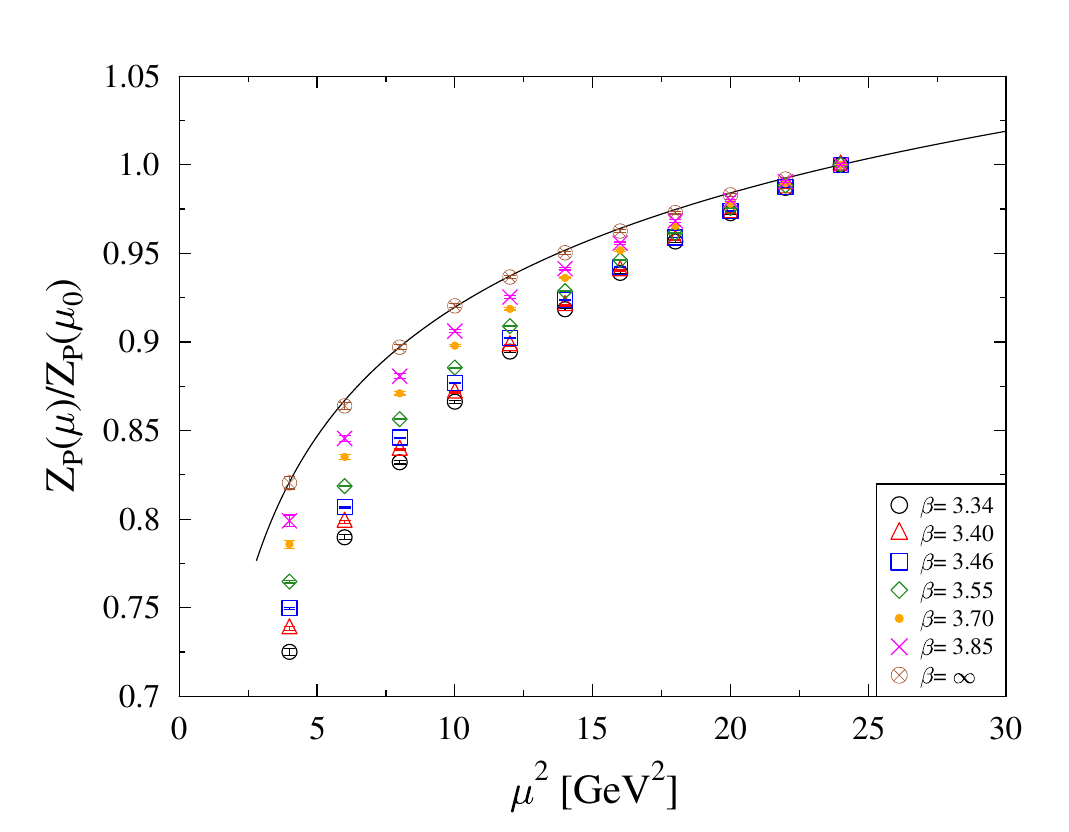}
\caption{\label{fig.g5run} The same as the lowest panel in 
Fig.~\ref{fig.tensumrun}, but for the pseudoscalar density and with the
$\MS$ curve computed in five-loop perturbation theory.}
\end{figure}

\begin{figure}
\includegraphics[width=.45\textwidth]{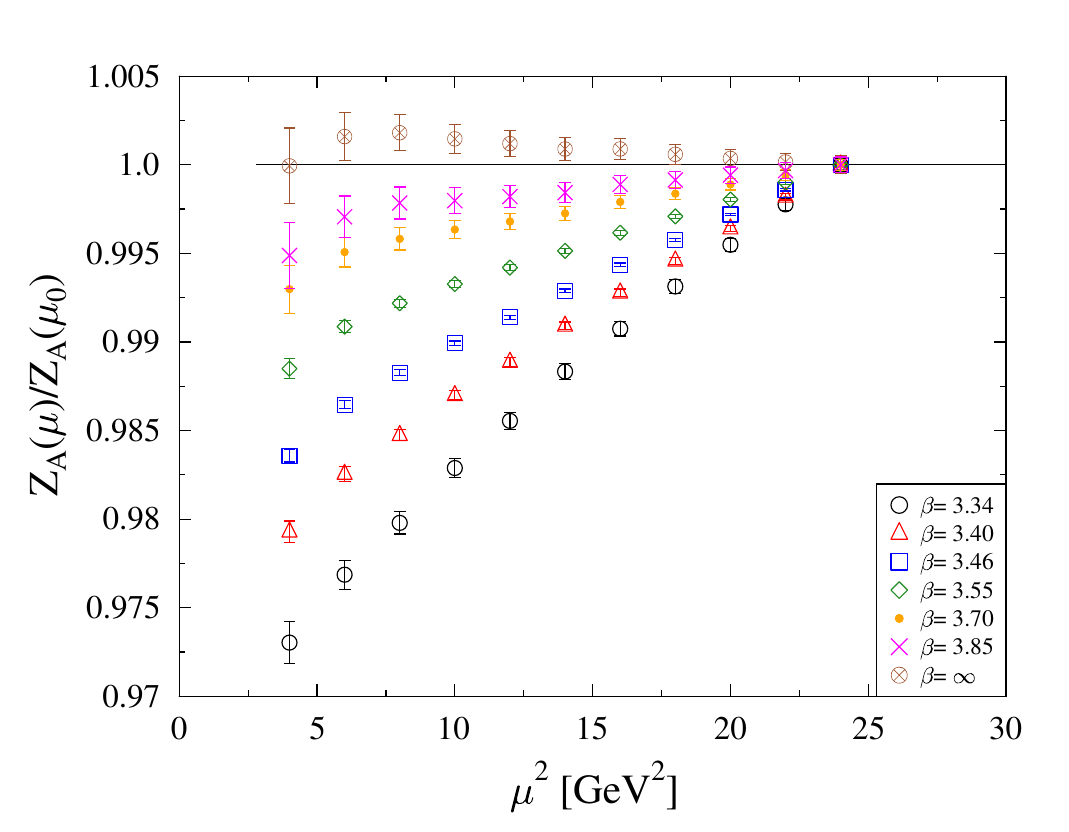}
\caption{\label{fig.vasumrun} The same as the lowest panel in 
Fig.~\ref{fig.tensumrun}, but for the axialvector with the renormalization
condition (\ref{eq.renco}).}
\end{figure}

In some cases the anomalous dimension is even known exactly, e.g.,
for (partially) conserved currents. For the axialvector current the
scale dependence, which we find from our nonperturbative calculations,
is displayed in Fig.~\ref{fig.vasumrun}, where the renormalization
condition (\ref{eq.renco}) has been employed. Using instead
Eq.~(\ref{eq.axvecwi}) leads to a less satisfactory agreement with the
expectations, although the conversion factor is exactly equal to 1 in
this case so that all sources of truncation errors disappear. While
in Fig.~\ref{fig.vasumrun} the deviation from 1 in the continuum limit
remains below 0.002, it reaches about 0.007 in the same range of scales
when Eq.~(\ref{eq.axvecwi}) is applied.

\begin{figure}
\includegraphics[width=.45\textwidth]{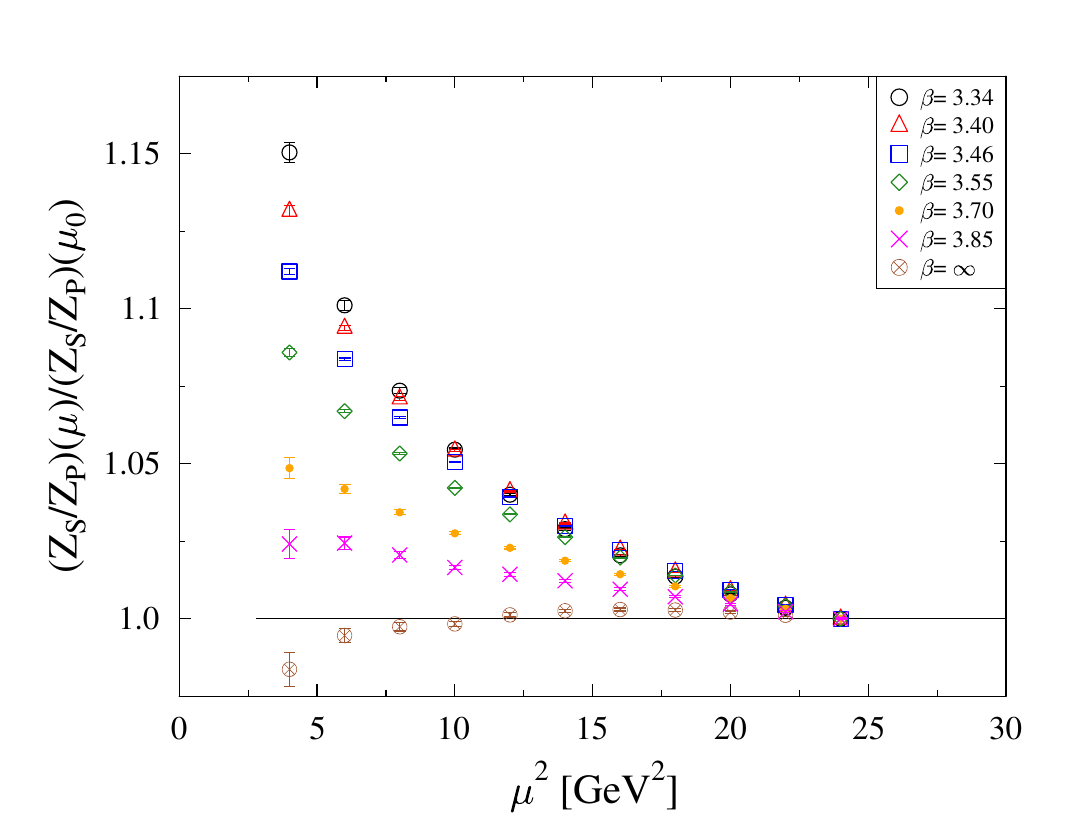}
\caption{\label{fig.zszprun} The same as the lowest panel in 
Fig.~\ref{fig.tensumrun}, but for the ratio $Z_S/Z_P$.}
\end{figure}

Another case where truncation errors are absent in the anomalous dimension
as well as in the conversion factor is given by the ratio $Z_S/Z_P$, see
Fig.~\ref{fig.zszprun}.

\begin{figure}
\includegraphics[width=.45\textwidth]{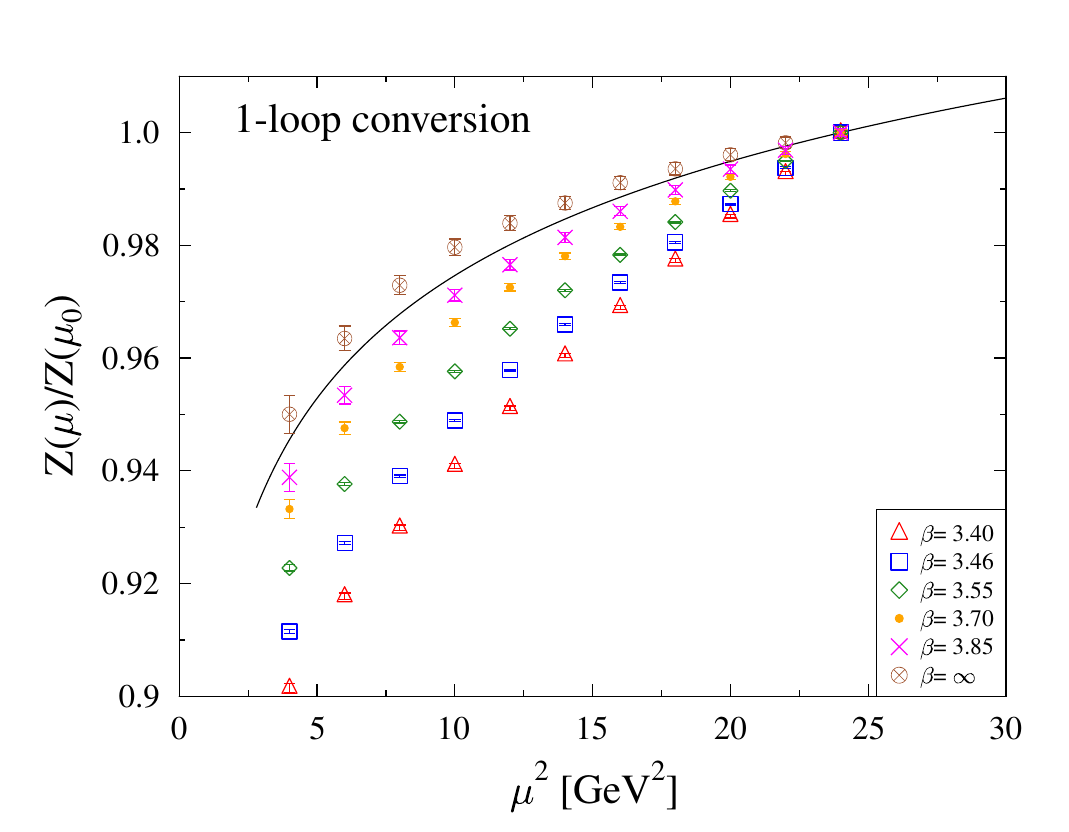}
\caption{\label{fig.3qoprun} Ratios of the renormalization factor of
the three-quark operator~(\ref{eq.s4.1}) evaluated in the \RI-SMOM
scheme at different scales with the scale in the denominator fixed
at $\mu_0^2 = 24 \, \mathrm {GeV}^2$. The nonperturbative results at
finite $\beta$ have been computed in the \RI-SMOM scheme and 
a global chiral extrapolation has been performed. Finally, the values have
been converted to the $\MS$ scheme using one-loop perturbation theory
and extrapolated to the continuum limit ($\beta = \infty$). The curve
shows the behavior calculated in three-loop perturbation theory in the
$\MS$ scheme.}
\end{figure}

For the three-quark operators the conversion matrices are only
known to one-loop accuracy, see Appendix~\ref{sec.confac3}. As an
example we consider the operator~(\ref{eq.s4.1}) and compare in   
Fig.~\ref{fig.3qoprun} $Z(\mu)/Z(\mu_0)$ with the scale dependence
obtained in three-loop perturbation theory. As above, we have used a
second-order polynomial in $a^2$ for the continuum extrapolation
of $Z(\mu)/Z(\mu_0)$.

\begin{figure}
\includegraphics[width=.45\textwidth]{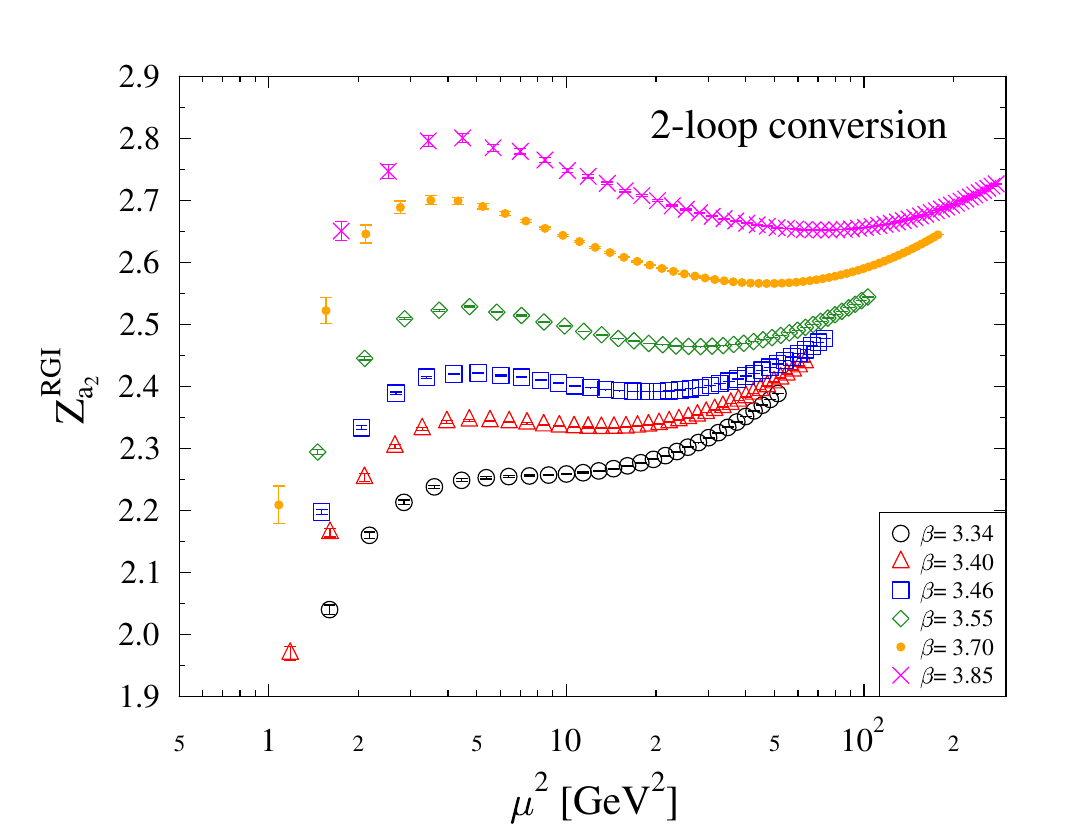}
\includegraphics[width=.45\textwidth]{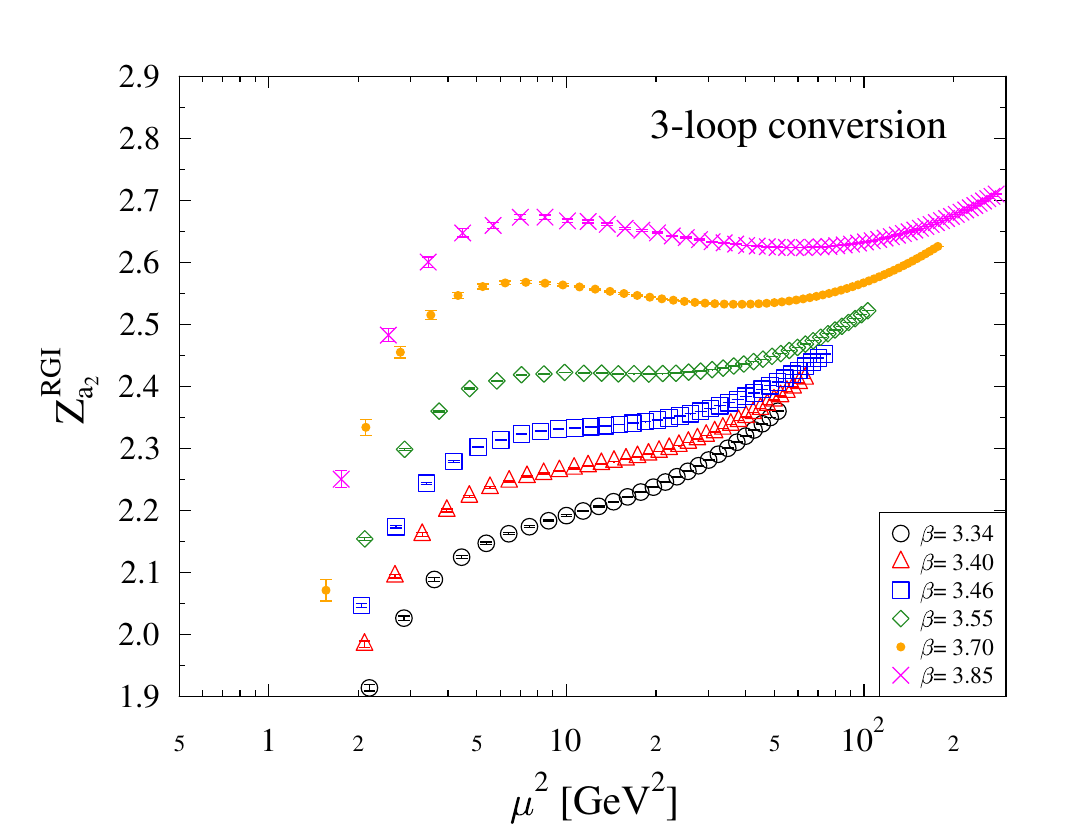}
\caption{\label{fig.a2_rgi} $Z^{\mathrm {RGI}}$ for $\cO_{a_2}$
computed from the locally chirally extrapolated \RI-MOM results with the
help of the two-loop (top panel) and the three-loop (bottom panel) conversion
factor. In both cases the five-loop anomalous dimension has been used.}
\end{figure}

As the figures show, there may be significant differences between the
perturbative predictions for the scale dependence and the values of
$Z(\mu) Z^{-1} (\mu_0)$ even after an extrapolation to the continuum limit.
When plotting $Z^{\mathrm {RGI}}$ rather than these ratios we should
therefore expect to see deviations from a constant,
not only due to the truncation of the perturbative series but
also due to lattice artifacts. As an example we show in
Fig.~\ref{fig.a2_rgi} $Z^{\mathrm {RGI}}_{a_2}$ computed with
the help of the two-loop and the three-loop conversion factors
from the \RI-MOM results. Using the locally chirally extrapolated
data for the purpose of these plots enables us to reach rather large
scales for the higher $\beta$ values, where the lattice artifacts
are particularly pronounced. 

Indeed for most of the curves there is no interval in $\mu^2$ where 
a plateau can be seen. There are several reasons for this behavior.
At very small scales finite size effects might not be negligible and
perturbation theory will probably break down completely. Moreover, the
chiral extrapolation is less stable in this region. Beyond 2 or 3 GeV$^2$
truncation effects compete with lattice artifacts. Both effects tend to
increase the $Z^{\mathrm {RGI}}$ values (at least in the case at hand).
The truncation effects are independent of $\beta$, decrease as a function
of $\mu^2$ and get smaller when higher orders in the perturbative expansion
are taken into account. The effect of changing the loop order of the
conversion factor can be seen by comparing the top and the bottom panels
in Fig.~\ref{fig.a2_rgi}. In both plots the scale dependence in the
$\MS$ scheme has been computed from the five-loop results for the anomalous
dimension and the $\beta$ function (see Appendix~\ref{sec.rgfunc}).
Using instead the four-loop approximation would only lead to hardly visible
changes.

The discretization artifacts, on the other hand, depend only on
$a^2 \mu^2$ for fixed $\beta$, being proportional to $a^2 \mu^2$ in a
first approximation. They increase as a function of $\mu^2$, but decrease
at a given $\mu^2$ when $\beta$ gets larger. The combination of both
effects produces a structure in the data, which moves to higher scales
and eventually becomes a minimum as $a$ gets smaller.
Depending on the loop order there may appear
``fake plateaus'', e.g., for $\beta = 3.34$ in the two-loop case or for
$\beta = 3.55$ when the three-loop conversion factor is employed.

\begin{figure}
\includegraphics[width=.45\textwidth]{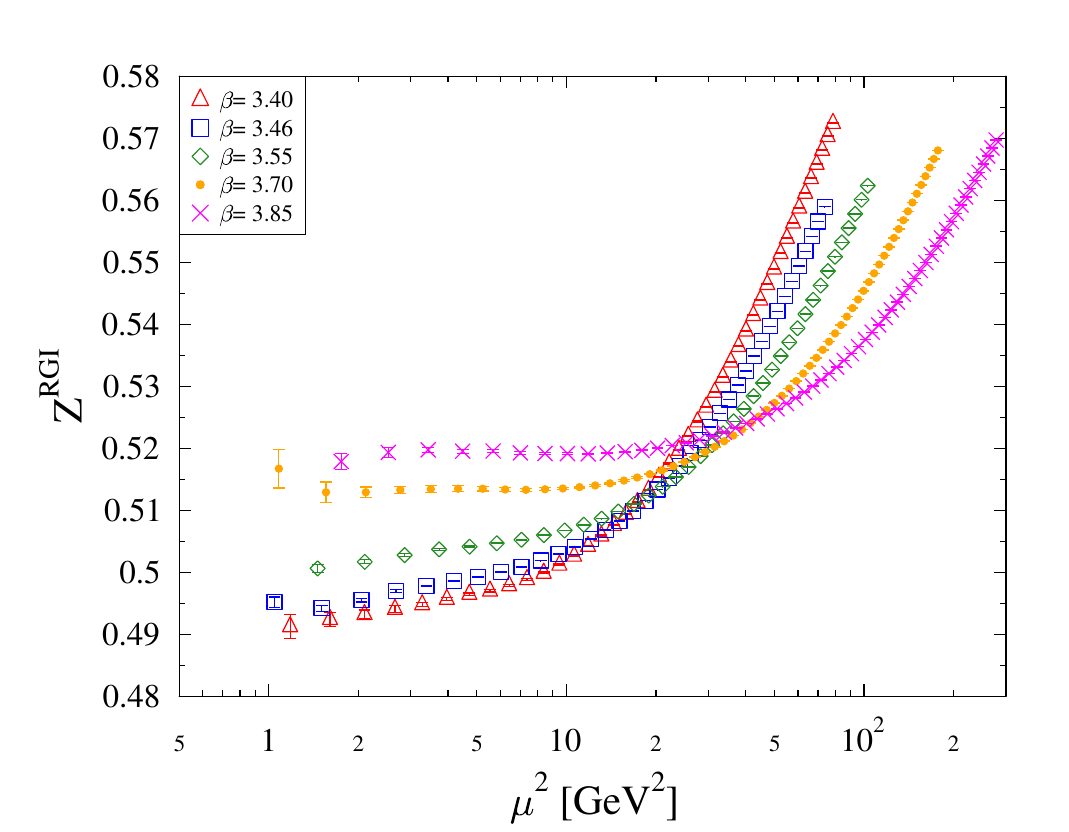}
\caption{\label{fig.3qop_rgi} $Z^{\mathrm {RGI}}$ for the
three-quark operator~(\ref{eq.s4.1}) computed from the locally chirally
extrapolated \RI-SMOM results with the help of the one-loop conversion
factor. For the anomalous dimension the three-loop approximation
has been used.}
\end{figure}

A plot of $Z^{\mathrm {RGI}}$ as a function of $\mu^2$ for the 
three-quark operator~(\ref{eq.s4.1}) is shown in Fig.~\ref{fig.3qop_rgi}.

\section{Towards final results}

In this section we present results for the renormalization and
mixing coefficients obtained by two different methods. The first method,
employed in our previous papers, e.g.,
Refs.~\cite{Braun:2016wnx,Bali:2019dqc,Bali:2019ecy}, makes use of fits
of the scale dependence. We call it the fit method in the following.
In the second method we evaluate the renormalization
factors at some suitable fixed scale and evolve them perturbatively to the
desired reference scale. This method will be referred to as the fixed-scale
method.

In the fit method we try to exploit as much of the available
nonperturbative information as possible by performing a joint fit of
the $\mu$-dependence of the (chirally extrapolated) renormalization
matrices $Z(\mu,a)_{\mathrm {MC}}$ for several $\beta$ values.
As explained above, the fit should describe (and hence suppress) the
lattice artifacts vanishing like a power of $a$.

The choice of the fitting procedure is motivated by the following 
considerations. From the matrix of anomalous dimensions (evaluated for
as many loops as possible) one can calculate a corresponding approximation
of $W(\mu, \mu_0) = Z(\mu) Z^{-1} (\mu_0)$, which should describe the
$\mu$-dependence for sufficiently large scales $\mu$ if there were no
discretization effects. As our renormalization conditions respect the
hypercubic symmetry of our lattices, the discretization artifacts 
must be H(4) invariant functions of the momenta defining the
renormalization scheme. Invariant polynomials in the momentum components
are, e.g., given by 
\begin{equation}
\sum_\mu p_\mu^2 \,, \; \sum_\mu q_\mu^2 \,, \; \sum_\mu p_\mu q_\mu \,, \;
  \sum_\mu p_\mu^4 \,, \; \sum_\mu p_\mu q_\mu^3 \,, \ldots
\end{equation}
Hence an ansatz for the description of the lattice artifacts can be
constructed from terms like
\begin{equation}
a^2 p^2 \,, \; \frac{a^2}{p^2} \sum_\mu p_\mu^4 \,, \;
  \frac{a^4}{p^2} \sum_\mu p_\mu^6 \,, \ldots
\end{equation}
see, e.g., Refs.~\cite{Boucaud:2003dx,deSoto:2007ht,Constantinou:2013ada}.
For the momenta $\hat{p}$, $\hat{q}$, {\ldots} chosen in our renormalization
conditions (see Eqs.~(\ref{eq.rimom}), (\ref{eq.rismom}),
and (\ref{eq.3qsmom})) all these terms reduce to powers of $a^2 \mu^2$.

Adding such an ansatz for the effective description of the lattice
artifacts, we arrive at the following fit function for the matrices
$Z(a,\mu)_{\mathrm {MC}}$:
\begin{equation} \label{eq.fitfun}
\begin{split}
Z(\mu,a)_{\mathrm {MC}} = {} & W(\mu, \mu_0) Z(\mu_0,a) + A_1 a^2 \mu^2 
  + A_2 (a^2 \mu^2)^2 \\
& + A_3 (a^2 \mu^2)^3 \,.
\end{split}
\end{equation}
The fit parameters are the entries of the renormalization matrices
$Z(\mu_0,a)$ at the reference scale $\mu_0$ and the entries of the three
matrices $A_i$ parametrizing the lattice artifacts. Performing a joint
fit for several $\beta$ values we neglect a possible dependence of $A_i$
on the gauge coupling $g^2$, which we only vary by 15\%.

Representing $W(\mu, \mu_0)$ as $\Delta Z (\mu)^{-1} \Delta Z (\mu_0)$
with $\Delta Z (\mu) = Z^{\mathrm {RGI}} Z^{-1} (\mu)$ we can write
\begin{equation} 
\begin{split}
Z(\mu,a)_{\mathrm {MC}} = {} & \Delta Z(\mu)^{-1} Z^{\mathrm {RGI}}(a)
  + A_1 a^2 \mu^2 + A_2 (a^2 \mu^2)^2 \\
& + A_3 (a^2 \mu^2)^3 \,.
\end{split}
\end{equation}
Therefore one has the equivalent options of using either $Z(\mu_0,a)$
or $Z^{\mathrm {RGI}}(a)$ as fit parameter. In the following we shall
give our results for $Z(\mu_0,a)$ with $\mu_0^2 = 4 \, \mathrm {GeV}^2$,
because these are more immediately useful, in particular in the
presence of mixing.

\begin{table} 
\caption{\label{tab.fits} Choices for the fits in the case
  $n_{\mathrm {loops}}^{\mathrm {max}} = 3$. Further explanations
  are given in the text.}
\begin{ruledtabular}
\begin{tabular}{ccccccc}
Fit & $\chi$ & $\mu_1^2 \, [\mbox{GeV}^2]$ & $n_{\mathrm {disc}}$ &
      $\lambda^2_{\mathrm {scale}}$ & $\Lambda_{\MS} \, [\mbox{MeV}]$ &
      $n_{\mathrm {loops}}$ \\ \hline
1 & g &  4 & $3$ & 1.0  & 341 & 3 \\
2 & g & 10 & $3$ & 1.0  & 341 & 3 \\
3 & g &  4 & $2$ & 1.0  & 341 & 3 \\
4 & g &  4 & $3$ & 1.03 & 341 & 3 \\
5 & g &  4 & $3$ & 1.0  & 353 & 3 \\ 
6 & g &  4 & $3$ & 1.0  & 341 & 2 \\
7 & l &  4 & $3$ & 1.0  & 341 & 3 
\end{tabular}
\end{ruledtabular} 
\end{table}

The fluctuations at different scales are correlated. Hence we estimate
the statistical uncertainty of the fit result $Z(\mu_0,a)$ by the
statistical error of the closest (in $\mu$) data point. In most
cases this procedure leads to errors that are still quite small.

The systematic uncertainties are much more important.
In order to estimate them we perform a number of fits varying exactly one
element of the analysis at a time, see Table~\ref{tab.fits}.
We choose two values of the lower limit $\mu_1$ of the fit range.
For the parametrization of the discretization artifacts we either take the
complete expression in Eq.~(\ref{eq.fitfun}) with $n_{\mathrm {disc}}=3$
terms or we set $A_3=0$ corresponding to $n_{\mathrm {disc}}=2$.
The uncertainty due to the scale setting is taken into account by
multiplying the values of $1/a^2$ given in Table~\ref{tab.spacings}
by $\lambda^2_{\mathrm {scale}}= 1.03$. This value contains the scale
uncertainty of~$8t_0^*=\left( \mu_{\mathrm{ref}}^{*} \right)^{-2}$ given in
Ref.~\cite{Bruno:2017gxd} and the largest error of our determination
of~$t_0^*/a^2$, added in quadrature. Also $\Lambda_{\MS}$ is varied
within its uncertainty~\cite{Bruno:2017gxd}. In order to estimate
the error caused by the truncation of the perturbative expansion of the
conversion factors we reduce the number of loops
$n_{\mathrm {loops}}$ used in the calculation of the conversion factors 
by one, compared to the maximal value $n_{\mathrm {loops}}^{\mathrm {max}}$
that is available. The size of the truncation error
is taken to be one half of the resulting difference, because going, e.g.,
from two loops to three or more loops in the perturbative expansion
is expected to lead to a smaller change than going from one loop to two
loops, at least for sufficiently large scales. 
For the quark-antiquark operators we have 
$n_{\mathrm {loops}}^{\mathrm {max}} = 3$ in the \RI-MOM scheme,
while in the \RI-SMOM scheme we have to be satisfied with
$n_{\mathrm {loops}}^{\mathrm {max}} = 2$
for the tensor operators with derivatives.
The symmetry properties of our operator multiplets can be found in
Table~\ref{tab.representations} in Appendix~\ref{sec.opmulti}, and
numerical values of the coefficients in the perturbative expansion
of the conversion factors are given in Table~\ref{tab.confac}
in Appendix~\ref{sec.confac}. For the three-quark operators
only one-loop results for the conversion factors are available so that
in these cases $n_{\mathrm {loops}}^{\mathrm {max}} = 1$.

Another source of systematic error is the chiral extrapolation, all the
more as on the two finest lattices we have only two different masses at
our disposal. We try to estimate the related uncertainty by comparing
results obtained by means of the local and the global chiral extrapolation,
cf.~Sec.~\ref{sec.chiex}. In the column of Table~\ref{tab.fits}
labeled by $\chi$ these two choices are indicated by the letters g and l
for the global and the local chiral extrapolation, respectively.  
It should however be noted that the upper limits
of the fit ranges differ for data which have been locally or globally
extrapolated. In both cases we include all available data points with
$\mu \geq \mu_1$ in the fit. In the case of the local chiral extrapolation
the upper limit is therefore determined by the largest momenta considered.
These have an approximately fixed value in lattice units corresponding to
$a^2 \mu^2 \approx 10$ in the \RI-MOM scheme and 
$a^2 \mu^2 \approx 5$ in the \RI-SMOM scheme for the quark-antiquark
operators and $a^2 \mu^2 \approx 11$ for the three-quark operators.
The global chiral extrapolation, on the other hand, requires the same scale
in physical units for all $\beta$ values. Thus the upper limit of the fit
range is essentially determined by the largest scale available at
the smallest $\beta$. In the case of the quark-antiquark operators,
where this is $\beta = 3.34$, the fit range extends up to
approximately $\mu^2 = 50 \, \mathrm {GeV}^2$
in the \RI-MOM scheme and $\mu^2 = 25 \, \mathrm {GeV}^2$ in the
\RI-SMOM scheme. For the three-quark operators we do not have data
at $\beta = 3.34$, and the fit range for the global chiral extrapolation
extends up to approximately $\mu^2 = 75 \, \mathrm {GeV}^2$.

As our central value we take the outcome of fit 1 with the statistical
error determined as explained above. The six differences due to the
discussed systematic uncertainties are added in quadrature to yield our
final estimate of the systematic error. In most cases, the error due to
the truncation of the perturbative expansion of the conversion factor
is the largest contribution to the total systematic uncertainty.

The results obtained in this way, based on the \RI-MOM scheme, are collected
in Tables~\ref{tab.momfit1}--\ref{tab.momfit3}, see Appendix~\ref{sec.tables}.
Table~\ref{tab.momfit1} (\ref{tab.momfit2}) contains results for
operators with less than two derivatives obtained without (with)
the perturbative subtraction of lattice artifacts. Here $Z^\prime_V$ and
$Z^\prime_A$ have been determined with the help of the renormalization
conditions (\ref{eq.vecwi}) and (\ref{eq.axvecwi}), respectively, while
for $Z_V$ and $Z_A$ the standard definition (\ref{eq.renco}) has been used.
In Table~\ref{tab.momfit3} we present the $Z$ factors for operators
with two derivatives, for which the perturbative subtraction of lattice
artifacts is not yet available.

Ideally, $Z$ factors determined with the help of the fit method
should agree within errors, whether or not lattice artifacts have been
perturbatively subtracted. A comparison of Tables~\ref{tab.momfit1}
and \ref{tab.momfit2} shows to which extent this expectation is fulfilled.
Note that the fit method, which tries to suppress discretization effects
as far as possible, is quite close in spirit to what is done in
lattice perturbation theory, where (power-like) lattice artifacts are
completely eliminated, cf.\ Sec.~\ref{sec.subtraction}.

The fixed-scale method for the determination of the renormalization and
mixing factors proceeds as follows. We first interpolate the chirally
extrapolated data by cubic splines in $\ln (a^2 \mu^2)$.
Using this interpolation we can read off $Z$ and its statistical error
at some scale $\mu^2$, which we choose to be $\mu^2 = 10 \, \mathrm {GeV}^2$,
and evolve the result perturbatively to the desired reference scale
$\mu_0^2 = 4 \, \mathrm {GeV}^2$. Thus the statistical error stems from
a higher scale than in the fit method and is therefore smaller.
Again, the systematic uncertainties are usually considerably larger than
the statistical error. The uncertainties due to the scale setting and
the error in $\Lambda_{\MS}$ are taken into account in the same way
as in the fit method, cf.\ the entries for fits 4 and 5 in
Table~\ref{tab.fits}. Also the error caused by the truncation of the
perturbative expansion of the conversion factors is estimated through
the same procedure as described above. The uncertainty due to the
chiral extrapolation is again determined from the difference between
the results obtained by means of the local and the global chiral
extrapolation. These four systematic errors
are added in quadrature to yield our final estimate of the total
systematic uncertainty. The central value and its statistical error are 
taken from the interpolation of the globally chirally extrapolated data.

In Tables~\ref{tab.momextract1}--\ref{tab.momextract3}
(see Appendix~\ref{sec.tables}) we display the results coming from the
fixed-scale method, based on the \RI-MOM scheme, separately for operators
with less than two and with two derivatives obtained with
and without the perturbative subtraction of lattice artifacts.
In the fixed-scale approach we keep the lattice artifacts as they are
generated in the simulations and try to get rid of them only when
performing the continuum limit of physical (renormalized) quantities.
Therefore, results obtained by this method with and without the
perturbative subtraction of lattice artifacts need not coincide. 
By definition, the choice $\mu^2 = 10 \, \mathrm {GeV}^2$
is fixed and a variation of this value does not enter the systematics.

The corresponding results for the \RI-SMOM scheme are presented in
Tables~\ref{tab.smomfit1}--\ref{tab.smomextract3}
in Appendix~\ref{sec.tables}, again separately for
the operators without derivatives (with and without perturbative
subtraction of lattice artifacts) and for operators with derivatives.
The results in Tables~\ref{tab.smomfit1}--\ref{tab.smomfit3} have been
obtained with the help of the fit method, while results coming from
the fixed-scale method can be found in
Tables~\ref{tab.smomextract1}--\ref{tab.smomextract3}. In the cases where
mixing occurs, i.e., for $\cO_{v_{3}}$, $\cO_{a_{2}}$, $\cO_{h_{2a}}$,
$\cO_{h_{2b}}$, and $\cO_{h_{2c}}$, we have given only the element
$Z_{11}$ of the renormalization and mixing matrix, which can be compared
with the \RI-MOM result. The full $2 \times 2$ mixing matrices can be
found in ancillary files. In the case of the three-quark operators,
where mixing matrices of size up to $4 \times 4$ appear, the results
are presented in ancillary files only.

When evaluating renormalized quantities one can take into account the
systematic uncertainties of the renormalization coefficients by error
propagation and subsequent continuum extrapolation. However, in the presence
of operator mixing, it might be more reasonable to use the various
determinations of $Z$ that go into the estimation of the systematic error,
e.g., the results from fits~2--7 in the fit method, in order to
compute the corresponding renormalized quantities and to use these
numbers to estimate the systematic uncertainty of the renormalized
quantity in the continuum limit. This is the procedure that we have
applied in Refs.~\cite{Braun:2016wnx,Bali:2019dqc,Bali:2019ecy}.
Note that in Ref.~\cite{Bali:2019dqc} the renormalization and mixing
coefficients were determined with the help of the fit method applied
to a smaller set of locally chirally extrapolated data. In spite of
these and a few further little differences the values used in
Ref.~\cite{Bali:2019dqc} are consistent with our present results
using two-loop conversion factors. An update of Ref.~\cite{Bali:2019dqc}
using the new three-loop conversion factors can be found in
Ref.~\cite{addendum}.

\section{Discussion} \label{sec.discussion}

\begin{figure}
\includegraphics[width=.45\textwidth]{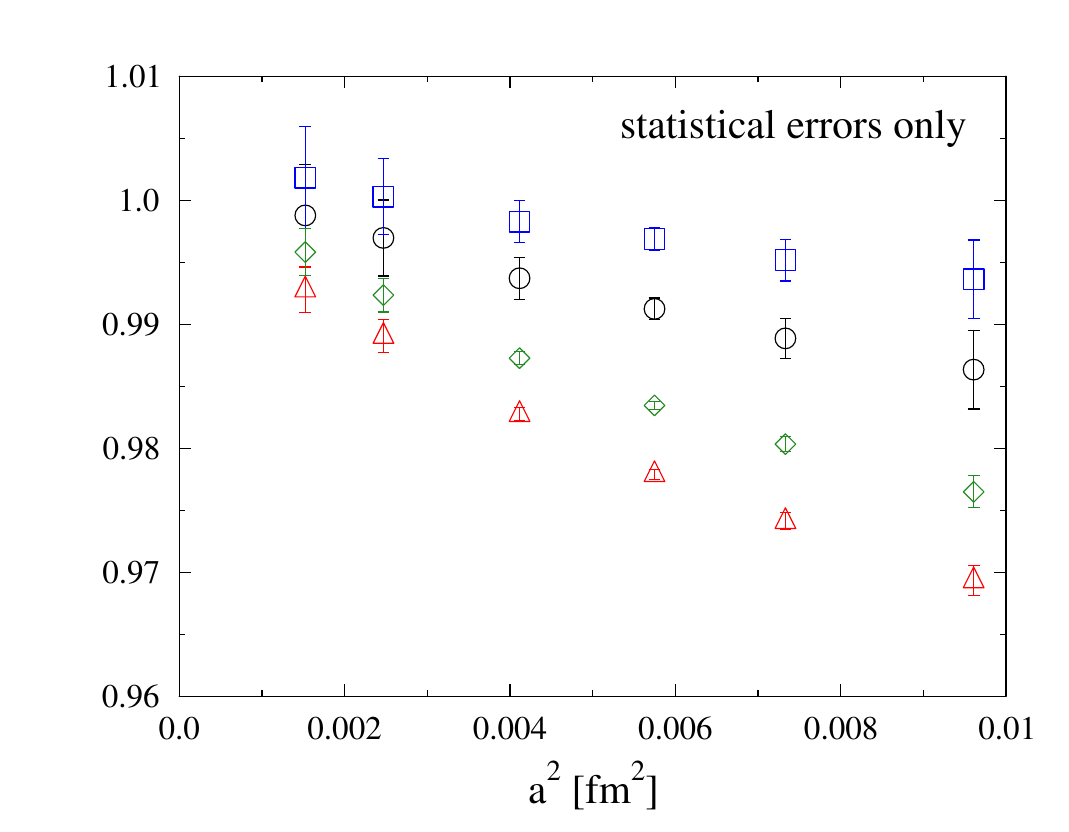}
\caption{\label{fig.ratios_axvecwi} Ratios of different determinations
of $Z'_A$ plotted against $a^2$. For the black circles (red triangles)
the fit (fixed-scale) method has been used, in the numerator within the
\RI-MOM scheme and in the denominator within the \RI-SMOM scheme. The
blue squares (green diamonds) represent results obtained in the \RI-MOM
(\RI-SMOM) scheme with the fit method employed for the numerator and
the fixed-scale method employed for the denominator.
The systematic errors of the \RI-MOM numbers amount to 2 \!{\permil}
at most, while the \RI-SMOM numbers obtained with the fit (fixed-scale) method
suffer from systematic uncertainties of up to 5 \!{\permil} (2 \!{\permil}).}
\end{figure}

\begin{figure}
\includegraphics[width=.45\textwidth]{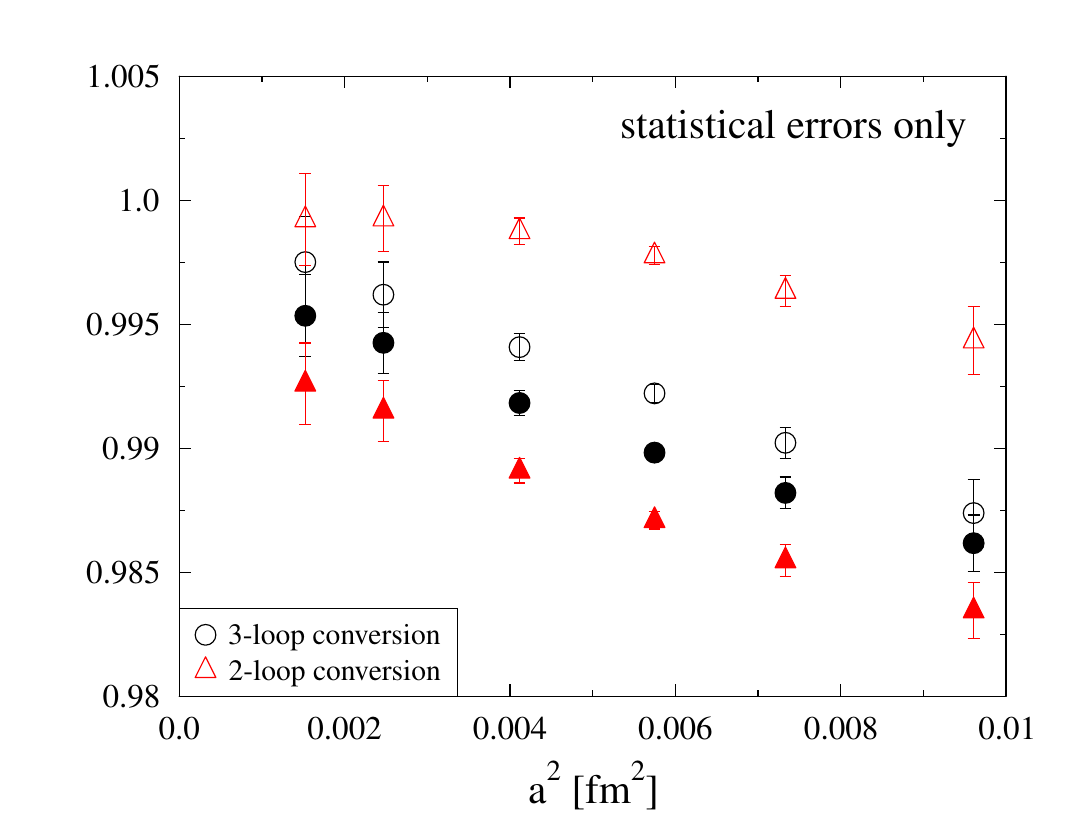}
\caption{\label{fig.ratios_vasum} Ratios of different determinations
of $Z_A$ plotted against $a^2$. For the filled symbols the fixed-scale
method has been used, in the numerator within the \RI-MOM scheme and in the
denominator within the \RI-SMOM scheme. The open symbols represent results
obtained in the \RI-SMOM scheme with the fit method employed for the
numerator and the fixed-scale method employed for the denominator.
The circles (triangles) show determinations using the three-loop
(two-loop) conversion factor. For the results obtained with the three-loop
conversion factors, the systematic uncertainty is up to 3 \!{\permil}
(7 \!{\permil}) in the case of the fixed-scale (fit) method.}
\end{figure}

\begin{figure}
\includegraphics[width=.45\textwidth]{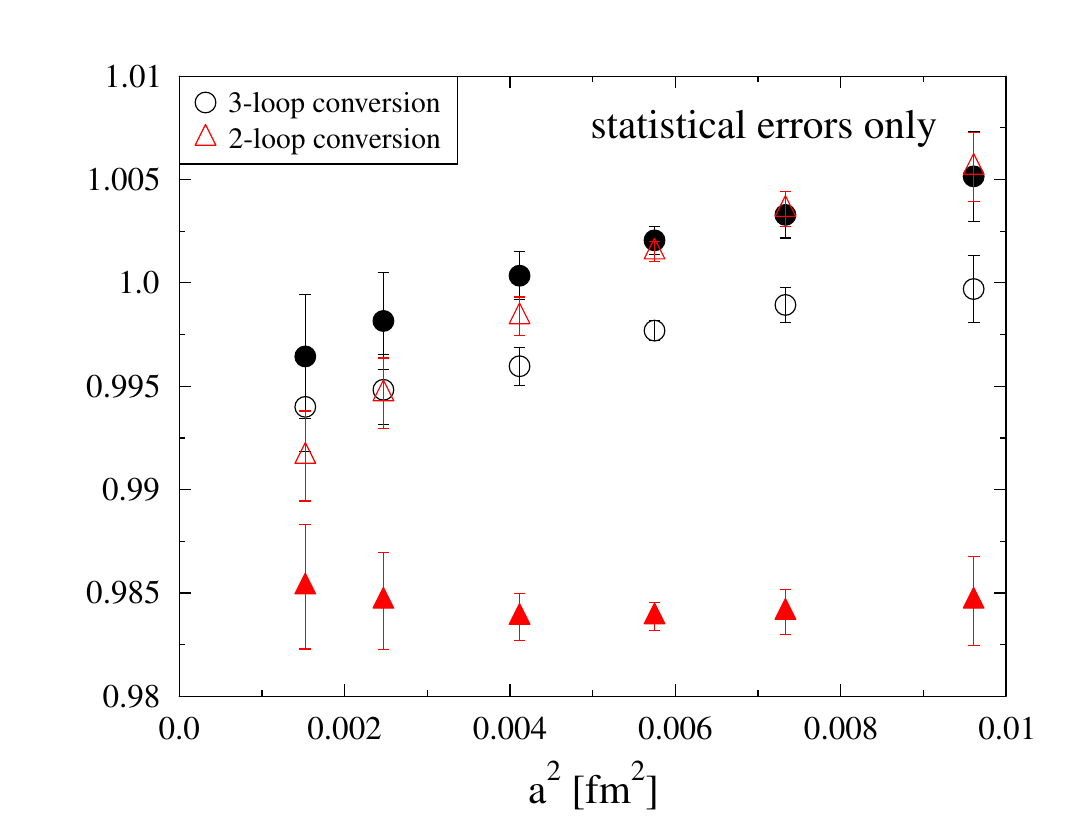}
\caption{\label{fig.ratios_v2b} Ratios of different determinations
of $Z_{v_{2b}}$ plotted against $a^2$. For the filled symbols the fit
method has been used, in the numerator within the \RI-MOM scheme and in the
denominator within the \RI-SMOM scheme. The open symbols represent results
obtained in the \RI-MOM scheme with the fit method employed for the
numerator and the fixed-scale method employed for the denominator.
The circles (triangles) show determinations using the three-loop
(two-loop) conversion factor. For the results obtained with the three-loop
conversion factors, the systematic uncertainty amounts to about
1\% (2--3\%) in the case of the \RI-MOM (\RI-SMOM) scheme.}
\end{figure}

\begin{figure}
\includegraphics[width=.45\textwidth]{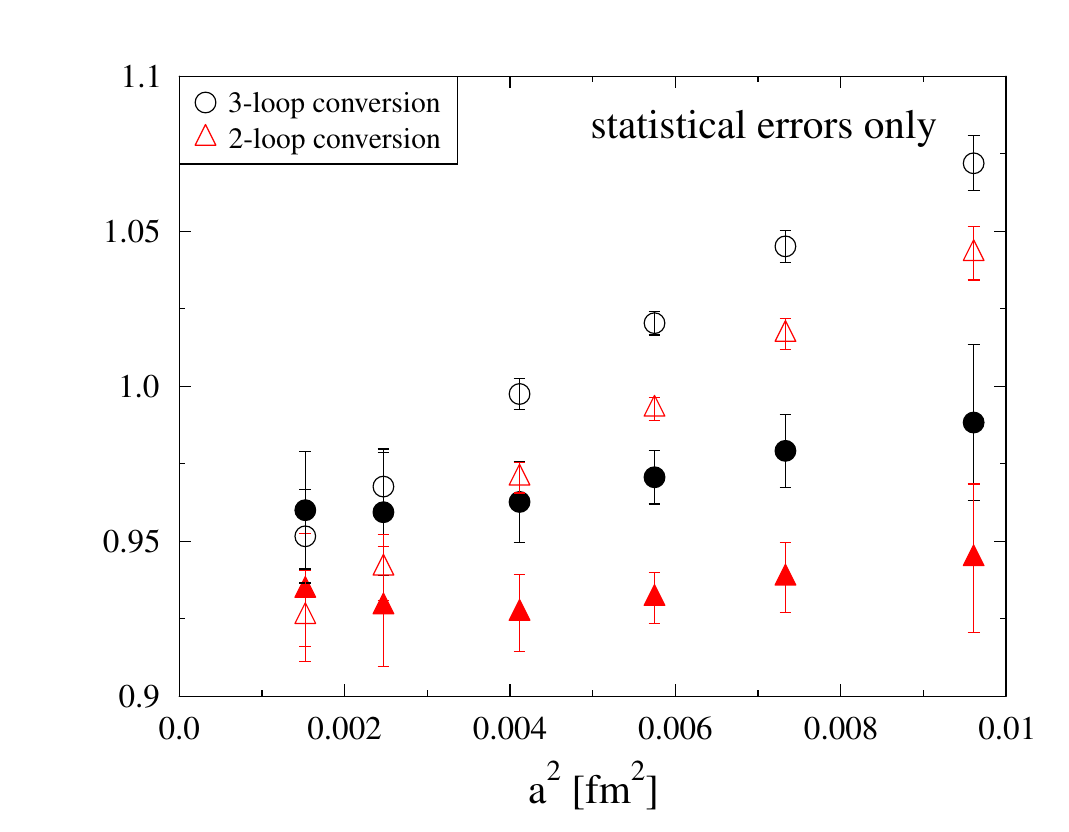}
\caption{\label{fig.ratios_I} Ratios of different determinations
of $Z_S$ plotted against $a^2$. For the filled (open) symbols the fit
(fixed-scale) method has been used, in the numerator within the \RI-MOM
scheme and in the denominator within the \RI-SMOM scheme. 
The circles (triangles) show determinations using the three-loop
(two-loop) conversion factor. For the results obtained with the three-loop
conversion factors, the systematic uncertainty amounts to 2\% (3--4\%)
in the case of the \RI-MOM scheme and the fixed-scale (fit) method,
it is about 0.4\% (0.5--1.5\%) in the case of the \RI-SMOM scheme
and the fixed-scale (fit) method.}
\end{figure}

For a large set of quark-antiquark operators as well as for
some three-quark operators we have presented renormalization factors
computed with the help of various methods. For each of the quark-antiquark
operators we end up with up to eight different
results. When multiplied with the corresponding bare matrix elements, all of
these renormalization factors should lead to the same continuum limit.
This means that for a given operator, ratios $Z^{(1)}(\mu,a)/Z^{(0)}(\mu,a)$
of two determinations $Z^{(1)}$ and $Z^{(0)}$ of the renormalization factor
must tend to one as $a \to 0$. However, one has to keep in mind that this
statement holds only up to errors due to the truncation of the perturbative
series entering the evaluation of $Z^{(1)}$ and $Z^{(0)}$.
The same considerations apply of course when one compares renormalization
factors of a given operator computed using entirely different methods.

In the following we shall discuss a selection of examples of such ratios.
In Figs.~\ref{fig.ratios_axvecwi} -- \ref{fig.ratios_I} we show for a
few operators ratios of renormalization factors
determined in this paper by different methods, plotted against $a^2$.
In all cases lattice artifacts have been subtracted perturbatively and
the errors have been computed by error propagation of the statistical
errors of the numerator and the denominator. These do not include
the systematics discussed above.

Ratios of different determinations of $Z'_A$ are displayed in
Fig.~\ref{fig.ratios_axvecwi}. As in this case the anomalous dimension
vanishes identically and the conversion factor is exactly equal to one,
there are no uncertainties due to truncations of perturbative expansions
and all ratios considered here should tend to one in the continuum limit.
The plot suggests that this is indeed the case within the statistical errors.

The next example (Fig.~\ref{fig.ratios_vasum}) is $Z_A$. In this case
the anomalous dimension is also known exactly, but the conversion factor
is only known to three loops. Therefore truncation errors are to be expected. 
Indeed, the results obtained with the three-loop conversion factor differ
significantly from those obtained with the two-loop conversion factor
and extrapolate to a value closer to one.

A similar behavior is found for operators with one derivative. For these
operators also the anomalous dimension is only known to a finite order
in perturbation theory. However, it turns out that the perturbative
expansion of the anomalous dimension converges much faster than the expansion
of the conversion factor. As an example we consider $Z_{v_{2b}}$ in
Fig.~\ref{fig.ratios_v2b}, where the five-loop expression for the anomalous
dimension is used throughout, but the number of loops taken into account
in the conversion factor is varied. The effect of this variation is clearly
visible and goes into the desired direction.

Finally, we show in Fig.~\ref{fig.ratios_I} ratios of different
determinations of $Z_S$, with the \RI-MOM (\RI-SMOM) scheme employed
in the numerator (denominator). For $a \to 0$ the value of
$Z_S$ determined with the help of the \RI-SMOM scheme is about
5\% larger than the number from the \RI-MOM scheme. Unusually large
differences between the two methods for the scalar density have already
been reported in Ref.~\cite{Hasan:2019noy} at two relatively large
values of the lattice spacing, using a different lattice action. In that case
the matching between the \RI-SMOM and the $\MS$ scheme could only be
carried out at two-loop order. In Fig.~\ref{fig.ratios_I} we only
display the statistical errors, however, also when including the
systematic uncertainties, the result is unsatisfactory.
This becomes even more apparent for the outcome of the
fixed-scale method, which carries smaller statistical and systematic
errors. In both cases (fit and fixed-scale method)
the results obtained with the three-loop conversion factor are closer
to unity than those obtained with the two-loop conversion factor.
In comparison to the other operators, the perturbative
coefficients for the matching between the $\MS$
and the \RI-SMOM schemes are quite small, see Table~\ref{tab.confac}.
Therefore, we suspect that our estimate of the perturbative
uncertainty as half of the difference between the results obtained
from two- and three-loop matching may be an underestimate
for this particular operator. Unfortunately, we
cannot carry out a similar comparison between the
\RI-SMOM and the \RI-MOM schemes for $Z_P$, which shares its
perturbative matching and its anomalous dimension with $Z_S$,
since the \RI-MOM scheme is unsuitable in this case.

\begin{figure}
\includegraphics[width=.45\textwidth]{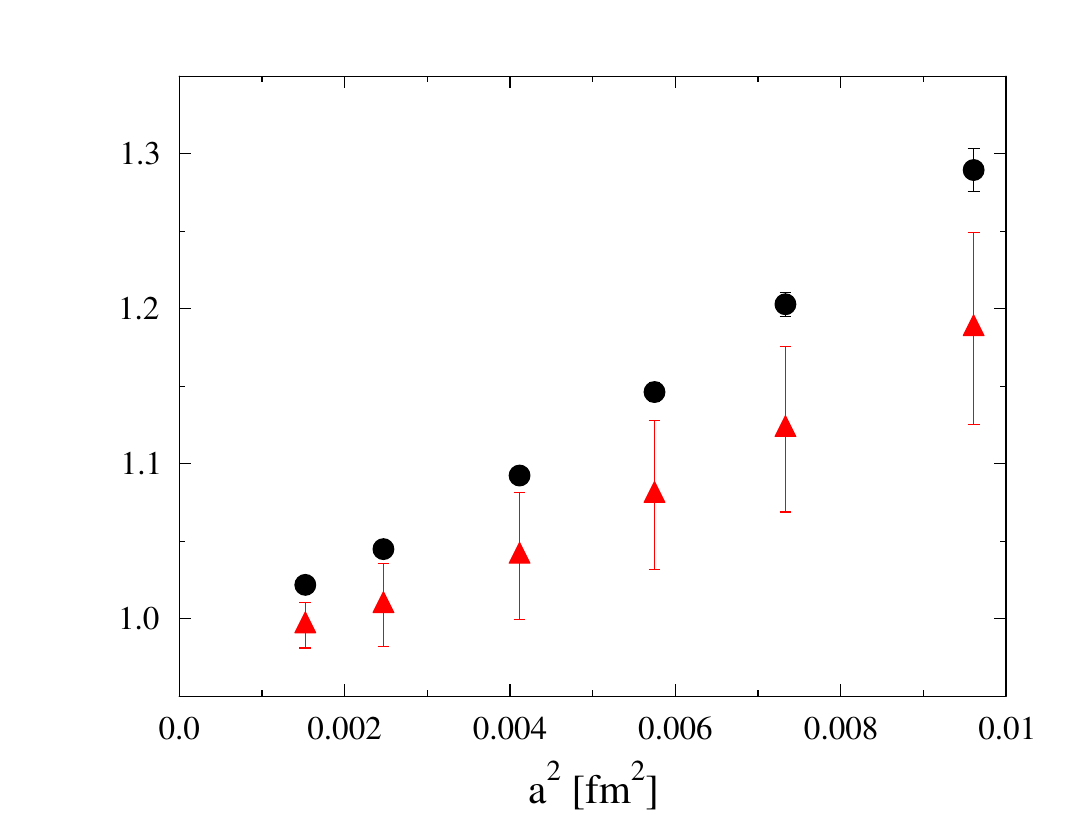}
\caption{\label{fig.ratios_alphaZSdZP} Ratios of different determinations
of $Z_S/Z_P$ plotted against $a^2$. The black circles
(red triangles) show the values of $Z_S/Z_P$ given in the column ``WI(1468)''
of Table~6 in Ref.~\cite{Heitger:2020mkp} divided by our results obtained
in the \RI-SMOM scheme with the fixed-scale (fit) method.}
\end{figure}

\begin{figure}
\includegraphics[width=.45\textwidth]{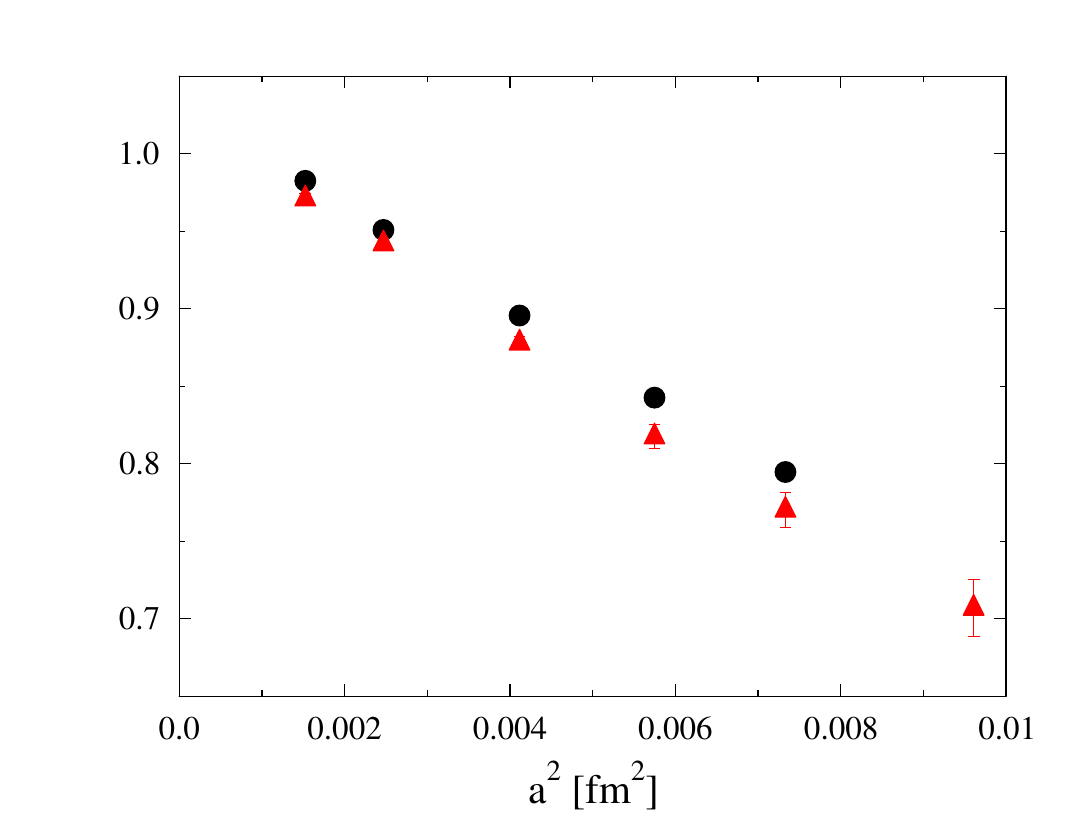}
\caption{\label{fig.ratios_alphaZPdZSZA} Ratios of different determinations
of $Z_P/(Z_S Z'_A)$ plotted against $a^2$. 
The black circles represent the numbers for $Z_P/(Z_S Z_A)$ in Table~4
(trajectory LCP-1) in Ref.~\cite{deDivitiis:2019xla} divided by our results
obtained in the \RI-SMOM scheme with the fixed-scale method.
The red triangles show data of the RQCD collaboration~\cite{rqcd} for
$Z_P/(Z_S Z_A)$ determined from axial Ward identity masses~\cite{Bali:2016umi}
divided by our results obtained in the \RI-SMOM scheme with the
fixed-scale method.}
\end{figure}

\begin{figure}
\includegraphics[width=.45\textwidth]{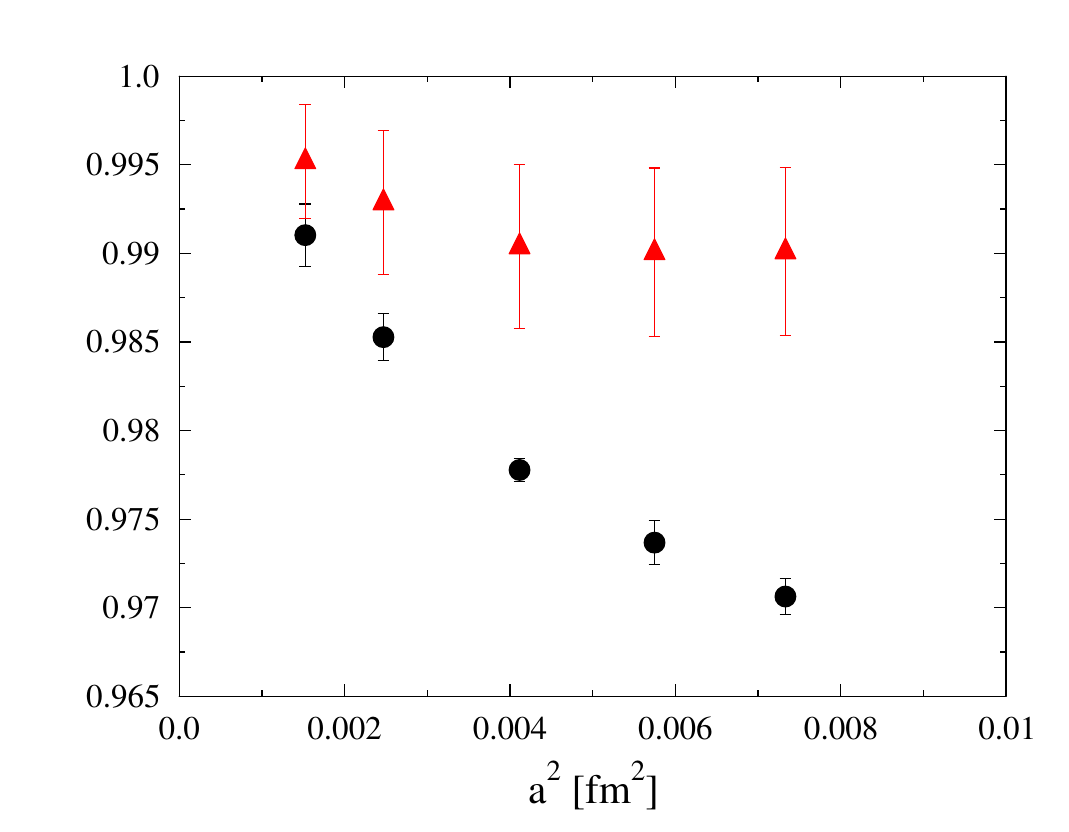}
\caption{\label{fig.ratios_alphaZA} Ratios of different determinations
of $Z'_A$ plotted against $a^2$. The black circles (red triangles) show
the values given in the column $Z^l_{A,\mathrm{sub}}$ of Table~7 in
Ref.~\cite{DallaBrida:2018tpn} divided by our results obtained in the
\RI-SMOM scheme with the fixed-scale (fit) method.}
\end{figure}

\begin{figure}
\includegraphics[width=.45\textwidth]{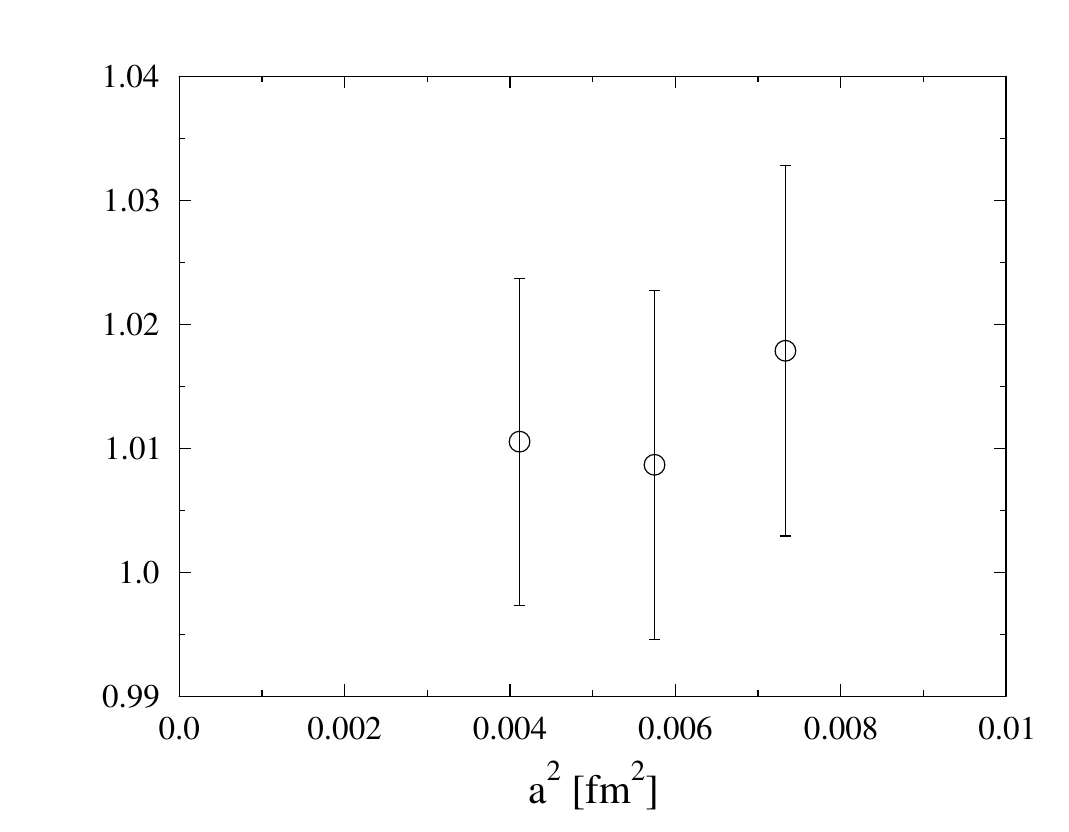}
\caption{\label{fig.ratios_mainzv2b} Ratios of different determinations
of $Z_{v_{2b}}$ plotted against $a^2$. The values given in Table~4
of Ref.~\cite{Harris:2019bih} have been divided by our results obtained
in the \RI-MOM scheme with the fit method.}
\end{figure}

Within CLS some renormalization factors have also been computed by
other groups, utilizing the \RI-MOM scheme~\cite{Harris:2019bih}
or Schr\"odinger functional techniques~\cite{Heitger:2017njs,DallaBrida:2018tpn,deDivitiis:2019xla,Heitger:2020mkp}. In
Figs.~\ref{fig.ratios_alphaZSdZP} -- \ref{fig.ratios_mainzv2b} we show for a few
examples ratios of such alternative results divided by our numbers.
For this comparison we employ our results with the perturbative subtraction
of lattice artifacts and add our statistical and systematic errors in
quadrature, since we cannot separate these two sources of uncertainties
in the case of the alternative determinations.

We begin with $Z_S/Z_P$ and $Z_P/(Z_S Z'_A)$. In these cases, our
results are free of perturbative ambiguities, because the anomalous
dimensions as well as the conversion factors cancel between numerator
and denominator. Nevertheless, systematic errors can be rather large,
as, e.g., Table~\ref{tab.smomfit1} shows. We denote the renormalization
factor of the axialvector current by $Z'_A$, because for our numbers
we use the data obtained with the help of the renormalization
condition (\ref{eq.axvecwi}). The ratio $Z_S/Z_P$ has
been studied by the ALPHA collaboration in Ref.~\cite{Heitger:2020mkp}.
We divide the values given in the column ``WI(1468)'' of Table~6 in
Ref.~\cite{Heitger:2020mkp} by our numbers determined with the help of
the fixed-scale and the fit method. The results are shown in
Fig.~\ref{fig.ratios_alphaZSdZP} as black circles and red triangles,
respectively, plotted against $a^2$. Indeed, these ratios of independent
determinations appear to approach unity in the continuum limit. However,
the curvature that we see when plotting the data as a function of $a^2$
indicates that within our range of lattice spacings the discretization
effects are not purely $O(a^2)$. It needs to be seen which of the three
methods (Ref.~\cite{Heitger:2020mkp}, fit method, fixed-scale method)
will result in the most convincing continuum limit extrapolation, once
$Z_S/Z_P$ is multiplied by corresponding ratios of matrix elements,
computed in lattice simulations.

In Fig.~\ref{fig.ratios_alphaZPdZSZA} we divide the combination
$Z_P/(Z_S Z_A)$ of the ALPHA collaboration given in Table~4 (trajectory LCP-1)
of Ref.~\cite{deDivitiis:2019xla} by our results on $Z_P/(Z_S Z_A')$
obtained in the \RI-SMOM scheme with the fixed-scale method (black circles).
The data appear to overshoot the continuum limit expectation by about
2\%. Repeating this exercise for the $Z_P/(Z_S Z_A)$ combination, determined
by the RQCD collaboration~\cite{rqcd} from a fit to axial Ward identity
quark masses~\cite{Bali:2016umi} (red triangles), some curvature of the
data becomes apparent. Still, the figure does not contradict the expectation
that the value of one will be reached in the continuum limit.

In Fig.~\ref{fig.ratios_alphaZA} we plot ratios of the renormalization factor
$Z_A$ determined by the ALPHA collaboration~\cite{DallaBrida:2018tpn},
divided by our results on $Z_A'$, obtained with the fixed-scale
and the fit methods. Both ratios tend to unity in the continuum
limit and again this approach is not linear in $a^2$ within
the range of lattice spacings covered. We can also compare the ratio
$Z_A/Z_P$ of Ref.~\cite{Campos:2018ahf} with
our results on $Z_A'/Z_P$ obtained from the \RI-SMOM scheme.
Using either the fit or the fixed-scale method,
in the continuum our results appear to be smaller by 2\%.
In the latter case this deviation exceeds the combined statistical and
systematic errors of our numbers by a factor of about three.
Given that the axialvector renormalization constants agree between different
determinations, as does $Z_S/Z_P$ (see Figs.~\ref{fig.ratios_alphaZA}
and \ref{fig.ratios_alphaZSdZP}), this tension is consistent
with the observation made in Fig.~\ref{fig.ratios_I} that $Z_S$ is
larger using the \RI-SMOM scheme than using the \RI-MOM scheme.
Therefore, we cannot exclude the possibility that our $Z_P$ is somewhat
overestimated, e.g., due to perturbation theory uncertainties.

As our last example we consider $Z_{v_{2b}}$, the renormalization factor
of an operator containing a covariant derivative. In
Fig.~\ref{fig.ratios_mainzv2b} our results are compared with the numbers
obtained in Ref.~\cite{Harris:2019bih}. These were determined in the 
\RI-MOM scheme at $\beta = 3.40$, $3.46$ and $3.55$ using an approach
similar to our fit method, though some details differ. Therefore we compare
them with our fit results obtained in the \RI-MOM scheme.
Within the errors the ratios are compatible with unity for all three $\beta$
values, although the numbers given in Ref.~\cite{Harris:2019bih} seem to
lie consistently above ours. Similar observations apply for the other
operators with one derivative studied in Ref.~\cite{Harris:2019bih}.

In view of the multitude of up to eight determinations of the
renormalization factors for a single operator given in this paper, the
question arises, which values should be used in a particular situation.
To answer this question a few criteria can be given. The decision in favor
of the \RI-MOM or \RI-SMOM intermediate scheme is obvious when nonforward
matrix elements of operators are required which mix with total-derivative
operators. In such cases the \RI-SMOM scheme is mandatory. It is also 
strongly preferred for the pseudoscalar density, because the pion pole makes
the chiral extrapolation problematic in the \RI-MOM scheme.
For all operators that are unaffected by the pion pole and for
which the kinematics of the \RI-MOM scheme is admissible, the choice of
the intermediate scheme is purely a matter of taste,
with the possible exception of the scalar density, where the \RI-SMOM
scheme may be more favorable because of the smaller coefficients in
the perturbative expansion (\ref{eq.matchexpand}) of the conversion factor
(see Table~\ref{tab.confac}).

Concerning the distinction between the fit method and the fixed-scale method
one can say that the fit method tries to suppress the power-like lattice
artifacts as far as possible. The fixed-scale method, on the other hand, 
does not care about lattice artifacts so that they will be taken into account
only in the continuum extrapolation of the physical observables. If the
data for the observables allow to perform such an extrapolation reliably,
the fixed-scale method could be preferable. However, if data are available
for a few lattice spacings only (maybe only for a single value) and a
decent continuum extrapolation is impossible, renormalization factors
determined with the help of the fit method might lead to estimated results
which are closer to the continuum value. Similarly, the perturbative
subtraction of lattice artifacts may be more useful when a careful
continuum limit is still out of reach, while it does not help that much
when the power-like lattice artifacts are taken care of by the continuum
extrapolation anyway.

\section{Summary} \label{sec.summary}

Using the Rome-Southampton method and variants thereof,
we have computed renormalization and mixing factors nonperturbatively
within the CLS setup, i.e., for $n_f=2+1$ flavors of nonperturbatively
improved Wilson quarks and the L\"uscher-Weisz action with tree-level
coefficients for the gluons. We have developed an approach that allows us to
include in our analysis also ensembles with open boundary conditions in
the time direction. Quark-antiquark operators as well as three-quark
operators have been considered with the help of a variety of methods.
The fit method has already been employed previously in our
papers~\cite{Bali:2019dqc,Bali:2019ecy}. Comparing the results obtained
with different methods allows us to check the consistency of the
employed procedures in the continuum limit. For applications we
recommend the numbers from the fixed-scale method given in
Tables~\ref{tab.smomextract2} and \ref{tab.smomextract3} in
Appendix~\ref{sec.tables}. The corresponding values from the fit method
are collected in Tables~\ref{tab.smomfit2} and \ref{tab.smomfit3}.
The $2 \times 2$ mixing matrices of quark-antiquark
operators and the results for the three-quark operators can be found in
ancillary files.

Regarding the renormalization factor $Z_S$, we see inconsistencies
on the several percent level that at present we cannot explain.
Similar problems have been reported in Ref.~\cite{Hasan:2019noy}.
Since the ratio $Z_S/Z_P$ appears to be consistent with
other determinations and the discrepancy reduces if the
matching to the $\MS$ scheme is carried out at a higher perturbative
order, we suspect that for the scalar and pseudoscalar densities, there
may be an unusually large perturbation theory uncertainty of about
2--3\% that we underestimated. By implication,
the results on $Z_A'/Z_P$ should be considered with caution.
In many cases, with our accuracy and within the range of lattice
spacings covered, the approach to the continuum limit
of ratios between determinations using different methods
is not a linear function of $a^2$. Therefore, it is advisable to employ
more than one set of renormalization factors in continuum extrapolations
of nonperturbative matrix elements.

At present, on the two finest lattices there are only two different
masses each available. This fact limits the accuracy of the required
chiral extrapolations. Generating additional ensembles with smaller masses
is desirable.

Another limitation of the accuracy of our results arises from the
necessity to use continuum perturbation theory. For many operators the
conversion or matching factors needed to obtain results in the
$\MS$ scheme are now known to three loops,
but the convergence of these perturbative expansions is generally not
too good, in contrast to the anomalous dimensions, which control the
scale dependence. Therefore further efforts in the perturbative evaluation 
of matching factors would be helpful, especially for the (pseudo)scalar
operator, but also for three-quark operators, for which the matching factors
are presently only known to one-loop accuracy.

\acknowledgments

This work was funded in part by the Deutsche Forschungsgemeinschaft 
(collaborative research centre SFB/TRR-55) and by the European Union’s
Horizon 2020 Research and Innovation programme under the Marie
Sk{\l}odowska-Curie grant agreement no. 813942 (ITN EuroPLEx).

The authors thank Oleg Veretin and John Gracey for their valuable
calculations in continuum perturbation theory.

We used a modified version of the {\sc Chroma}~\cite{Edwards:2004sx}
software package along with improved
inverters~\cite{Luscher:2012av,Nobile:2010zz,Frommer:2013fsa,Heybrock:2015kpy}.
The configurations were generated as part of the CLS
effort~\cite{Bruno:2014jqa,Bali:2016umi} using
{\sc openQCD}~\cite{LuscherOpenQCD,Luscher:2012av}. We thank all
our CLS colleagues for the joint generation of the gauge ensembles.
Additional ensembles with degenerate light and strange quark masses
were generated with {\sc openQCD} by members of the Mainz group
on HPC Clusters of IKP Mainz as well as by RQCD on
the QPACE systems of the SFB/TRR 55 using the BQCD code~\cite{Nakamura:2010qh}.
Correlation functions were computed on the QPACE~3 machine of
the SFB/TRR 55.

Computer time on the DFG-funded Ara cluster at the
Friedrich-Schiller-University Jena is acknowledged.
A. Sternbeck acknowledges
support by the BMBF under Grant No.\ 05P15SJFAA (FAIR-APPA-SPARC) and by
the DFG Research Training Group GRK1523.

The authors gratefully acknowledge the Gauss Centre for Supercomputing
(GCS) for providing computing time for GCS Large-Scale Projects on the
GCS share of the SuperMUC system at Leibniz Supercomputing Centre
(LRZ, \url{https://www.lrz.de})
and JUWELS~\cite{juwels} at J\"ulich
Supercomputing Centre (JSC, \url{http://www.fz-juelich.de/ias/jsc/}).
GCS is the alliance of the three national
supercomputing centres HLRS (Universit\"at Stuttgart),
JSC (For\-schungs\-zen\-trum J\"ulich), and LRZ (Bayerische Akademie der
Wissenschaften), funded by the German Federal Ministry of Education and
Research (BMBF) and the German State Ministries for Research of
Baden-W{\"u}rttemberg (MWK), Bayern (StMWFK) and Nordrhein-Westfalen (MIWF).

\newpage

\begin{appendix}

\section{Quark-antiquark operator multiplets}
\label{sec.opmulti}

In this Appendix we list the multiplets of quark-antiquark operators
that we consider. Flavor indices are omitted for simplicity. Color and spinor
indices are suppressed. As usual, $\{ \cdots \}$ will denote the
symmetrization of all enclosed indices. While a universal factor multiplying
all members of a multiplet is irrelevant for the renormalization, it is
important to ensure that the ratios of the individual normalization factors
are such that the operators transform under H(4) according to a unitary
representation.

We start with the operators without derivatives. Their mass dimension equals
three and they do not mix with other operators (operator multiplets). The
scalar density 
\begin{equation}
\mathcal S (x) = \bar{\psi} (x) \mathds{1} \psi(x)
\end{equation}
and the pseudoscalar density
\begin{equation}
\mathcal P (x) = \bar{\psi} (x) \gamma_5 \psi(x)
\end{equation}
form multiplets of dimension one. In the cases of the local vector current
\begin{equation} 
\mathcal V_\mu (x) = \bar{\psi}(x) \gamma_\mu \psi(x)
\end{equation}
and the local axialvector current
\begin{equation} 
\mathcal A_\mu (x) = \bar{\psi}(x) \gamma_\mu \gamma_5 \psi(x) 
\end{equation}
we have multiplets of dimension four ($\mu = 1,2,3,4$). The components of the
tensor density (or tensor current) 
\begin{equation} 
\mathcal T_{\mu \nu} (x) =
  \bar{\psi} (x) \mathrm i [ \gamma_\mu , \gamma_\nu ] \psi(x)
\end{equation}
with $1 \leq \mu < \nu \leq 4$ make up a six-dimensional multiplet.

The operators with a single covariant derivative have mass dimension four.
We consider the following multiplets, which in the flavor-nonsinglet case
do not mix with any other operators. The names we give to the operators
and to the corresponding renormalization factors (matrices) are motivated
by the nomenclature for the moments of the parton distribution functions
of the nucleon.

The $v_{2a}$ operators 
\begin{equation} 
\bar{\psi} (x) \gamma_{\{ \mu} \Dd{\nu \}} \psi(x) \,, 
\end{equation}
where $1 \leq \mu < \nu \leq 4$, form a multiplet of dimension six, while
the $v_{2b}$ operators
\begin{equation} \label{eq.v2b}
\begin{split}
(1/2) \bar{\psi} (x) \Big(  & 
  \gamma_1 \Dd{1} + \gamma_2 \Dd{2} - \gamma_3 \Dd{3} - \gamma_4 \Dd{4} 
  \Big) \psi(x) \,, \\
(1/ \sqrt{2}) \bar{\psi} (x) \Big(  & 
  \gamma_3 \Dd{3} - \gamma_4 \Dd{4} \Big) \psi(x) \,, \\
(1/ \sqrt{2}) \bar{\psi} (x) \Big(  & 
  \gamma_1 \Dd{1} - \gamma_2 \Dd{2} \Big) \psi(x) 
\end{split}
\end{equation}
span a three-dimensional multiplet.

Analogously we define the $r_{2a}$ operators
\begin{equation} 
\bar{\psi} (x) \gamma_{\{ \mu} \Dd{\nu \}} \gamma_5 \psi(x) \,, 
\end{equation}
with $1 \leq \mu < \nu \leq 4$, and the $r_{2b}$ operators
\begin{equation} 
\begin{split}
(1/2) \bar{\psi} (x) \Big(  & 
  \gamma_1 \Dd{1} + \gamma_2 \Dd{2} - \gamma_3 \Dd{3} - \gamma_4 \Dd{4} 
  \Big) \gamma_5 \psi(x) \,, \\
(1/ \sqrt{2}) \bar{\psi} (x) \Big(  & 
  \gamma_3 \Dd{3} - \gamma_4 \Dd{4} \Big) \gamma_5 \psi(x) \,, \\
(1/ \sqrt{2}) \bar{\psi} (x) \Big(  & 
  \gamma_1 \Dd{1} - \gamma_2 \Dd{2} \Big) \gamma_5 \psi(x) \,.
\end{split}
\end{equation}
With the help of the abbreviation
\begin{equation} 
\cO^T_{\nu \mu_1 \mu_2}(x) = 
  \bar{\psi} (x) \mathrm i [ \gamma_\nu , \gamma_{\mu_1}] \Dd{\mu_2} \psi(x)
\end{equation}
we can write the eight $h_{1a}$ operators as 
\begin{equation} 
\begin{split}
(1/ \sqrt{6}) & \left( \cO^T_{123}(x) + \cO^T_{231}(x) 
                                       - 2 \cO^T_{312}(x) \right) \,, \\
(1/ \sqrt{6}) & \left( \cO^T_{124}(x) + \cO^T_{241}(x) 
                                       - 2 \cO^T_{412}(x) \right) \,, \\
(1/ \sqrt{6}) & \left( \cO^T_{134}(x) + \cO^T_{341}(x) 
                                       - 2 \cO^T_{413}(x) \right) \,, \\
(1/ \sqrt{6}) & \left( \cO^T_{234}(x) + \cO^T_{342}(x) 
                                       - 2 \cO^T_{423}(x) \right) \,, \\
(1/ \sqrt{2}) & \left( \cO^T_{213}(x) + \cO^T_{231}(x) \right) \,, \\
(1/ \sqrt{2}) & \left( \cO^T_{214}(x) + \cO^T_{241}(x) \right) \,, \\
(1/ \sqrt{2}) & \left( \cO^T_{314}(x) + \cO^T_{341}(x) \right) \,, \\
(1/ \sqrt{2}) & \left( \cO^T_{324}(x) + \cO^T_{342}(x) \right) \,.
\end{split}
\end{equation}
The $h_{1b}$ operators
\begin{equation} 
\begin{split}
(1/ \sqrt{6}) & \left( \cO^T_{122}(x) + \cO^T_{133}(x) 
                                       - 2 \cO^T_{144}(x) \right) \,, \\
(1/ \sqrt{6}) & \left( \cO^T_{211}(x) + \cO^T_{233}(x) 
                                       - 2 \cO^T_{244}(x) \right) \,, \\
(1/ \sqrt{6}) & \left( \cO^T_{311}(x) + \cO^T_{322}(x) 
                                       - 2 \cO^T_{344}(x) \right) \,, \\
(1/ \sqrt{6}) & \left( \cO^T_{411}(x) + \cO^T_{422}(x) 
                                       - 2 \cO^T_{433}(x) \right) \,, \\
(1/ \sqrt{2}) & \left( \cO^T_{122}(x) - \cO^T_{133}(x) \right) \,, \\
(1/ \sqrt{2}) & \left( \cO^T_{211}(x) - \cO^T_{233}(x) \right) \,, \\
(1/ \sqrt{2}) & \left( \cO^T_{311}(x) - \cO^T_{322}(x) \right) \,, \\
(1/ \sqrt{2}) & \left( \cO^T_{411}(x) - \cO^T_{422}(x) \right)
\end{split}
\end{equation}
form an eight-dimensional multiplet as well.

The operators with two derivatives are of mass dimension five. In the
flavor-nonsinglet case at leading twist one can choose them such that
they do not mix with operators of lower mass dimension. However, when
nonforward matrix elements are needed, as they appear, e.g., in the
\RI-SMOM scheme, we cannot avoid the mixing with the so-called
total-derivative operators, which occurs already in the continuum.
Therefore we consider in the case of the $v_3$ operators the two
four-dimensional multiplets 
\begin{equation} 
\begin{split}
& \cO^{(\rho)}_1(x) \\
& = \bar{\psi} (x) \left( \Dr{\{ \mu} \Dr{\nu} +  \Dl{\{ \mu} \Dl{\nu} 
   - 2 \, \Dl{\{ \mu} \Dr{\nu} \right) \gamma_{\lambda \}} \psi(x) \,, \\
& \cO^{(\rho)}_2(x) \\
& = \bar{\psi} (x) \left( \Dr{\{ \mu} \Dr{\nu} +  \Dl{\{ \mu} \Dl{\nu} 
   + 2 \, \Dl{\{ \mu} \Dr{\nu} \right) \gamma_{\lambda \}} \psi(x) 
\end{split}
\end{equation}
with $\{ \mu , \nu , \lambda , \rho \} = \{ 1,2,3,4 \}$. They transform
identically under H(4). Note that in the continuum the operator
$\cO^{(\rho)}_1$ can be written as 
\begin{equation} 
\cO^{(\rho)}_1(x) =
\bar{\psi}(x) \gamma_{\{ \mu} \Dd{\nu} \Dd{\lambda \}} \psi(x) \,,
\end{equation}
while $\cO^{(\rho)}_2$ is the second derivative of the vector current:
\begin{equation} 
\cO^{(\rho)}_2(x) =
    \partial_{\{ \mu} \partial_\nu \mathcal V_{\lambda \}} (x) \,.
\end{equation}
On the lattice, this relation is violated because of lattice artifacts in
the derivatives. Analogous remarks hold also for the remaining operators
with two derivatives. These comprise the $a_2$ operators
\begin{equation} 
\begin{split}
& \cO^{(\rho)}_1 (x) \\
& = \bar{\psi} (x) \left( \Dr{\{ \mu} \Dr{\nu} +  \Dl{\{ \mu} \Dl{\nu} 
  - 2 \, \Dl{\{ \mu} \Dr{\nu} \right) \gamma_{\lambda \}} \gamma_5
    \psi(x) \,, \\
& \cO^{(\rho)}_2 (x) \\
& = \bar{\psi} (x) \left( \Dr{\{ \mu} \Dr{\nu} +  \Dl{\{ \mu} \Dl{\nu} 
  + 2 \, \Dl{\{ \mu} \Dr{\nu} \right) \gamma_{\lambda \}} \gamma_5
    \psi(x) \,,
\end{split}
\end{equation}
again with $\{ \mu , \nu , \lambda , \rho \} = \{ 1,2,3,4 \}$, as well as
the $h_{2a}$, $h_{2b}$, and $h_{2c}$ operators.

In order to define the latter we use the abbreviation 
\begin{equation} 
\begin{split}
& \cO^{T \pm}_{\nu \mu_1 \mu_2 \mu_3}(x) \\
&  =  \bar{\psi} (x) \mathrm i [ \gamma_\nu , \gamma_{\mu_1}]\left( 
  \Dr{\mu_2} \Dr{\mu_3} + \Dl{\mu_2} \Dl{\mu_3} \pm 2 \Dl{\mu_2} \Dr{\mu_3}
  \right) \psi (x) \,.
\end{split}
\end{equation}
The first multiplet of the $h_{2a}$ operators is then given by
\begin{equation} 
\begin{split}
\cO^{(1)}_1 (x) & = - \sqrt{6} \, \cO^{T-}_{4 \{123\}} (x) \,, \\
\cO^{(2)}_1 (x) & = (\sqrt{3} /2) \left( -3 {\cO}^{T-}_{3 \{124\}} (x) 
                         - {\cO}^{T-}_{4 \{123\}} (x) \right) \,, \\
\cO^{(3)}_1 (x) & = (3/2) \left( {\cO}^{T-}_{1 \{234\}} (x) 
                     - {\cO}^{T-}_{2 \{134\}} (x) \right) \,.
\end{split}
\end{equation}
Replacing $\cO^{T-}$ by $\cO^{T+}$ yields the corresponding second multiplet
$\cO_2$ with the identical transformation behavior under H(4). In the
case of the $h_{2b}$ operators we have the first multiplet
\begin{equation} 
\begin{split}
\cO^{(1)}_1 (x) = & \sqrt{3/2} \Big( \cO^{T-}_{1 \{122\}} (x)
  - \cO^{T-}_{1 \{133\}} (x) + \cO^{T-}_{2 \{233\}} (x) \Big) \,, 
\\
\cO^{(2)}_1 (x)  = & (\sqrt{3} /4) \Big( -2 \cO^{T-}_{1 \{122\}} (x)
  - \cO^{T-}_{1 \{133\}} (x) \\
  & + 3 \cO^{T-}_{1 \{144\}} (x)
    + \cO^{T-}_{2 \{233\}} (x) - 3 \cO^{T-}_{2 \{244\}} (x) \Big) \,, 
\\
\cO^{(3)}_1 (x) = & (3/4) \Big( - \cO^{T-}_{1 \{133\}} (x)
  + \cO^{T-}_{1 \{144\}} (x) - \cO^{T-}_{2 \{233\}} (x) \\
  & + \cO^{T-}_{2 \{244\}} (x) - 2 \cO^{T-}_{3 \{344\}} (x) \Big) \,.
\end{split}
\end{equation}
Again, the second multiplet $\cO_2$ is obtained by replacing
$\cO^{T-}$ with $\cO^{T+}$. Finally, as the first multiplet of the $h_{2c}$
operators we take
\begin{equation} 
\begin{split}
&\cO^{(1)}_1 (x) \\ & = 
    \cO^{T-}_{13 \{32\}} (x) + \cO^{T-}_{23 \{31\}} (x) 
  - \cO^{T-}_{14 \{42\}} (x) - \cO^{T-}_{24 \{41\}} (x) \,, 
\\ 
&\cO^{(2)}_1 (x) \\ & = 
    \cO^{T-}_{12 \{23\}} (x) + \cO^{T-}_{32 \{21\}} (x) 
  - \cO^{T-}_{14 \{43\}} (x) - \cO^{T-}_{34 \{41\}} (x) \,, 
\\ 
&\cO^{(3)}_3 (x) \\ & = 
    \cO^{T-}_{12 \{24\}} (x) + \cO^{T-}_{42 \{21\}} (x) 
  - \cO^{T-}_{13 \{34\}} (x) - \cO^{T-}_{43 \{31\}} (x) \,, 
\\ 
&\cO^{(4)}_1 (x) \\ & = 
    \cO^{T-}_{21 \{13\}} (x) + \cO^{T-}_{31 \{12\}} (x) 
  - \cO^{T-}_{24 \{43\}} (x) - \cO^{T-}_{34 \{42\}} (x) \,, 
\\ 
&\cO^{(5)}_1 (x) \\ & = 
    \cO^{T-}_{21 \{14\}} (x) + \cO^{T-}_{41 \{12\}} (x) 
  - \cO^{T-}_{23 \{34\}} (x) - \cO^{T-}_{43 \{32\}} (x) \,, 
\\ 
&\cO^{(6)}_1 (x) \\ & = 
    \cO^{T-}_{31 \{14\}} (x) + \cO^{T-}_{41 \{13\}} (x) 
  - \cO^{T-}_{32 \{24\}} (x) - \cO^{T-}_{42 \{23\}} (x) \,.
\end{split}
\end{equation}
The corresponding multiplet $\cO_2$ is constructed as in the two
previous cases.

\begin{table}[h]
\caption{Operator multiplets and their transformation behavior under H(4).
The charge conjugation parity is denoted by $C$.}  
\label{tab.representations}
\begin{ruledtabular}
\begin{tabular}{cccccc}
Operator  & Representation & $C$  & Operator  & Representation & $C$ \\ \hline
$\mathcal S$    & $\tau^{(1)}_1$  & $+1$ &
                       $\cO_{r_{2b}}$  & $\tau^{(3)}_4$  & $-1$ \\
$\mathcal P$    & $\tau^{(1)}_4$  & $+1$ &
                       $\cO_{h_{1a}}$  & $\tau^{(8)}_2$  & $+1$ \\
$\mathcal V$    & $\tau^{(4)}_1$  & $-1$ &
                       $\cO_{h_{1b}}$  & $\tau^{(8)}_1$  & $+1$ \\
$\mathcal A$    & $\tau^{(4)}_4$  & $+1$ &
                       $\cO_{v_{3}}$   & $\tau^{(4)}_2$  & $-1$ \\
$\mathcal T$    & $\tau^{(6)}_1$  & $-1$ &
                       $\cO_{a_{2}}$   & $\tau^{(4)}_3$  & $+1$ \\
$\cO_{v_{2a}}$  & $\tau^{(6)}_3$  & $+1$ &
                       $\cO_{h_{2a}}$  & $\tau^{(3)}_2$  & $-1$ \\
$\cO_{v_{2b}}$  & $\tau^{(3)}_1$  & $+1$ &
                       $\cO_{h_{2b}}$  & $\tau^{(3)}_3$  & $-1$ \\
$\cO_{r_{2a}}$  & $\tau^{(6)}_4$  & $-1$ &
                       $\cO_{h_{2c}}$  & $\tau^{(6)}_2$  & $-1$ 
\end{tabular}
\end{ruledtabular}

\end{table}

In the \RI-MOM case the respective second multiplets $\cO_2$ do not
contribute because their forward matrix elements vanish.

In Table~\ref{tab.representations} we give the H(4) irreducible
representations in the notation of Ref.~\cite{Baake:1981qe} and the charge
conjugation parity for our operator multiplets.

\section{Three-quark operator multiplets}
\label{sec.3qops}

In the case of three-quark operators it is convenient to employ operator
multiplets that transform irreducibly not only with respect to the
spinorial hypercubic group $\overline{\mathrm {H(4)}}$ but also with
respect to the group $\mathcal S_3$ of permutations of the three quark
flavors. The latter group has three nonequivalent irreducible
representations, which we label by the names of the corresponding
ground state particle multiplets in a flavor symmetric world. Therefore,
the one-dimensional trivial representation is labeled by $\mathscr D$
in Appendix~\ref{sec.confac3}, the one-dimensional totally antisymmetric
representation by $\mathscr S$ and the two-dimensional representation
by $\mathscr O$.

Following Ref.~\cite{Bali:2015ykx} we construct multiplets with the
desired transformation properties from the multiplets defined in
Ref.~\cite{Kaltenbrunner:2008pb}. For operators without derivatives in
the representation $\tau^{\underbar{$\scriptstyle 12$}}_1$ of
$\overline{\mathrm {H(4)}}$ we have one doublet of
operator multiplets transforming according to the two-dimensional
representation of $\mathcal S_3$,
\begin{equation} \label{eq.o12.1}
\mathscr O_1^{\underbar{$\scriptstyle 12$}} = 
\begin{Bmatrix}
\frac{1}{\sqrt{6}}(\cO_7+\cO_8-2\cO_9) \\
\frac{1}{\sqrt{2}}(\cO_7-\cO_8) 
\end{Bmatrix} \,,
\end{equation}
(with the first multiplet being mixed-symmetric and the second one being
mixed-anti-symmetric) and one operator multiplet transforming trivially
under $\mathcal S_3$:
\begin{equation} \label{eq.d12.1}
\mathscr D_1^{\underbar{$\scriptstyle 12$}} =   
\tfrac{1}{\sqrt{3}}(\cO_7+\cO_8+\cO_9) \,.
\end{equation}
For operators without derivatives in the $\overline{\mathrm {H(4)}}$
representation $\tau^{\underbar{$\scriptstyle 4$}}_1$ we have one
multiplet that is totally antisymmetric under flavor permutations,
\begin{equation} \label{eq.s4.1}
\mathscr S_1^{\underbar{$\scriptstyle 4$}} = 
\tfrac{1}{\sqrt{3}}(\cO_3-\cO_4-\cO_5) \,,
\end{equation}
and two doublets of operator multiplets transforming according to the
two-dimensional representation of $\mathcal S_3$:
\begin{align} \label{eq.o4.1a}
( \mathscr O_1^{\underbar{$\scriptstyle 4$}} ) \rule[-1mm]{0mm}{10mm}_1
  =&\begin{Bmatrix}
  \frac{1}{\sqrt{2}}(\cO_3+\cO_4)\\
  \frac{1}{\sqrt{6}}(-\cO_3+\cO_4-2\cO_5)
 \end{Bmatrix}\,,
\\  \label{eq.o4.1b}
( \mathscr O_1^{\underbar{$\scriptstyle 4$}} ) \rule[-1mm]{0mm}{10mm}_2
  =&\begin{Bmatrix}
  \cO_2\\
  \frac{1}{\sqrt{3}}(2\cO_1+\cO_2)
 \end{Bmatrix}\,.
\end{align}

In the case of operators with one derivative we restrict ourselves to 
multiplets that transform according to the $\overline{\mathrm {H(4)}}$
representation $\tau^{\underbar{$\scriptstyle 12$}}_2$, because only these
are safe from mixing with lower-dimensional operators. There are twelve
linearly independent multiplets with this transformation
behavior~\cite{Kaltenbrunner:2008pb}. In Appendix A.2 of
Ref.~\cite{Kaltenbrunner:2008pb} one can find explicit expressions
for four multiplets, labeled $\cO_{D5}$, $\cO_{D6}$, $\cO_{D7}$, $\cO_{D8}$,
where the derivative acts on the third quark field. As the transformation
properties of the operators do not depend on the position of the derivative,
the remaining eight multiplets can be constructed by moving the derivative
to the second or to the first quark field. In the following we replace
the $D$ by $f$, $g$, or $h$ in order to indicate on which quark field the
covariant derivative acts: $f$ ($g$, $h$) means that the derivative acts
on the first (second, third) quark field. 

In this way we get one multiplet that is totally antisymmetric under
$\mathcal S_3$,
\begin{equation} \label{eq.s12.2}
\mathscr S_2^{\underbar{$\scriptstyle 12$}}
  = \tfrac{1}{\sqrt6} \bigl[(\cO_{g5}-\cO_{h5})
  +(\cO_{h6}-\cO_{f6}) +(\cO_{f7}-\cO_{g7})\bigr] \,.
\end{equation}
Additionally there are four doublets of operator multiplets corresponding
to the two-dimensional representation of $\mathcal S_3$,
\begin{widetext}
\begin{align} \label{eq.o12.2a}
(\mathscr O_2^{\underbar{$\scriptstyle 12$}}) \rule[-1mm]{0mm}{10mm}_1
  =&\begin{Bmatrix}
  \frac{1}{3\sqrt2}\bigl[(\cO_{f5}+\cO_{g5}+\cO_{h5})
  +(\cO_{f6}+\cO_{g6}+\cO_{h6})-2(\cO_{f7}
  +\cO_{g7}+\cO_{h7})\bigr]\\
  \frac{1}{\sqrt6}\bigl[(\cO_{f5}+\cO_{g5}+\cO_{h5})
  -(\cO_{f6}+\cO_{g6}+\cO_{h6})\bigr]
 \end{Bmatrix}\,,
\\ \label{eq.o12.2b}
(\mathscr O_2^{\underbar{$\scriptstyle 12$}}) \rule[-1mm]{0mm}{10mm}_2
=&\begin{Bmatrix}
  \frac{1}{6}\bigl[(-2\cO_{f5}+\cO_{g5}+\cO_{h5})
  +(\cO_{f6}-2\cO_{g6}+\cO_{h6})-2(\cO_{f7}
  +\cO_{g7}-2\cO_{h7})\bigr]\\
  \frac{1}{2\sqrt3}\bigl[(-2\cO_{f5}+\cO_{g5}
  +\cO_{h5})-(\cO_{f6}-2\cO_{g6}+\cO_{h6})\bigr]
 \end{Bmatrix}\,,
\\ \label{eq.o12.2c}
(\mathscr O_2^{\underbar{$\scriptstyle 12$}}) \rule[-1mm]{0mm}{10mm}_3
=&\begin{Bmatrix}
  \frac{1}{2}\bigl[(\cO_{g5}-\cO_{h5})-(\cO_{h6}
  -\cO_{f6})\bigr]\\
  \frac{1}{2\sqrt3}\bigl[(\cO_{h5}-\cO_{g5})
  +(\cO_{f6}-\cO_{h6})-2(\cO_{g7}
  -\cO_{f7})\bigr]
 \end{Bmatrix}\,,
\\ \label{eq.o12.2d}
(\mathscr O_2^{\underbar{$\scriptstyle 12$}}) \rule[-1mm]{0mm}{10mm}_4
=&\begin{Bmatrix}
  \frac{1}{\sqrt6}(\cO_{f8}+\cO_{g8}-2\cO_{h8})\\
  \frac{1}{\sqrt2}(\cO_{f8}-\cO_{g8})
 \end{Bmatrix}\,,
\end{align}
and three operator multiplets transforming trivially under flavor permutations:
\begin{align} \label{eq.d12.2a}
(\mathscr D_2^{\underbar{$\scriptstyle 12$}}) \rule[-1mm]{0mm}{5mm}_1
= {} &\tfrac{1}{3}\bigl[(\cO_{f5}+\cO_{g5}+\cO_{h5})
  +(\cO_{f6}+\cO_{g6}+\cO_{h6})+(\cO_{f7}
  +\cO_{g7}+\cO_{h7})\bigr] \,, \\  \label{eq.d12.2b}
(\mathscr D_2^{\underbar{$\scriptstyle 12$}}) \rule[-1mm]{0mm}{5mm}_2
= {} &\tfrac{1}{3\sqrt2}\bigl[(-2\cO_{f5}+\cO_{g5}
  +\cO_{h5})+(\cO_{f6}-2\cO_{g6}+\cO_{h6})
  +(\cO_{f7}+\cO_{g7}-2\cO_{h7})\bigr] \,,
  \\ \label{eq.d12.2c}
(\mathscr D_2^{\underbar{$\scriptstyle 12$}}) \rule[-1mm]{0mm}{5mm}_3
= {} &\tfrac{1}{\sqrt3}(\cO_{f8}+\cO_{g8}+\cO_{h8}) \,.
\end{align}
\end{widetext}

\section{Lattice spacing artifacts in one-loop lattice perturbation
    theory}
\label{sec.LPT}

\begin{figure*}
\includegraphics[width=0.9\textwidth]{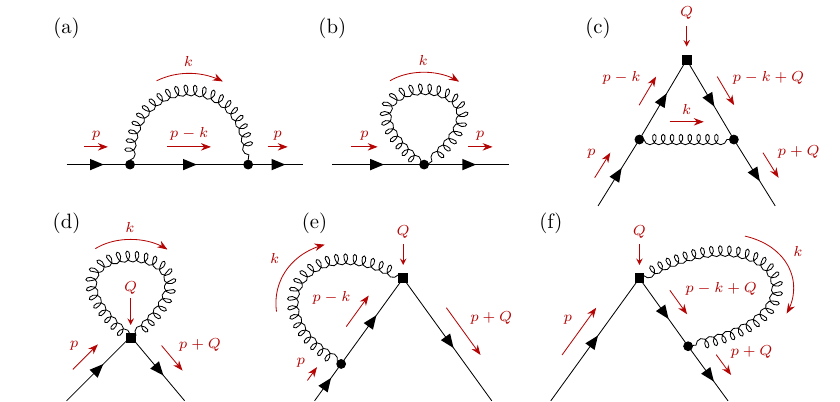}
\caption{One-loop diagrams for the quark self-energy (a,b) and quark bilinears 
(c,d,e,f) with momentum transfer at the operator insertion (solid box). For
operators without derivative the last row of diagrams is absent.}
\label{fig.LPTdiagrams}
\end{figure*}

We summarize in this Appendix our results for 
the term $F(a^2\mu^2)$ in Eq.~(\ref{eq.Zmua}). This term is calculated
numerically in lattice perturbation theory for the quark propagator and
all quark-antiquark operators with less than two derivatives
for a set of renormalization scales $a^2\mu^2$. To be consistent with
our nonperturbative calculation, we use the Landau gauge, apply exactly
the \RI-(S)MOM renormalization conditions described in
section~\ref{sec.schemes} and use the lattice expressions for the
propagators and vertices corresponding to our action.
In particular, the improved gauge action~\cite{Weisz:1983bn} leads to 
a complicated gluon propagator consisting of many terms, which slows
down the calculation considerably.

The analytical part of the calculation is performed in FORM
(v4.2.1)~\cite{Kuipers:2012rf}. In FORM we implement the \RI-(S)MOM
renormalization conditions and insert the one-loop expressions for
the respective correlation functions. The corresponding diagrams are shown
in Fig.~\ref{fig.LPTdiagrams}. For the quark self-energy there are two
diagrams, a \emph{sunset} and a \emph{tadpole} diagram, and for all
operator insertions a \emph{vertex} diagram. In addition there is a
\emph{tadpole} and a \emph{left-} and a \emph{right-sail}
diagram for operators with one derivative.

\begin{figure*}
  \mbox{\includegraphics[width=8cm]{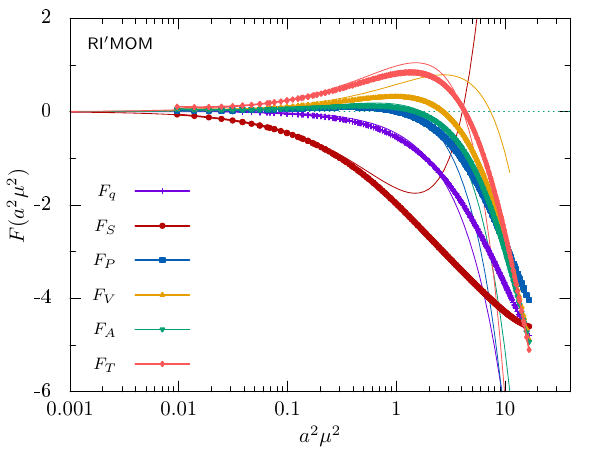}
        \includegraphics[width=8cm]{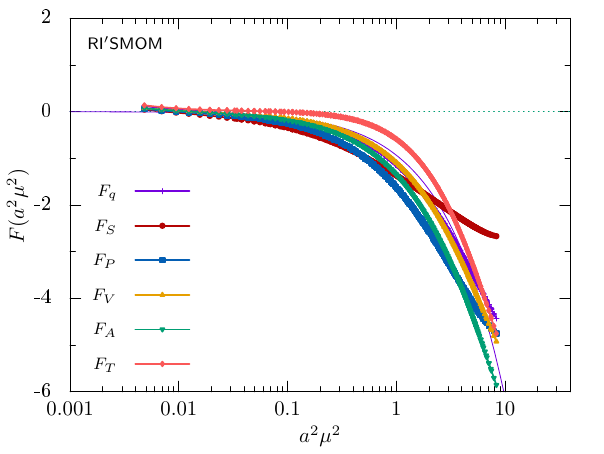}}
  \mbox{\includegraphics[width=8cm]{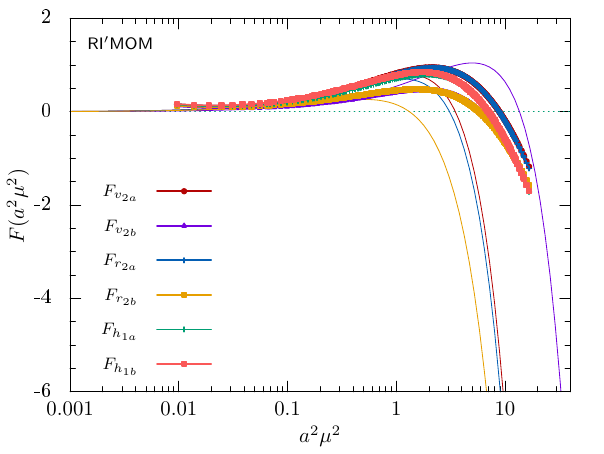}
        \includegraphics[width=8cm]{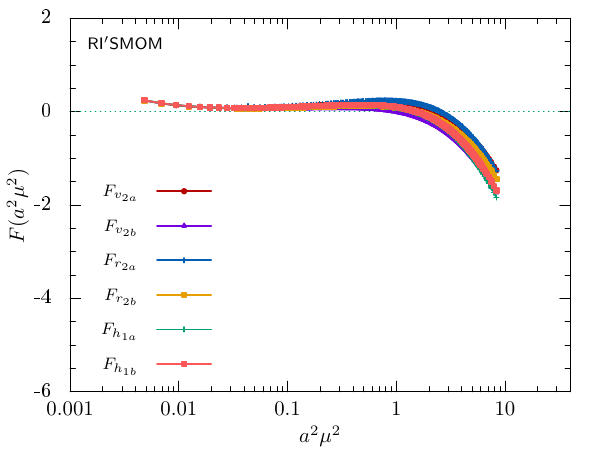}}
\caption{Numerical results for $F(a^2\mu^2)$ versus $a^2\mu^2$ for
operators with (bottom) and without (top) derivative in the \RI-MOM
scheme (left panels) and in the \RI-SMOM scheme (right panels), displayed
as data points connected by lines. The top panels also
show $F(a^2\mu^2)$ for the quark propagator where the quark momentum
direction matches that of the respective \RI-(S)MOM condition.
Analytical results for $F(a^2\mu^2)$ valid up to $O(a^2)$ are shown as
solid lines. The deviations from zero of the data points at small $a^2\mu^2$
are numerical artifacts and can be ignored.}
\label{fig.F_1loop}
\end{figure*}

The propagators and the quark-gluon vertex on the lattice can be found
in standard textbooks or in the review~\cite{Capitani:2002mp}. 
This review contains many useful expressions, for example, the gluon
propagator and the three vertices for the operator
insertion $\bar{\psi}(x)\gamma_\mu \Dr{\nu} \psi(x)$
at zero momentum transfer, which enter the calculation
of the \RI-MOM renormalization constants $Z_{v_{2a}}$ and $Z_{v_{2b}}$. 

For the \RI-SMOM scheme the operator vertices at finite momentum transfer
are required. A straightforward calculation, e.g., for the operator
$\bar{\psi}(x)\gamma_\mu \Dd{\nu}\psi(x)$
yields contributions of order $g^0$, $g^1$ and $g^2$: 
\begin{widetext}
\begin{align}
 \label{eq:vertex_V0_finiteQ}
   V^{(0)}_{\mu\nu}(p,p^\prime,Q) &= + i\gamma_\mu\, \frac{2}{a}\sin \left(\frac{a(p^\prime+p)_\nu}{2}\right)\cos\left(\frac{aQ_\nu}{2}\right) \,,
\\
 \label{eq:vertex_V1_finiteQ}
   V^{(1),a}_{\mu\nu}(p,p^\prime,Q) &= + i\gamma_\mu \,  2g T^a \cos\left(\frac{a(p^\prime+p)_\nu}{2}\right) \cos\left(\frac{aQ_\nu}{2}\right)\,,
\\
 \label{eq:vertex_V2_finiteQ}
   V^{(2),ab}_{\mu\nu}(p,p^\prime,Q) &= - i\gamma_\mu \,\frac{ag^2}{2} \left\{T^a,T^b\right\} \; \sin\left(\frac{a(p^\prime+p)_\nu}{2}\right) \cos\left(\frac{aQ_\nu}{2}\right) \,.
\end{align}
\end{widetext}
For brevity, the factor $(2\pi)^4$ and the momentum conserving
$\delta$ function have been omitted. The incoming and
outgoing quark momenta are denoted by $p$ and $p^\prime$, respectively,
and $Q=p^\prime-p$ is the quark momentum transfer at the vertex.
The matrices $T^a$ are the standard SU(3) generators.
For operators without derivatives, there is only $V^{(0)}$ and this equals
the gamma matrix structure of the respective operator insertion, because the
sine and cosine functions in Eq.~(\ref{eq:vertex_V0_finiteQ})
are due to the derivative in the operator.

Taking the traces in the renormalization conditions, FORM generates very
long expressions for the one-loop part of the renormalization constants.
These expressions have to be integrated over the
internal gluon loop momentum (see Fig.~\ref{fig.LPTdiagrams}). This is
performed numerically using a separate \texttt{C++} program and the library
\texttt{cubature}~\cite{cubature}. The external momenta are kept fixed,
i.e., for each setup of external quark momenta a separate numerical integration
is performed. The quark momenta are chosen to exactly match those for the
nonperturbative calculation. 

To speed up the integration, we use FORM's output optimization feature 
(\texttt{Format O3}) to generate compact integration kernels as input for our
\texttt{C++} program. Compared to the other FORM operations, this final output
optimization is CPU-time intensive and uses a Monte Carlo Tree Search to find
an optimal Horner scheme for a kernel. In this way, the final expressions
for the integration kernels become compact multivariate polynomials
whose numerical evaluation (for each integration step) is less CPU-time
intensive than that of the original kernel.

After integration the one-loop contribution to $Z$ for each operator
can be written as (cf.\ Eq.~(\ref{eq.Zmua}))
\begin{equation}
 z(a^2\mu^2) = (\gamma_0/2) \ln (a^2 \mu^2) + \Delta + F(a^2 \mu^2) \,,
\end{equation}
where the first two terms are known from lattice perturbation theory in the 
limit $a\to0$ and $F(0)=0$. Subtracting these terms from the numerical
results for $z(a^2\mu^2)$ gives $F(a^2 \mu^2)$. 
Note that $F$ at fixed $a^2\mu^2$ would also depend on the directions of
the external momenta if they are varied.
For our \RI-(S)MOM renormalization conditions, however, the momentum directions
are fixed, cf.\ Eqs.~(\ref{eq.rimom}) and (\ref{eq.rismom}).
Hence there is only a single curve $F(a^2 \mu^2)$ for each operator.

Fig.~\ref{fig.F_1loop} shows the numerical results for $F(a^2\mu^2)$,
separately for the \RI-MOM and \RI-SMOM schemes as well as for operators
without derivatives and with one derivative. In the case of the \RI-MOM
scheme, analytical expressions for $F(a^2 \mu^2)$ valid up to $O(a^2)$
can be derived from results in the
literature~\cite{Constantinou:2009tr,Alexandrou:2012mt}.
They are included in Fig.~\ref{fig.F_1loop} for comparison (solid lines)
and agree well with our numerical results for $a^2\mu^2\lesssim 0.4$. Beyond
this point deviations are clearly seen and corrections of higher order in $a$
seem to become important.

\section{Renormalization group functions for quark-antiquark operators}
\label{sec.rgfunc}

For the reader's convenience we collect in this Appendix perturbative
results for the QCD $\beta$ function and the anomalous dimension of the
quark-antiquark operators considered in this paper.

In the $\MS$ scheme the coefficients of the $\beta$ function, as defined
in Eqs.~(\ref{eq.betadef}) and (\ref{eq.betaexpansion}), are given
by~\cite{vanRitbergen:1997va,Baikov:2016tgj,Herzog:2017ohr}
\begin{align}
\beta_0 = {} & 11 - \frac{2}{3} n_f \,, \label{eq.beta0}\\
\beta_1 = {} & 102 - \frac{38}{3} n_f \,, \\
\beta_2 = {} & \frac{2857}{2} - \frac{5033}{18} n_f 
              + \frac{325}{54} n_f^2 \,, \\
\beta_3 = {} & \frac{149753}{6} + 3564 \zeta_3 
          - \left( \frac{1078361}{162} + \frac{6508}{27} \zeta_3  \right) n_f
        \nonumber \\ & {}
           + \left( \frac{50065}{162} + \frac{6472}{81} \zeta_3 \right) n_f^2
                       + \frac{1093}{729} n_f^3 \,, \\
\beta_4 = {} & \frac{8157455}{16} + \frac{621885}{2} \zeta_3
               -\frac{88209}{2} \zeta_4  - 288090 \, \zeta_5
        \nonumber \\ & {}
   + \left( -\frac{336460813}{1944} - \frac{4811164}{81} \zeta_3
                       +\frac{33935}{6} \zeta_4 \right.
        \nonumber \\ & \hspace*{12.0em} \left. {}
                       +\frac{1358995}{27} \zeta_5 \right) n_f
        \nonumber \\ & {}
   + \left( \frac{25960913}{1944} + \frac{698531}{81} \zeta_3
                       -\frac{10526}{9} \zeta_4 \right.
        \nonumber \\ & \hspace*{12.0em} \left. {}
                       - \frac{381760}{81} \zeta_5 \right) n_f^2
        \nonumber \\ & {}
   + \left( -\frac{630559}{5832} - \frac{48722}{243} \zeta_3
            +\frac{1618}{27} \zeta_4 + \frac{460}{9} \zeta_5 \right) n_f^3
        \nonumber \\ & {}
   + \left( \frac{1205}{2916} - \frac{152}{81} \zeta_3 \right) n_f^4 \,.
\label{eq.beta4}
\end{align}
Here and in the following, $n_f$ denotes the number of flavors and $\zeta_n$
is the value of Riemann's $\zeta$-function at $n$.

We now turn to the coefficients of the anomalous dimension in the
$\MS$ scheme employing the conventions given in Eqs.~(\ref{eq.gammadef})
and (\ref{eq.gammaexpansion}). An anticommuting $\gamma_5$ is assumed.

Of course, the vector current and the (nonsinglet) axialvector current have
vanishing anomalous dimension. For $\mathcal T_{\mu \nu}$ we
have~\cite{Gracey:2000am}
\begin{align}
\gamma_0 = {} & \frac{8}{3} \,, \\
\gamma_1 = {} & \frac{724}{9} - \frac{104}{27} n_f \,, \\
\gamma_2 = {} & \frac{105110}{81} - \frac{1856}{27} \zeta_3
             - \left( \frac{10480}{81} 
             + \frac{320}{9} \zeta_3 \right) n_f - \frac{8}{9} n_f^2 \,.
\end{align}
For $\mathcal S$ and $\mathcal P$ one
finds~\cite{Vermaseren:1997fq,Chetyrkin:1997dh}
\begin{align}
\gamma_0 = {} & -8 \,, \\
\gamma_1 = {} & - \frac{404}{3} + \frac{40}{9} n_f \,, \\
\gamma_2 = {} & - 2498 + \left( \frac{4432}{27} 
           + \frac{320}{3} \zeta_3 \right) n_f + \frac{280}{81} n_f^2 \,, \\
\gamma_3 = {} & - \frac{4603055}{81} - \frac{271360}{27}\zeta_3
                  + 17600 \, \zeta_5
              \nonumber \\ & {}
              + \left( \frac{183446}{27} + \frac{68384}{9} \zeta_3 
                - 1760 \, \zeta_4 - \frac{36800}{9} \zeta_5 \right) n_f
              \nonumber \\ & {}
              + \left( -\frac{10484}{243} - \frac{1600}{9} \zeta_3 
                   + \frac{320}{3} \zeta_4 \right) n_f^2
              \nonumber \\ & {}
       + \left( \frac{664}{243} - \frac{128}{27} \zeta_3 \right) n_f^3 \,.
\end{align}
The five-loop coefficient has been calculated as well~\cite{Baikov:2014qja}:
\begin{widetext}
\begin{align}
\gamma_4 = {} & - \frac{99512327}{81} - \frac{92804932}{243} \zeta_3 
  - 193600 \, \zeta_3^2 + \frac{1396252}{9} \zeta_4
  + \frac{463514320}{243} \zeta_5 - 484000 \, \zeta_6 - 825440 \, \zeta_7 
\nonumber \\ & {}
    + \left( \frac{150736283}{729} + \frac{25076032}{81} \zeta_3
    + \frac{151360}{9} \zeta_3^2 - \frac{4077484}{27} \zeta_4
    - \frac{99752360}{243} \zeta_5 + \frac{1276000}{9} \zeta_6
    + \frac{3640000}{27} \zeta_7 \right) n_f 
\nonumber \\ & {}
    + \left( - \frac{2641484}{729} - \frac{4021648}{243} \zeta_3
    - \frac{92800}{27} \zeta_3^2 + \frac{332600}{27} \zeta_4
    + \frac{528080}{81} \zeta_5 - \frac{184000}{27} \zeta_6 \right) n_f^2 
\nonumber \\ & {}
    + \left( - \frac{91865}{729} - \frac{25696}{81} \zeta_3
    - \frac{896}{9} \zeta_4 + \frac{10240}{27} \zeta_5 \right) n_f^3
       + \left( \frac{520}{243} + \frac{640}{243} \zeta_3
    - \frac{128}{27} \zeta_4 \right) n_f^4 \,.
\end{align}
The $v_{2a}$, $v_{2b}$, $r_{2a}$, and $r_{2b}$ operators have the same
anomalous dimension. From Ref.~\cite{Larin:1993vu} we obtain
\begin{align}
\gamma_0 = {} & \frac{64}{9} \,, \\
\gamma_1 = {} & \frac{23488}{243} - \frac{512}{81} n_f \,, \\
\gamma_2 = {} & \frac{11028416}{6561} + \frac{2560}{81} \zeta_3
             - \left( \frac{334400}{2187} 
             + \frac{2560}{27} \zeta_3 \right) n_f - \frac{1792}{729} n_f^2 \,.
\end{align}
The coefficient $\gamma_3$ can be found from Ref.~\cite{Velizhanin:2011es}:
\begin{align}
\gamma_3 = {} & \frac{6200738288}{177147} + \frac{52121728}{6561} \zeta_3
                - \frac{14080}{27} \zeta_4 - \frac{2498560}{243} \zeta_5 
      + \left( - \frac{334439344}{59049} - \frac{12645952}{2187} \zeta_3 
      + \frac{129280}{81} \zeta_4 + \frac{29440}{9} \zeta_5 \right) n_f
  \nonumber \\
 & {} + \left( \frac{2169808}{19683} + \frac{5120}{27} \zeta_3
          - \frac{2560}{27} \zeta_4 \right) n_f^2
   + \left( - \frac{8192}{6561} + \frac{1024}{243} \zeta_3 \right) n_f^3 \,.
\end{align}
The five-loop contribution has been calculated in Ref.~\cite{Herzog:2018kwj}:
\begin{align}
& \gamma_4 = \frac{3364807978412}{4782969}
      + \frac{40209657632}{177147} \zeta_3 - \frac{276459616}{2187} \zeta_4
      - \frac{20140392320}{19683} \zeta_5 + \frac{234874880}{2187} \zeta_3^2
  \nonumber \\ & \hspace{9.0cm} {}
      + \frac{68710400}{243} \zeta_6 + \frac{139807808}{243} \zeta_7
  \nonumber \\ & {}
  + \left( - \frac{227462023672}{1594323} - \frac{3959139616}{19683} \zeta_3
      + \frac{772997248}{6561} \zeta_4 + \frac{1803282880}{6561} \zeta_5
      - \frac{9598976}{729} \zeta_3^2 - \frac{78070400}{729} \zeta_6
      - \frac{8655808}{81} \zeta_7 \right) n_f
  \nonumber \\ & {}
  + \left( \frac{808174192}{177147}  + \frac{96501760}{6561} \zeta_3            
      - \frac{22962688}{2187} \zeta_4 - \frac{12238720}{2187} \zeta_5
      + \frac{26624}{9} \zeta_3^2 + \frac{147200}{27} \zeta_6 \right) n_f^2
  \nonumber \\ & {}
  + \left( \frac{20641064}{177147} + \frac{1234688}{6561} \zeta_3
      + \frac{9728}{81} \zeta_4 - \frac{81920}{243} \zeta_5 \right) n_f^3
  + \left( - \frac{44032}{59049} - \frac{8192}{2187} \zeta_3
           + \frac{1024}{243} \zeta_4 \right) n_f^4 \,.
\end{align}
For the $v_3$ and $a_2$ operators we extract from
Refs.~\cite{Moch:2004pa,Gracey:2006zr}
\begin{align}
\gamma_0 = {} & \frac{100}{9} \,, \\
\gamma_1 = {} & \frac{34450}{243} - \frac{830}{81} n_f \,, \\
\gamma_2 = {} & \frac{64486199}{26244} + \frac{2200}{81} \zeta_3
             - \left( \frac{967495}{4374} 
             + \frac{4000}{27} \zeta_3 \right) n_f - \frac{2569}{729} n_f^2 \,.
\end{align}
Note that the coefficient of $n_f$ in $\gamma_2$ was incorrectly given in
Ref.~\cite{Gockeler:2010yr} due to a misinterpretation of the results of
Ref.~\cite{Retey:2000nq}. The ensuing difference in the anomalous dimension
is however small, because the coefficient changes just from 392.948 to 
399.275. The four-loop coefficient $\gamma_3$ can be found
in Ref.~\cite{Moch:2017uml}:
\begin{align}
\gamma_3 = {} & \frac{69231923065}{1417176} + \frac{73641835}{6561} \zeta_3
                - \frac{12100}{27} \zeta_4 - \frac{3669100}{243} \zeta_5 
  \nonumber \\ & {}
      + \left( - \frac{1978909951}{236196} - \frac{19276720}{2187} \zeta_3 
      + \frac{200200}{81} \zeta_4 + \frac{46000}{9} \zeta_5 \right) n_f
  \nonumber \\ & {}
      + \left( \frac{3466612}{19683} + \frac{24400}{81} \zeta_3
          - \frac{4000}{27} \zeta_4 \right) n_f^2
   + \left( - \frac{23587}{13122} + \frac{1600}{243} \zeta_3 \right) n_f^3 \,.
\end{align}
From Ref.~\cite{Herzog:2018kwj} we get the five-loop result
\begin{align}
\gamma_4 = & \frac{309669533018351}{306110016}
      + \frac{281169024521}{708588} \zeta_3 - \frac{763282685}{4374} \zeta_4
      - \frac{31249942865}{19683} \zeta_5 + \frac{338652080}{2187} \zeta_3^2
  \nonumber \\ & \hspace{8.5cm} {}
      + \frac{100900250}{243} \zeta_6 + \frac{383634265}{486} \zeta_7
  \nonumber \\ & {}
  + \left( - \frac{2579650329389}{12754584} - \frac{12276163259}{39366} \zeta_3
      + \frac{1173351925}{6561} \zeta_4 + \frac{2845162870}{6561} \zeta_5
      - \frac{14484320}{729} \zeta_3^2 \right.
  \nonumber \\ & \hspace{8.7cm} \left. {}
      - \frac{120810500}{729} \zeta_6
      - \frac{13101340}{81} \zeta_7 \right) n_f
  \nonumber \\ & {}
  + \left( \frac{9569184149}{1417176}  + \frac{5525302}{243} \zeta_3        
      - \frac{35574820}{2187} \zeta_4 - \frac{19976200}{2187} \zeta_5
      + \frac{41600}{9} \zeta_3^2 + \frac{230000}{27} \zeta_6 \right) n_f^2
  \nonumber \\ & {}
  + \left( \frac{238595185}{1417176} + \frac{1769720}{6561} \zeta_3
      + \frac{5200}{27} \zeta_4 - \frac{128000}{243} \zeta_5 \right) n_f^3
  + \left( - \frac{259993}{236196} - \frac{13280}{2187} \zeta_3
           + \frac{1600}{243} \zeta_4 \right) n_f^4 \,.
\end{align}
The three-loop anomalous dimension of the $h_{1a}$ and $h_{1b}$ operators
can be found in Refs.~\cite{Gracey:2003mr,Gracey:2009da}:
\begin{align}
\gamma_0 = {} & 8 \,, \\
\gamma_1 = {} & 124 - 8 n_f \,, \\
\gamma_2 = {} & \frac{19162}{9} - \left( \frac{5608}{27} 
       + \frac{320}{3} \zeta_3 \right) n_f - \frac{184}{81} n_f^2 \,.
\end{align}
For the $h_{2a}$, $h_{2b}$, and $h_{2c}$ operators we obtain from
Ref.~\cite{Gracey:2006zr}
\begin{align}
\gamma_0 = {} &  \frac{104}{9} \,, \\
\gamma_1 = {} & \frac{38044}{243} - \frac{904}{81} n_f  \,, \\
\gamma_2 = {} & \frac{17770162}{6561} + \frac{1280}{81} \zeta_3
             - \left( \frac{552308}{2187} 
       + \frac{4160}{27} \zeta_3 \right) n_f - \frac{2408}{729} n_f^2 \,.
\end{align}
Finally, we can take the anomalous dimension of the quark field
(see Eq.~(\ref{eq.qfanodim})) in Landau gauge to four loops from
Ref.~\cite{Chetyrkin:1999pq}:
\begin{align}
\gamma_0 = {} & 0 \,, \\
\gamma_1 = {} & \frac{134}{3} - \frac{8}{3} n_f \,, \\
\gamma_2 = {} & \frac{20729}{18} - 79 \zeta_3 - \frac{1100}{9} n_f
               + \frac{40}{27} n_f^2 \,, \\
\gamma_3 = {} & \frac{2109389}{81} - \frac{565939}{162}\zeta_3 
              + \frac{2607}{2} \zeta_4 - \frac{761525}{648} \zeta_5
              - \left( \frac{324206}{81} + \frac{4582}{27} \zeta_3 
                + 79 \zeta_4 + \frac{320}{3} \zeta_5 \right) n_f
              \nonumber \\ & {}
           + \left( \frac{7706}{81} + \frac{320}{9} \zeta_3 \right) n_f^2
           + \frac{280}{243} n_f^3 \,.
\end{align}

\end{widetext}

In the case of the \RI-SMOM scheme we have to take into account that
operators with two derivatives mix with total-derivative operators.
This concerns the $v_3$ and $a_2$ operators as well as the $h_{2a}$,
$h_{2b}$, and $h_{2c}$ operators. The anomalous dimension becomes a
$2 \times 2$ matrix, whose 1-1 element coincides with the anomalous
dimension given above. For the $v_3$ and $a_2$ operators the only other 
nonzero entry is the 1-2 element. This matrix element can be calculated
from the results given in Refs.~\cite{Gracey:2009da,Gracey:2011zg}.
Transforming these results to our operator basis one finds
\begin{align}
(\gamma_0)_{12} = {} & - \frac{20}{9} \,, \\
(\gamma_1)_{12} = {} & - \frac{5954}{243} + \frac{118}{81} n_f  \,.
\end{align}
In our computations we use also the three-loop coefficient $(\gamma_2)_{12}$,
which can be evaluated with the help of unpublished results of
John Gracey~\cite{Gracey:private}. Recently, the three-loop coefficients
missing in Ref.~\cite{Gracey:2011zg} have been calculated
numerically~\cite{Kniehl:2020nhw}. Utilizing these numbers one finds
\begin{equation}
(\gamma_2)_{12} = - 417.165 + 64.0972 \, n_f + 1.09319 \, n_f^2 \,.
\end{equation}

For the $h_{2a}$, $h_{2b}$, and $h_{2c}$ operators the anomalous dimension
matrix has three nonvanishing entries. The 1-1 element is identical to the
anomalous dimension of these operators given above, while the 2-2 element
coincides with the anomalous dimension of the tensor density
$\mathcal T_{\mu \nu}$. Transforming the results given in
Ref.~\cite{Braun:2016wnx} to our operator basis one obtains for the
last nonvanishing matrix element
\begin{align}
(\gamma_0)_{12} = {} & - \frac{16}{9} \,, \\
(\gamma_1)_{12} = {} & - \frac{2720}{243} + \frac{80}{81} n_f  \,, \\
(\gamma_2)_{12} = {} & - \frac{6826684}{32805} - \frac{6848}{405} \zeta_3
             \nonumber \\ & {}
             + \left( \frac{28660}{2187} 
             + \frac{640}{27} \zeta_3 \right) n_f + \frac{544}{729} n_f^2 \,.
\end{align}

\section{Anomalous dimensions of three-quark operators}
\label{sec.anodim3q}

In this Appendix we present the anomalous dimensions of the
three-quark operators that we consider. For the operators without
derivatives three-loop results have been obtained in Ref.~\cite{Gracey:2012gx}.

For the multiplets (\ref{eq.o12.1}) and (\ref{eq.d12.1}) 
transforming according to the $\overline{\mathrm {H(4)}}$
representation $\tau^{\underbar{$\scriptstyle 12$}}_1$ we have
\begin{align}
\gamma_0 = {} & \frac{4}{3} \,, \\
\gamma_1 = {} & \frac{236}{3} - \frac{112}{27} n_f \,, \\
\gamma_2 = {} & \frac{18496}{9} - \frac{544}{3} \zeta_3
                - \left( \frac{16168}{81} 
       + \frac{160}{9} \zeta_3 \right) n_f + \frac{128}{81} n_f^2 \,.
\end{align}
In the case of the multiplets (\ref{eq.o4.1a}) and (\ref{eq.o4.1b})
one finds a diagonal matrix of anomalous dimensions with the entries
\begin{align}
(\gamma_0)_{11} = {} & -4 \,, \\
(\gamma_1)_{11} = {} & \frac{100}{3} - \frac{32}{9} n_f \,, \\
(\gamma_2)_{11} = {} & \frac{10988}{9}  + \left( - \frac{4264}{27} 
       + \frac{160}{3} \zeta_3 \right) n_f + \frac{112}{27} n_f^2 \,, \\
(\gamma_0)_{22} = {} & -4 \,, \\
(\gamma_1)_{22} = {} & \frac{140}{3} - \frac{32}{9} n_f \,, \\
(\gamma_2)_{22} = {} & \frac{10784}{9} + 32 \, \zeta_3 + \left( - 160
       + \frac{160}{3} \zeta_3 \right) n_f + \frac{112}{27} n_f^2 \,.
\end{align}
In the case of the multiplet (\ref{eq.s4.1}) we have
\begin{align}
\gamma_0 = {} & -4 \,, \\
\gamma_1 = {} & \frac{100}{3} - \frac{32}{9} n_f \,, \\
\gamma_2 = {} & \frac{10988}{9}  + \left( - \frac{4264}{27} 
       + \frac{160}{3} \zeta_3 \right) n_f + \frac{112}{27} n_f^2 \,.
\end{align}
For the operators with one derivative, only one-loop anomalous dimensions
are available. For the multiplet (\ref{eq.s12.2}) we have
\begin{equation}
\gamma_0 = \frac{52}{9} 
\end{equation}
The $4 \times 4$ anomalous dimension matrix of the octet multiplets
(\ref{eq.o12.2a}) - (\ref{eq.o12.2d}) is diagonal in one-loop order:
\begin{equation}
\gamma_0 = \text{diag} (4/3 \,,\, 20/3 \,,\, 52/9 \,,\, 8) \,.
\end{equation}
In the case of the decuplet multiplets (\ref{eq.d12.2a}) - (\ref{eq.d12.2c})
we get a $3 \times 3$ anomalous dimension matrix:
\begin{equation}
\gamma_0 = \text{diag} (4/3 \,,\, 20/3 \,,\, 4) \,.
\end{equation}

\section{Conversion factors for quark-antiquark operators}
\label{sec.confac}

In this Appendix we compile results for the expansion coefficients of the
conversion matrices (factors) of our operators.
The conversion matrices $C$ are defined in Eq.~(\ref{eq.match})
and their perturbative expansion is given in Eq.~(\ref{eq.matchexpand}).

We start with the \RI-MOM scheme. From Ref.~\cite{Chetyrkin:1999pq} we
get for $\mathcal S$ and $\mathcal P$
\begin{widetext}
\begin{equation} 
\begin{split}
c_1 = {} &  \frac{16}{3}\,, \\
c_2 = {} &  \frac{4291}{18} - \frac{152}{3}\zeta_3 - \frac{83}{9} n_f \,, \\
c_3 = {} &  \frac{3890527}{324} - \frac{224993}{54}\zeta_3 
                                              + \frac{2960}{9} \zeta_5
         - \left( \frac{241294}{243} - \frac{4720}{27} \zeta_3 
                                     + \frac{80}{3} \zeta_4 \right) n_f 
         + \left( \frac{7514}{729} + \frac{32}{27} \zeta_3 \right) n_f^2 \,.
\end{split}
\end{equation}
For $\mathcal V_\mu$ and $\mathcal A_\mu$ one finds from 
Ref.~\cite{Gracey:2003yr}
\begin{equation} 
\begin{split}
c_1 = {} &  0 \,, \\
c_2 = {} &  -\frac{67}{6} + \frac{2}{3} n_f \,, \\
c_3 = {} &  -\frac{52321}{72} + \frac{607}{4} \zeta_3 
   + \left( \frac{2236}{27} - 8 \zeta_3 \right) n_f - \frac{52}{27} n_f^2  \,.
\end{split}
\end{equation}
With the help of Ref.~\cite{Gracey:2003yr} we find for $\mathcal T_{\mu \nu}$
(for a correction, see Eq.~(\ref{eq.corrTconv}) of the Addendum)
\begin{equation} 
\begin{split}
c_1  = {} &  0 \,, \\
c_2  = {} &  -\frac{3847}{54} + \frac{184}{9}\zeta_3 + \frac{313}{81} n_f \,, \\
c_3  = {} &   -\frac{9858659}{2916} + \frac{678473}{486} \zeta_3 
                   + \frac{1072}{81} \zeta_4 - \frac{10040}{27} \zeta_5
         + \left( \frac{286262}{729} - \frac{2096}{27} \zeta_3 
                                     + \frac{80}{9} \zeta_4 \right) n_f
         - \left( \frac{13754}{2187} + \frac{32}{81} \zeta_3 \right) n_f^2 \,.
\end{split}
\end{equation}
In the case of the $v_{2a}$ and $r_{2a}$ operators one extracts from
Ref.~\cite{Gracey:2003mr}
\begin{equation} 
\begin{split}
c_1  = {} &  -\frac{130}{27} \,, \\
c_2  = {} &  -\frac{113084}{729} + \frac{86}{3}\zeta_3
             + \frac{2938}{243} n_f \,, \\
c_3  = {} &  -\frac{2105418469}{314928} + \frac{18986323}{8748} \zeta_3 
                   - \frac{640}{81} \zeta_4 - \frac{47335}{81} \zeta_5
       + \left( \frac{17190598}{19683} - \frac{4492}{81} \zeta_3 
                                     + \frac{640}{27} \zeta_4 \right) n_f \\
     & - \left( \frac{127772}{6561} + \frac{256}{243} \zeta_3 \right) n_f^2 \,.
\end{split}
\end{equation}
For the $v_{2b}$ and $r_{2b}$ operators we get from Ref.~\cite{Gracey:2003mr}
\begin{equation} 
\begin{split}
c_1  = {} &  -\frac{124}{27} \,, \\
c_2  = {} &  -\frac{98072}{729} + \frac{268}{9}\zeta_3
             + \frac{2668}{243} n_f \,, \\
c_3  = {} &  -\frac{849683327}{157464} + \frac{7809041}{4374} \zeta_3 
                   - \frac{640}{81} \zeta_4 - \frac{36410}{81} \zeta_5
       + \left( \frac{14433520}{19683} - \frac{4184}{81} \zeta_3 
                                     + \frac{640}{27} \zeta_4 \right) n_f \\
     & - \left( \frac{105992}{6561} + \frac{256}{243} \zeta_3 \right) n_f^2 \,.
\end{split}
\end{equation}
For the $h_{1a}$ and $h_{1b}$ operators one obtains from
Ref.~\cite{Gracey:2003mr}
\begin{equation} 
\begin{split}
c_1  = {} &  -\frac{14}{3} \,, \\
c_2  = {} &  -\frac{985}{6} + \frac{98}{3}\zeta_3 + 13 n_f \,, \\
c_3  = {} &  -\frac{8834075}{1296} + \frac{235505}{108} \zeta_3 
                   - \frac{17545}{27} \zeta_5
       + \left( \frac{449921}{486} - \frac{562}{9} \zeta_3 
                                     + \frac{80}{3} \zeta_4 \right) n_f 
       - \left( \frac{5050}{243} + \frac{32}{27} \zeta_3 \right) n_f^2 \,.
\end{split}
\end{equation}
Similarly, Ref.~\cite{Gracey:2006zr} yields for the $h_{2a}$, $h_{2b}$,
and $h_{2c}$ operators the coefficients
\begin{equation} 
\begin{split}
c_1  = {} &  -\frac{218}{27} \,, \\
c_2  = {} &  -\frac{814597}{3645} + \frac{632}{15}\zeta_3
             + \frac{4961}{243} n_f \,, \\
c_3  = {} &  -\frac{1396663105}{157464} + \frac{113197753}{43740} \zeta_3 
                   - \frac{320}{81} \zeta_4 - \frac{22919}{27} \zeta_5
       + \left( \frac{126822281}{98415} - \frac{46688}{1215} \zeta_3 
                                     + \frac{1040}{27} \zeta_4 \right) n_f \\
     & - \left( \frac{220360}{6561} + \frac{416}{243} \zeta_3 \right) n_f^2 \,.
\end{split}
\end{equation}
For the $v_3$ and $a_2$ operators we have the coefficients
\begin{equation} 
\begin{split}
c_1  = {} &  -\frac{223}{27} \,, \\
c_2  = {} &  -\frac{6644237}{29160} + \frac{198}{5}\zeta_3
             + \frac{39599}{1944} n_f \,, \\
c_3  = {} &  -\frac{3720946031}{393660} + \frac{251485339}{87480} \zeta_3 
                   - \frac{550}{81} \zeta_4 - \frac{5575}{6} \zeta_5
       + \left( \frac{2084393177}{1574640} - \frac{5219}{135} \zeta_3 
                                     + \frac{1000}{27} \zeta_4 \right) n_f \\
  & - \left( \frac{1793923}{52488} + \frac{400}{243} \zeta_3 \right) n_f^2 \,
\end{split}
\end{equation}
extracted from Ref.~\cite{Gracey:2006zr}.
For the conversion factor of the quark wave function renormalization $C_q$
one obtains from Ref.~\cite{Chetyrkin:1999pq} in Landau gauge
\begin{equation} 
\begin{split}
c_1  = {} &  0 \,, \\
c_2  = {} &  -\frac{359}{9} + 12 \zeta_3 + \frac{7}{3} n_f \,, \\
c_3  = {} &  -\frac{439543}{162} + \frac{8009}{6} \zeta_3
       + \frac{79}{4} \zeta_4 - \frac{1165}{3} \zeta_5
       + \left( \frac{24722}{81} - \frac{440}{9} \zeta_3 \right) n_f 
       - \frac{1570}{243} n_f^2 \,.
\end{split}
\end{equation}

\end{widetext}

Perturbative expressions for the \RI-SMOM vertex functions are available
in analytic form for up to two
loops~\cite{Gracey:2011fb,Gracey:2011zn,Gracey:2011zg,Braun:2016wnx}.
As they are rather complicated, also the corresponding results for
the coefficients of the conversion factors as derived from Eq.~(\ref{eq.match})
are quite lengthy. Therefore we give them only in numerical form for
$n_f=3$, see Table~\ref{tab.confac}. Three-loop results for the operators
without derivatives have recently been obtained in Ref.~\cite{Kniehl:2020sgo}
and for some operators with derivatives in Ref.~\cite{Kniehl:2020nhw}. In these
papers the required integrals have been evaluated with the help of numerical
integration. The corresponding results for $c_3$
computed from the given central values are also included in
Table~\ref{tab.confac}. Their numerical uncertainty is calculated by means
of error propagation from the errors of the individual contributions
given in Refs.~\cite{Kniehl:2020sgo,Kniehl:2020nhw}. For our final
renormalization factors this uncertainty is completely irrelevant.
In the case of the scalar density, there is even an analytic expression
for $c_3$~\cite{Bednyakov:2020ugu}, which agrees perfectly with the
numerical result.

Note that here the mixing with total-derivative operators has been ignored.
Hence the numbers in Table~\ref{tab.confac} for the operators with two
derivatives are sufficient only when forward matrix elements are considered.
In order to facilitate the comparison between the \RI-SMOM and the \RI-MOM
scheme we include also the numerical values of the coefficients in the
\RI-MOM scheme. While for the scalar and the pseudoscalar densities the
coefficients in the \RI-SMOM scheme are substantially smaller than in
the \RI-MOM scheme, they are quite similar in the other cases.

\begin{table}[h]
\caption{Coefficients of conversion factors for $n_f=3$. In the cases where
loop integrals have been evaluated numerically estimates of the resulting
uncertainties have been included. For $\mathcal T$ in the MOM scheme
$c_3$ should read $- 1194.56$ instead of $- 1207.96$, see the Addendum.}
\label{tab.confac}
\begin{ruledtabular}
\begin{tabular}{ll@{\hspace{-1mm}}ddd}
Operators & Scheme & c_1 & c_2 & c_3 \\ \hline
$\mathcal S$, $\mathcal P$ & SMOM &  0.645519 &  10.9838  & 399.63(7) \\ 
$\mathcal S$, $\mathcal P$ & MOM  &  5.33333  & 149.818   & 5010.89 \\ 
$\mathcal V$, $\mathcal A$ & SMOM & -1.33333  & -26.8410  & -704.85(14) \\ 
$\mathcal V$, $\mathcal A$ & MOM  &  0.0      &  -9.16667 & -342.007 \\
$\mathcal T$               & SMOM & -0.215173 & -31.0741  & -1084.85(11) \\ 
$\mathcal T$               & MOM  &  0.0      & -35.0728  & -1207.96 \\ 
$v_{2a}$, $r_{2a}$         & SMOM & -5.12954  & -78.0195  & -2220.96(2) \\ 
$v_{2a}$, $r_{2a}$         & MOM  & -4.81481  & -84.3915  & -2380.58 \\
$v_{2b}$, $r_{2b}$         & SMOM & -5.12954  & -78.0195  & -2220.96(2) \\ 
$v_{2b}$, $r_{2b}$         & MOM  & -4.59259  & -65.7966  & -1790.84 \\
$h_{1a}$, $h_{1b}$         & SMOM & -4.75682  & -80.6120  & - \\ 
$h_{1a}$, $h_{1b}$         & MOM  & -4.66667  & -85.8995  & -2430.19 \\
$v_{3}$, $a_{2}$           & SMOM & -8.30885  &-103.787   & -2939.92(11)\\ 
$v_{3}$, $a_{2}$           & MOM  & -8.25926  &-119.143   & -3340.75 \\
$h_{2a}$, $h_{2b}$         & SMOM & -8.12940  &-106.004   & - \\ 
$h_{2a}$, $h_{2b}$         & MOM  & -8.07407  &-111.590   & -3111.68 \\
$h_{2c}$                   & SMOM & -8.14087  &-105.975   & - \\ 
$h_{2c}$                   & MOM  & -8.07407  &-111.590   & -3111.68 \\
\end{tabular}
\end{ruledtabular}
\end{table}

When nonforward matrix elements are to be investigated, we have to
take into account that operators with two derivatives mix with 
total-derivative operators. In this case we must use the \RI-SMOM scheme,
and the conversion factors become conversion matrices of size
$2 \times 2$ whose 1-1 entry coincides with what one gets from
Table~\ref{tab.confac}. This concerns the $v_3$ and $a_2$ operators as
well as the $h_{2a}$, $h_{2b}$, and $h_{2c}$ operators. One finds for
the $v_3$ and $a_2$ operators the matrices
\begin{align}
c_1 &= 
\begin{pmatrix}
-8.30885 & 1.34861 \\ 0.0382119 & -1.73957
\end{pmatrix} \,, \\
c_2 &= 
\begin{pmatrix}
-103.787 & 31.4819 \\ 2.62254 & -33.1654
\end{pmatrix} \,.
\end{align}
For the $h_{2a}$ and $h_{2b}$ operators we have
\begin{align}
c_1 &= 
\begin{pmatrix}
-8.12940 & 1.35805 \\ 0.0191060 & -0.196067
\end{pmatrix} \,, \\
c_2 &= 
\begin{pmatrix}
-106.004 & 13.5855 \\ 0.0500372 & -31.0241
\end{pmatrix} \,.
\end{align}
In the case of the $h_{2c}$ operators one gets
\begin{align}
c_1 &= 
\begin{pmatrix}
-8.14087 & 1.29815 \\ -0.0191060 & -0.234279
\end{pmatrix} \,, \\
c_2 &= 
\begin{pmatrix}
-105.975 & 14.3534 \\ -0.0500372 & -31.1242
\end{pmatrix} \,.
\end{align}

\section{Conversion factors for three-quark operators}
\label{sec.confac3}

In this Appendix we present the conversion matrices (factors) of the
three-quark operators that we consider, as obtained in
Ref.~\cite{Gruber:2017ozo} to one-loop accuracy. Here we need the
trigamma function
\begin{equation}
\psi_1 (x) = \frac{d^2}{dx^2} \ln (\Gamma (x)) =
  \int_0^\infty \!\! dt \frac{t \mathrm e^{-tx}}{1 - \mathrm e^{-t}}
\end{equation}
with the special values $\psi_1 (1/3) \approx 10.0956$ and 
$\psi_1 (1/4) \approx 17.1973$.

For the multiplets (\ref{eq.o12.1}) and (\ref{eq.d12.1}) of operators
without derivatives transforming according to the $\overline{\mathrm {H(4)}}$
representation $\tau^{\underbar{$\scriptstyle 12$}}_1$ we have
\begin{equation}
C(\mathscr O_1^{\underbar{$\scriptstyle 12$}})
  = C(\mathscr D_1^{\underbar{$\scriptstyle 12$}}) 
\end{equation}
with
\begin{equation}
\begin{split}
C(\mathscr D_1^{\underbar{$\scriptstyle 12$}})
  = 1 + \frac{\gc^2}{16 \pi^2} \Big( &
  - \frac{5}{9} - \frac{5}{81} \pi^2 + \frac{10}{27} \ln(2) \\ & {}
  + \frac{4}{27} \psi_1(1/3) - \frac{1}{27} \psi_1(1/4) \Big) \,.  
\end{split}
\end{equation}
In the case of the multiplets (\ref{eq.s4.1}), (\ref{eq.o4.1a}),
and (\ref{eq.o4.1b}) one finds
\begin{equation}
\begin{split}
C(\mathscr S_1^{\underbar{$\scriptstyle 4$}})
  =  1 + \frac{\gc^2}{16 \pi^2} \Big( &
  \frac{17}{3} + \frac{14}{27} \pi^2 + \frac{2}{9} \ln(2) \\ & {} 
  - \frac{4}{9} \psi_1(1/3) - \frac{2}{9} \psi_1(1/4) \Big) 
\end{split}
\end{equation}
and
\begin{equation}
C(\mathscr O_1^{\underbar{$\scriptstyle 4$}})
  = C(\mathscr S_1^{\underbar{$\scriptstyle 4$}}) \mathds{1}_2 \,. 
\end{equation}
For the multiplet (\ref{eq.s12.2}) of operators with one derivative the
conversion factor reads
\begin{equation}
\begin{split}
C(\mathscr S_2^{\underbar{$\scriptstyle 12$}})
  =  1 + \frac{\gc^2}{16 \pi^2} \Big( &
  - \frac{707}{162} - \frac{3625}{23328} \pi^2 + \frac{17}{54} \ln(2) \\ & {} 
  + \frac{131}{486} \psi_1(1/3) - \frac{7}{288} \psi_1(1/4) \Big) \,.
\end{split}
\end{equation}
The octet multiplets (\ref{eq.o12.2a}) - (\ref{eq.o12.2d}) have a
$4 \times 4$ mixing matrix with the diagonal entries
\begin{equation}
\begin{split}
C(\mathscr O_2^{\underbar{$\scriptstyle 12$}})_{11}
  =  1 + \frac{\gc^2}{16 \pi^2} \Big( &
  - \frac{53}{81} - \frac{149}{1458} \pi^2 + \frac{8}{27} \ln(2) \\ & {} 
  + \frac{44}{243} \psi_1(1/3) - \frac{1}{54} \psi_1(1/4) \Big) \,,
\end{split}
\end{equation}
\begin{equation}
\begin{split}
C(\mathscr O_2^{\underbar{$\scriptstyle 12$}})_{22}
  =  1 + \frac{\gc^2}{16 \pi^2} \Big( &
  - \frac{845}{162} - \frac{5413}{23328} \pi^2 + \frac{11}{54} \ln(2) \\ & {} 
  + \frac{143}{486} \psi_1(1/3) + \frac{31}{864} \psi_1(1/4) \Big) \,,
\end{split}
\end{equation}
\begin{equation}
\begin{split}
C(\mathscr O_2^{\underbar{$\scriptstyle 12$}})_{33}
  =  1 + \frac{\gc^2}{16 \pi^2} \Big( &
  - \frac{707}{162} - \frac{3625}{23328} \pi^2 + \frac{17}{54} \ln(2) \\ & {} 
  + \frac{131}{486} \psi_1(1/3) - \frac{7}{288} \psi_1(1/4) \Big) \,,
\end{split}
\end{equation}
\begin{equation}
\begin{split}
C(\mathscr O_2^{\underbar{$\scriptstyle 12$}})_{44}
  =  1 + \frac{\gc^2}{16 \pi^2} \Big( &
  - 7 - \frac{115}{288} \pi^2 + \frac{4}{9} \ln(2) \\ & {} 
  + \frac{1}{2} \psi_1(1/3) + \frac{19}{288} \psi_1(1/4) \Big) \,.
\end{split}
\end{equation}
The only nonvanishing off-diagonal entries are
\begin{equation}
\begin{split}
& C(\mathscr O_2^{\underbar{$\scriptstyle 12$}})_{21} =
4 C(\mathscr O_2^{\underbar{$\scriptstyle 12$}})_{12} 
= \sqrt{2} \frac{\gc^2}{16 \pi^2} \Big(
- \frac{4}{81} - \frac{95}{2916} \pi^2 \\ & {}
- \frac{2}{27} \ln(2)  
  - \frac{5}{243} \psi_1(1/3) + \frac{5}{108} \psi_1(1/4) \Big) \,.
\end{split}
\end{equation}
In the case of the decuplet multiplets (\ref{eq.d12.2a}) - (\ref{eq.d12.2c})
we get a $3 \times 3$ mixing matrix, whose nonzero entries are given by
\begin{equation}
C(\mathscr D_2^{\underbar{$\scriptstyle 12$}})_{mn} = 
C(\mathscr O_2^{\underbar{$\scriptstyle 12$}})_{mn}
\end{equation}
for $m,n \in \{1,2\}$ and
\begin{equation}
\begin{split}
C(\mathscr D_2^{\underbar{$\scriptstyle 12$}})_{33}
  =  1 + \frac{\gc^2}{16 \pi^2} \Big( &
  - \frac{11}{3} - \frac{19}{54} \pi^2 + \frac{4}{9} \ln(2) \\ & {} 
  + \frac{4}{9} \psi_1(1/3) + \frac{1}{18} \psi_1(1/4) \Big) \,.
\end{split}
\end{equation}

\section{Tables of results}
\label{sec.tables}

In this Appendix we present tables of our results obtained with the help
of the methods discussed above.

\begin{table*}[h]
\caption{Results from fits for operators with less than two derivatives
based on the \RI-MOM scheme without the perturbative subtraction of
lattice artifacts. The chiral extrapolation has been performed globally.
The first number in parentheses gives the statistical error, while the
second number is an estimate of the systematic uncertainty. All values
refer to the $\MS$ scheme at the scale $\mu_0^2 = 4 \, \mathrm {GeV}^2$.
See Table~\ref{tab.Nmomfit1} of the Addendum for updated results.}
\label{tab.momfit1}
\begin{ruledtabular}
\begin{tabular}{cllllll}
{ } &  \multicolumn{1}{c}{$\beta=3.34$} &  \multicolumn{1}{c}{$\beta=3.40$}
    &  \multicolumn{1}{c}{$\beta=3.46$} &  \multicolumn{1}{c}{$\beta=3.55$}
    &  \multicolumn{1}{c}{$\beta=3.70$} &  \multicolumn{1}{c}{$\beta=3.85$} \\
\hline
$Z_q$                &     0.7821(11)(65) &     0.7899(5)(60) &     0.7970(2)(52) &     0.8069(4)(47) &     0.8220(10)(34) &     0.8357(13)(30) \\
$Z_S$                &     0.6322(154)(300) &     0.6202(71)(255) &     0.6089(52)(261) &     0.5964(78)(267) &     0.5831(119)(233) &     0.5729(138)(183) \\
$Z_V$                &     0.6915(22)(78) &     0.7011(13)(61) &     0.7102(7)(53) &     0.7229(11)(52) &     0.7419(24)(48) &     0.7589(33)(54) \\
$Z^\prime_V$         &     0.6968(21)(63) &     0.7065(11)(47) &     0.7156(7)(39) &     0.7279(11)(38) &     0.7465(24)(37) &     0.7628(32)(45) \\
$Z_A$                &     0.7372(24)(74) &     0.7449(12)(56) &     0.7525(7)(55) &     0.7628(13)(59) &     0.7782(22)(58) &     0.7919(29)(57) \\
$Z^\prime_A$         &     0.7440(22)(42) &     0.7514(12)(30) &     0.7582(6)(27) &     0.7678(13)(29) &     0.7823(22)(30) &     0.7952(30)(36) \\
$Z_T$                &     0.8063(21)(126) &     0.8202(10)(99) &     0.8337(6)(97) &     0.8519(11)(102) &     0.8794(23)(99) &     0.9048(31)(98) \\
$Z_{v_{2a}}$         &     1.0527(18)(179) &     1.0786(9)(167) &     1.1020(6)(164) &     1.1335(10)(162) &     1.1798(21)(161) &     1.2228(27)(187) \\
$Z_{v_{2b}}$         &     1.0672(17)(136) &     1.0916(9)(128) &     1.1136(5)(122) &     1.1431(10)(118) &     1.1868(19)(116) &     1.2269(25)(134) \\
$Z_{r_{2a}}$         &     1.0514(20)(185) &     1.0776(10)(172) &     1.1012(6)(171) &     1.1329(11)(171) &     1.1791(23)(167) &     1.2217(30)(183) \\
$Z_{r_{2b}}$         &     1.0907(19)(142) &     1.1146(10)(135) &     1.1358(5)(129) &     1.1645(10)(120) &     1.2067(20)(107) &     1.2452(26)(113) \\
$Z_{h_{1a}}$         &     1.0794(19)(183) &     1.1083(8)(172) &     1.1344(5)(169) &     1.1696(11)(167) &     1.2218(22)(166) &     1.2702(30)(194) \\
$Z_{h_{1b}}$         &     1.0896(20)(186) &     1.1182(11)(175) &     1.1440(6)(170) &     1.1788(11)(163) &     1.2305(22)(158) &     1.2787(30)(186) \\
$Z_A/Z_V$            &     1.0643(32)(47) &     1.0613(15)(42) &     1.0585(11)(36) &     1.0546(16)(28) &     1.0486(30)(10) &     1.0433(41)(9) \\
$Z'_A/Z'_V$          &     1.0660(32)(40) &     1.0624(17)(29) &     1.0589(10)(24) &     1.0543(16)(21) &     1.0477(30)(19) &     1.0424(40)(24) \\
\end{tabular}
\end{ruledtabular}

\end{table*}
 
\begin{table*}[h]
\caption{Results from fits for operators with less than two derivatives
based on the \RI-MOM scheme with the perturbative subtraction of lattice
artifacts. The chiral extrapolation has been performed globally. The first
number in parentheses gives the statistical error, while the second number
is an estimate of the systematic uncertainty. All values refer to the
$\MS$ scheme at the scale $\mu_0^2 = 4 \, \mathrm {GeV}^2$. See
Table~\ref{tab.Nmomfit2} of the Addendum for updated results.}
\label{tab.momfit2}
\begin{ruledtabular}
\begin{tabular}{cllllll}
{ } &  \multicolumn{1}{c}{$\beta=3.34$} &  \multicolumn{1}{c}{$\beta=3.40$}
    &  \multicolumn{1}{c}{$\beta=3.46$} &  \multicolumn{1}{c}{$\beta=3.55$}
    &  \multicolumn{1}{c}{$\beta=3.70$} &  \multicolumn{1}{c}{$\beta=3.85$} \\
\hline
$Z_q$                &     0.7807(11)(47) &     0.7892(5)(44) &     0.7968(2)(40) &     0.8074(4)(37) &     0.8230(10)(29) &     0.8369(13)(25) \\
$Z_S$                &     0.6041(154)(242) &     0.5939(71)(211) &     0.5846(52)(214) &     0.5744(78)(216) &     0.5639(119)(188) &     0.5560(109)(148) \\
$Z_V$                &     0.6995(22)(35) &     0.7098(13)(28) &     0.7190(7)(22) &     0.7317(11)(17) &     0.7506(24)(13) &     0.7669(33)(20) \\
$Z^\prime_V$         &     0.7027(21)(25) &     0.7130(11)(17) &     0.7222(7)(12) &     0.7347(11)(8) &     0.7533(24)(11) &     0.7690(32)(17) \\
$Z_A$                &     0.7432(24)(28) &     0.7516(12)(23) &     0.7593(7)(24) &     0.7698(13)(25) &     0.7852(22)(25) &     0.7983(29)(24) \\
$Z^\prime_A$         &     0.7474(22)(6) &     0.7555(12)(4) &     0.7626(6)(5) &     0.7724(13)(5) &     0.7870(22)(9) &     0.7995(30)(11) \\
$Z_T$                &     0.8182(22)(68) &     0.8330(10)(61) &     0.8465(6)(59) &     0.8649(11)(56) &     0.8922(23)(52) &     0.9166(31)(49) \\
$Z_{v_{2a}}$         &     1.0632(18)(167) &     1.0894(9)(160) &     1.1128(6)(154) &     1.1444(10)(145) &     1.1909(21)(137) &     1.2334(27)(155) \\
$Z_{v_{2b}}$         &     1.0727(17)(132) &     1.0974(9)(127) &     1.1193(5)(121) &     1.1489(10)(112) &     1.1927(19)(102) &     1.2323(25)(110) \\
$Z_{r_{2a}}$         &     1.0615(20)(169) &     1.0880(10)(161) &     1.1116(6)(155) &     1.1433(11)(148) &     1.1897(23)(139) &     1.2317(30)(148) \\
$Z_{r_{2b}}$         &     1.0970(19)(141) &     1.1211(10)(136) &     1.1424(5)(129) &     1.1712(10)(119) &     1.2135(20)(102) &     1.2517(26)(100) \\
$Z_{h_{1a}}$         &     1.0890(19)(173) &     1.1182(8)(166) &     1.1443(5)(160) &     1.1797(11)(151) &     1.2322(22)(142) &     1.2801(30)(158) \\
$Z_{h_{1b}}$         &     1.1002(20)(182) &     1.1291(11)(175) &     1.1550(6)(166) &     1.1900(11)(154) &     1.2420(22)(141) &     1.2896(29)(154) \\
$Z_A/Z_V$            &     1.0622(32)(52) &     1.0592(15)(46) &     1.0565(11)(41) &     1.0527(16)(33) &     1.0468(31)(16) &     1.0418(41)(9) \\
$Z'_A/Z'_V$          &     1.0636(31)(37) &     1.0600(17)(28) &     1.0566(10)(21) &     1.0521(16)(14) &     1.0456(30)(8) &     1.0406(40)(15) \\
\end{tabular}
\end{ruledtabular}

\end{table*}
 
\begin{table*}[h]
\caption{Results from fits for operators with two derivatives based on the
\RI-MOM scheme without the perturbative subtraction of lattice artifacts.
The chiral extrapolation has been performed globally. The
first number in parentheses gives the statistical error, while the second
number is an estimate of the systematic uncertainty. All values refer to the
$\MS$ scheme at the scale $\mu_0^2 = 4 \, \mathrm {GeV}^2$.}
\label{tab.momfit3}
\begin{ruledtabular}
\begin{tabular}{cllllll}
{ } &  \multicolumn{1}{c}{$\beta=3.34$} &  \multicolumn{1}{c}{$\beta=3.40$}
    &  \multicolumn{1}{c}{$\beta=3.46$} &  \multicolumn{1}{c}{$\beta=3.55$}
    &  \multicolumn{1}{c}{$\beta=3.70$} &  \multicolumn{1}{c}{$\beta=3.85$} \\
\hline
$Z_{v_{3}}$          &     1.3573(35)(407) &     1.3976(18)(394) &     1.4333(11)(379) &     1.4810(19)(349) &     1.5523(42)(290) &     1.6192(56)(284) \\
$Z_{a_{2}}$          &     1.3607(37)(405) &     1.4011(19)(391) &     1.4369(11)(377) &     1.4848(20)(348) &     1.5561(45)(290) &     1.6227(61)(278) \\
$Z_{h_{2a}}$         &     1.3774(35)(391) &     1.4198(18)(378) &     1.4571(10)(365) &     1.5072(19)(335) &     1.5824(42)(276) &     1.6527(57)(269) \\
$Z_{h_{2b}}$         &     1.3839(36)(402) &     1.4259(18)(388) &     1.4627(11)(376) &     1.5123(20)(344) &     1.5863(43)(283) &     1.6556(58)(265) \\
$Z_{h_{2c}}$         &     1.3854(35)(391) &     1.4274(18)(379) &     1.4644(11)(366) &     1.5141(20)(336) &     1.5884(44)(277) &     1.6577(58)(266) \\
\end{tabular}
\end{ruledtabular}

\end{table*}
 
\begin{table*}[h]
\caption{Results for operators with less than two derivatives based on the
\RI-MOM scheme, obtained by means of the fixed-scale method without the
perturbative subtraction of lattice artifacts.
The chiral extrapolation has been performed globally. The first
number in parentheses gives the statistical error, while the second number
is an estimate of the systematic uncertainty. All values refer to the
$\MS$ scheme at the scale $\mu_0^2 = 4 \, \mathrm {GeV}^2$. See
Table~\ref{tab.Nmomextract1} of the Addendum for updated results.}
\label{tab.momextract1}
\begin{ruledtabular}
\begin{tabular}{cllllll}
{ } &  \multicolumn{1}{c}{$\beta=3.34$} &  \multicolumn{1}{c}{$\beta=3.40$}
    &  \multicolumn{1}{c}{$\beta=3.46$} &  \multicolumn{1}{c}{$\beta=3.55$}
    &  \multicolumn{1}{c}{$\beta=3.70$} &  \multicolumn{1}{c}{$\beta=3.85$} \\
\hline
$Z_q$                &     0.8089(4)(24) &     0.8111(2)(24) &     0.8144(1)(24) &     0.8204(1)(24) &     0.8313(4)(28) &     0.8426(5)(27) \\
$Z_S$                &     0.6978(54)(134) &     0.6713(31)(113) &     0.6473(23)(110) &     0.6220(30)(100) &     0.5880(62)(102) &     0.5648(85)(118) \\
$Z_V$                &     0.7005(8)(12) &     0.7074(5)(16) &     0.7156(2)(13) &     0.7275(3)(10) &     0.7472(9)(10) &     0.7639(13)(18) \\
$Z^\prime_V$         &     0.7070(8)(15) &     0.7137(4)(14) &     0.7213(3)(12) &     0.7322(3)(6) &     0.7501(9)(3) &     0.7648(13)(18) \\
$Z_A$                &     0.7456(8)(18) &     0.7509(4)(15) &     0.7565(3)(14) &     0.7660(4)(10) &     0.7812(9)(12) &     0.7947(12)(23) \\
$Z^\prime_A$         &     0.7561(9)(17) &     0.7604(5)(12) &     0.7648(3)(15) &     0.7725(4)(9) &     0.7851(10)(3) &     0.7965(14)(13) \\
$Z_T$                &     0.8083(7)(37) &     0.8214(4)(38) &     0.8345(2)(39) &     0.8541(3)(40) &     0.8853(8)(43) &     0.9132(11)(46) \\
$Z_{v_{2a}}$         &     1.0482(4)(111) &     1.0760(3)(114) &     1.1023(2)(118) &     1.1389(3)(124) &     1.1923(7)(140) &     1.2399(10)(137) \\
$Z_{v_{2b}}$         &     1.0666(4)(86) &     1.0920(3)(88) &     1.1155(2)(91) &     1.1478(3)(95) &     1.1944(7)(108) &     1.2364(9)(106) \\
$Z_{r_{2a}}$         &     1.0438(5)(111) &     1.0726(3)(115) &     1.0993(2)(119) &     1.1363(3)(124) &     1.1893(8)(142) &     1.2363(11)(138) \\
$Z_{r_{2b}}$         &     1.0898(5)(88) &     1.1147(3)(91) &     1.1374(2)(93) &     1.1685(3)(97) &     1.2135(7)(108) &     1.2537(10)(105) \\
$Z_{h_{1a}}$         &     1.0761(5)(116) &     1.1069(3)(121) &     1.1357(2)(125) &     1.1756(3)(131) &     1.2341(8)(149) &     1.2868(11)(147) \\
$Z_{h_{1b}}$         &     1.0887(5)(117) &     1.1185(3)(122) &     1.1465(2)(126) &     1.1857(3)(132) &     1.2433(8)(150) &     1.2956(11)(147) \\
$Z_A/Z_V$            &     1.0636(10)(5) &     1.0605(6)(7) &     1.0566(4)(7) &     1.0524(5)(10) &     1.0451(11)(7) &     1.0397(17)(13) \\
$Z'_A/Z'_V$          &     1.0688(11)(7) &     1.0648(6)(6) &     1.0597(4)(5) &     1.0541(6)(8) &     1.0456(13)(3) &     1.0399(19)(7) \\
\end{tabular}
\end{ruledtabular}

\end{table*}
 
\begin{table*}[h]
\caption{Results for operators with less than two derivatives based on the
\RI-MOM scheme, obtained by means of the fixed-scale method with the
perturbative subtraction of lattice artifacts.
The chiral extrapolation has been performed globally. The first
number in parentheses gives the statistical error, while the second number
is an estimate of the systematic uncertainty. All values refer to the
$\MS$ scheme at the scale $\mu_0^2 = 4 \, \mathrm {GeV}^2$. See
Table~\ref{tab.Nmomextract2} of the Addendum for updated results.}
\label{tab.momextract2}
\begin{ruledtabular}
\begin{tabular}{cllllll}
{ } &  \multicolumn{1}{c}{$\beta=3.34$} &  \multicolumn{1}{c}{$\beta=3.40$}
    &  \multicolumn{1}{c}{$\beta=3.46$} &  \multicolumn{1}{c}{$\beta=3.55$}
    &  \multicolumn{1}{c}{$\beta=3.70$} &  \multicolumn{1}{c}{$\beta=3.85$} \\
\hline
$Z_q$                &     0.7893(4)(23) &     0.7961(2)(23) &     0.8027(1)(23) &     0.8121(1)(24) &     0.8266(4)(28) &     0.8398(5)(28) \\
$Z_S$                &     0.6507(54)(127) &     0.6298(31)(105) &     0.6105(23)(104) &     0.5914(30)(94) &     0.5655(64)(98) &     0.5483(87)(117) \\
$Z_V$                &     0.7038(8)(11) &     0.7128(5)(15) &     0.7220(2)(13) &     0.7341(3)(10) &     0.7529(9)(10) &     0.7683(13)(19) \\
$Z^\prime_V$         &     0.7080(8)(14) &     0.7168(4)(14) &     0.7254(3)(12) &     0.7368(3)(6) &     0.7541(9)(3) &     0.7680(13)(19) \\
$Z_A$                &     0.7446(8)(18) &     0.7525(5)(15) &     0.7594(3)(14) &     0.7697(4)(10) &     0.7848(9)(12) &     0.7977(12)(24) \\
$Z^\prime_A$         &     0.7522(9)(17) &     0.7591(5)(12) &     0.7650(3)(15) &     0.7737(4)(9) &     0.7868(10)(3) &     0.7981(14)(13) \\
$Z_T$                &     0.8185(7)(39) &     0.8335(4)(39) &     0.8470(2)(40) &     0.8661(3)(40) &     0.8951(8)(43) &     0.9207(11)(46) \\
$Z_{v_{2a}}$         &     1.0616(5)(112) &     1.0892(3)(115) &     1.1149(2)(119) &     1.1501(3)(124) &     1.2011(7)(140) &     1.2465(10)(137) \\
$Z_{v_{2b}}$         &     1.0730(4)(87) &     1.0986(2)(89) &     1.1219(2)(92) &     1.1536(3)(95) &     1.1989(7)(108) &     1.2397(9)(106) \\
$Z_{r_{2a}}$         &     1.0571(5)(112) &     1.0857(3)(116) &     1.1117(2)(119) &     1.1473(3)(124) &     1.1979(8)(142) &     1.2428(11)(138) \\
$Z_{r_{2b}}$         &     1.0961(5)(89) &     1.1215(3)(91) &     1.1442(2)(94) &     1.1749(3)(97) &     1.2187(7)(108) &     1.2579(10)(105) \\
$Z_{h_{1a}}$         &     1.0874(5)(118) &     1.1182(3)(122) &     1.1466(2)(126) &     1.1855(3)(131) &     1.2419(8)(149) &     1.2927(11)(147) \\
$Z_{h_{1b}}$         &     1.1004(5)(119) &     1.1305(3)(123) &     1.1582(2)(127) &     1.1965(3)(132) &     1.2521(8)(150) &     1.3024(11)(147) \\
$Z_A/Z_V$            &     1.0593(10)(6) &     1.0566(6)(7) &     1.0531(4)(8) &     1.0495(5)(10) &     1.0430(12)(7) &     1.0383(17)(13) \\
$Z'_A/Z'_V$          &     1.0640(11)(8) &     1.0604(6)(7) &     1.0557(5)(5) &     1.0508(6)(8) &     1.0433(13)(3) &     1.0383(21)(7) \\
\end{tabular}
\end{ruledtabular}

\end{table*}
 
\begin{table*}[h]
\caption{Results for operators with two derivatives based on the
\RI-MOM scheme, obtained by means of the fixed-scale method without the
perturbative subtraction of lattice artifacts.
The chiral extrapolation has been performed globally. The first
number in parentheses gives the statistical error, while the second number
is an estimate of the systematic uncertainty. All values refer to the
$\MS$ scheme at the scale $\mu_0^2 = 4 \, \mathrm {GeV}^2$.}
\label{tab.momextract3}
\begin{ruledtabular}
\begin{tabular}{cllllll}
{ } &  \multicolumn{1}{c}{$\beta=3.34$} &  \multicolumn{1}{c}{$\beta=3.40$}
    &  \multicolumn{1}{c}{$\beta=3.46$} &  \multicolumn{1}{c}{$\beta=3.55$}
    &  \multicolumn{1}{c}{$\beta=3.70$} &  \multicolumn{1}{c}{$\beta=3.85$} \\
\hline
$Z_{v_{3}}$          &     1.3590(13)(225) &     1.4007(8)(234) &     1.4401(5)(243) &     1.4945(7)(255) &     1.5731(19)(291) &     1.6449(27)(284) \\
$Z_{a_{2}}$          &     1.3587(14)(226) &     1.4011(8)(234) &     1.4409(5)(244) &     1.4958(7)(256) &     1.5741(20)(296) &     1.6447(28)(286) \\
$Z_{h_{2a}}$         &     1.3820(13)(213) &     1.4250(8)(222) &     1.4655(4)(230) &     1.5211(7)(242) &     1.6010(19)(280) &     1.6741(27)(271) \\
$Z_{h_{2b}}$         &     1.3913(14)(215) &     1.4335(8)(223) &     1.4729(5)(232) &     1.5274(7)(243) &     1.6058(20)(283) &     1.6778(28)(272) \\
$Z_{h_{2c}}$         &     1.3895(14)(214) &     1.4324(8)(223) &     1.4725(5)(232) &     1.5277(7)(243) &     1.6071(19)(280) &     1.6794(27)(272) \\
\end{tabular}
\end{ruledtabular}

\end{table*}
 
\begin{table*}[h]
\caption{Results from fits for operators with less than two derivatives
based on the \RI-SMOM scheme without the perturbative subtraction of
lattice artifacts. The chiral extrapolation has been performed globally. 
The first number in parentheses gives the statistical error, while the
second number is an estimate of the systematic uncertainty. All values
refer to the $\MS$ scheme at the scale $\mu_0^2 = 4 \, \mathrm {GeV}^2$.
See Table~\ref{tab.Nsmomfit1} of the Addendum for updated results.}
\label{tab.smomfit1}
\begin{ruledtabular}
\begin{tabular}{cllllll}
{ } &  \multicolumn{1}{c}{$\beta=3.34$} &  \multicolumn{1}{c}{$\beta=3.40$}
    &  \multicolumn{1}{c}{$\beta=3.46$} &  \multicolumn{1}{c}{$\beta=3.55$}
    &  \multicolumn{1}{c}{$\beta=3.70$} &  \multicolumn{1}{c}{$\beta=3.85$} \\
\hline
$Z_q$                &     0.7859(9)(91) &     0.7938(4)(84) &     0.8007(2)(79) &     0.8105(3)(71) &     0.8253(9)(53) &     0.8389(12)(43) \\
$Z_S$                &     0.6233(14)(79) &     0.6179(7)(73) &     0.6128(4)(66) &     0.6062(5)(55) &     0.5962(11)(37) &     0.5864(16)(28) \\
$Z_P$                &     0.4958(16)(255) &     0.4968(6)(229) &     0.4978(3)(205) &     0.5006(5)(176) &     0.5056(13)(116) &     0.5092(19)(66) \\
$Z_V$                &     0.7151(11)(96) &     0.7236(5)(87) &     0.7310(3)(79) &     0.7417(4)(67) &     0.7581(10)(45) &     0.7730(15)(40) \\
$Z^\prime_V$         &     0.7104(13)(86) &     0.7192(6)(72) &     0.7268(3)(63) &     0.7378(5)(53) &     0.7550(13)(38) &     0.7702(18)(49) \\
$Z_A$                &     0.7464(10)(66) &     0.7543(4)(62) &     0.7612(3)(58) &     0.7711(4)(49) &     0.7857(10)(37) &     0.7988(14)(34) \\
$Z^\prime_A$         &     0.7625(9)(74) &     0.7679(4)(63) &     0.7724(2)(62) &     0.7795(4)(60) &     0.7909(10)(46) &     0.8017(14)(35) \\
$Z_T$                &     0.8313(12)(101) &     0.8447(5)(94) &     0.8565(3)(87) &     0.8731(4)(77) &     0.8981(12)(60) &     0.9214(16)(62) \\
$Z_{v_{2a}}$         &     1.0707(17)(304) &     1.0985(9)(290) &     1.1226(6)(278) &     1.1552(8)(251) &     1.2026(21)(203) &     1.2451(28)(177) \\
$Z_{v_{2b}}$         &     1.0655(16)(315) &     1.0920(8)(300) &     1.1150(6)(288) &     1.1465(8)(260) &     1.1927(20)(209) &     1.2347(28)(180) \\
$Z_{r_{2a}}$         &     1.0640(18)(298) &     1.0918(9)(286) &     1.1160(6)(273) &     1.1486(8)(248) &     1.1959(21)(202) &     1.2384(30)(172) \\
$Z_{r_{2b}}$         &     1.0842(18)(314) &     1.1106(9)(301) &     1.1334(6)(288) &     1.1644(9)(261) &     1.2096(22)(210) &     1.2505(30)(177) \\
$Z_{h_{1a}}$         &     1.1388(18)(374) &     1.1677(9)(361) &     1.1934(6)(351) &     1.2274(9)(334) &     1.2756(23)(299) &     1.3194(32)(270) \\
$Z_{h_{1b}}$         &     1.1529(18)(379) &     1.1817(10)(365) &     1.2071(6)(356) &     1.2408(10)(338) &     1.2884(24)(302) &     1.3315(32)(273) \\
$Z_S/Z_P$            &     1.2119(39)(635) &     1.2065(17)(586) &     1.2001(10)(546) &     1.1880(14)(477) &     1.1657(33)(318) &     1.1437(44)(160) \\
$Z_A/Z_V$            &     1.0412(7)(59) &     1.0407(4)(55) &     1.0400(2)(51) &     1.0387(2)(45) &     1.0360(5)(29) &     1.0333(8)(18) \\
$Z'_A/Z'_V$          &     1.0705(12)(60) &     1.0655(6)(46) &     1.0609(3)(35) &     1.0550(4)(25) &     1.0466(9)(22) &     1.0400(13)(29) \\
$Z_A/Z_P$      &     1.4529(39)(462) &     1.4732(16)(421) &     1.4908(8)(412) &     1.5102(12)(384) &     1.5335(35)(281) &     1.5543(51)(212) \\
$Z'_A/Z_P$     &     1.4582(47)(335) &     1.4753(15)(316) &     1.4891(9)(297) &     1.5045(15)(258) &     1.5239(41)(169) &     1.5422(59)(149) \\
$Z_P/(Z_S Z_A)$      &     1.0893(34)(439) &     1.0847(16)(418) &     1.0823(10)(399) &     1.0816(12)(339) &     1.0860(31)(204) &     1.0916(43)(126) \\
$Z_P/(Z_S Z'_A)$     &     1.0661(32)(377) &     1.0655(14)(351) &     1.0666(10)(330) &     1.0702(13)(265) &     1.0790(34)(137) &     1.0876(46)(94) \\
\end{tabular}
\end{ruledtabular}

\end{table*}
 
\begin{table*}[h]
\caption{Results from fits for operators with less than two derivatives
based on the \RI-SMOM scheme with the perturbative subtraction of lattice
artifacts. The chiral extrapolation has been performed globally. The first
number in parentheses gives the statistical error, while the second number
is an estimate of the systematic uncertainty. All values refer to the
$\MS$ scheme at the scale $\mu_0^2 = 4 \, \mathrm {GeV}^2$. See
Table~\ref{tab.Nsmomfit2} of the Addendum for updated results.}
\label{tab.smomfit2}
\begin{ruledtabular}
\begin{tabular}{cllllll}
{ } &  \multicolumn{1}{c}{$\beta=3.34$} &  \multicolumn{1}{c}{$\beta=3.40$}
    &  \multicolumn{1}{c}{$\beta=3.46$} &  \multicolumn{1}{c}{$\beta=3.55$}
    &  \multicolumn{1}{c}{$\beta=3.70$} &  \multicolumn{1}{c}{$\beta=3.85$} \\
\hline
$Z_q$                &     0.7822(9)(70) &     0.7908(4)(66) &     0.7982(2)(63) &     0.8087(3)(57) &     0.8241(9)(45) &     0.8379(12)(36) \\
$Z_S$                &     0.6112(14)(92) &     0.6065(7)(85) &     0.6022(4)(79) &     0.5966(5)(66) &     0.5878(11)(44) &     0.5791(16)(31) \\
$Z_P$                &     0.4879(15)(215) &     0.4899(6)(194) &     0.4919(3)(172) &     0.4956(5)(145) &     0.5018(13)(91) &     0.5062(19)(49) \\
$Z_V$                &     0.7136(11)(76) &     0.7227(5)(69) &     0.7307(3)(64) &     0.7419(4)(56) &     0.7587(10)(41) &     0.7737(15)(34) \\
$Z^\prime_V$         &     0.7071(13)(59) &     0.7166(6)(49) &     0.7249(3)(43) &     0.7365(5)(37) &     0.7543(13)(27) &     0.7697(19)(33) \\
$Z_A$                &     0.7456(10)(57) &     0.7540(4)(54) &     0.7613(3)(51) &     0.7715(4)(46) &     0.7863(10)(37) &     0.7994(14)(30) \\
$Z^\prime_A$         &     0.7578(9)(41) &     0.7640(4)(36) &     0.7693(2)(36) &     0.7773(4)(36) &     0.7894(10)(30) &     0.8005(14)(21) \\
$Z_T$                &     0.8321(11)(93) &     0.8462(5)(88) &     0.8585(3)(83) &     0.8755(4)(75) &     0.9009(12)(61) &     0.9241(16)(52) \\
$Z_{v_{2a}}$         &     1.0731(17)(307) &     1.1010(9)(293) &     1.1251(6)(282) &     1.1578(8)(256) &     1.2053(21)(207) &     1.2476(28)(177) \\
$Z_{v_{2b}}$         &     1.0672(16)(318) &     1.0938(8)(303) &     1.1170(6)(290) &     1.1485(8)(263) &     1.1949(20)(212) &     1.2367(28)(179) \\
$Z_{r_{2a}}$         &     1.0666(18)(303) &     1.0946(9)(290) &     1.1188(6)(279) &     1.1515(8)(254) &     1.1989(22)(207) &     1.2412(30)(172) \\
$Z_{r_{2b}}$         &     1.0862(18)(316) &     1.1127(9)(304) &     1.1355(6)(293) &     1.1666(9)(267) &     1.2119(22)(215) &     1.2528(30)(177) \\
$Z_{h_{1a}}$         &     1.1412(18)(373) &     1.1704(9)(360) &     1.1961(6)(351) &     1.2302(9)(333) &     1.2786(23)(297) &     1.3222(32)(268) \\
$Z_{h_{1b}}$         &     1.1554(18)(377) &     1.1843(10)(365) &     1.2099(6)(355) &     1.2436(10)(337) &     1.2914(24)(301) &     1.3344(32)(270) \\
$Z_S/Z_P$            &     1.2086(39)(616) &     1.2026(17)(567) &     1.1960(10)(530) &     1.1837(13)(464) &     1.1608(33)(306) &     1.1392(45)(159) \\
$Z_A/Z_V$            &     1.0422(7)(50) &     1.0414(4)(45) &     1.0405(2)(42) &     1.0390(2)(36) &     1.0362(6)(23) &     1.0333(8)(13) \\
$Z'_A/Z'_V$          &     1.0692(12)(63) &     1.0643(6)(48) &     1.0599(3)(38) &     1.0542(4)(26) &     1.0459(9)(16) &     1.0395(13)(26) \\
$Z_A/Z_P$      &     1.4582(39)(426) &     1.4785(16)(392) &     1.4958(8)(380) &     1.5150(12)(350) &     1.5379(35)(247) &     1.5582(51)(190) \\
$Z'_A/Z_P$     &     1.4600(46)(325) &     1.4773(16)(309) &     1.4910(9)(286) &     1.5065(14)(245) &     1.5262(41)(158) &     1.5442(59)(148) \\
$Z_P/(Z_S Z_A)$      &     1.0945(34)(446) &     1.0897(16)(423) &     1.0871(10)(404) &     1.0862(12)(342) &     1.0903(31)(203) &     1.0954(43)(127) \\
$Z_P/(Z_S Z'_A)$     &     1.0751(33)(401) &     1.0740(14)(371) &     1.0746(10)(349) &     1.0775(13)(284) &     1.0856(33)(152) &     1.0934(46)(102) \\
\end{tabular}
\end{ruledtabular}

\end{table*}
 
\begin{table*}[h]
\caption{Results from fits for operators with two derivatives based on the
\RI-SMOM scheme without the perturbative subtraction of lattice artifacts.
The chiral extrapolation has been performed globally. The
first number in parentheses gives the statistical error, while the second
number is an estimate of the systematic uncertainty. All values refer to the
$\MS$ scheme at the scale $\mu_0^2 = 4 \, \mathrm {GeV}^2$.}
\label{tab.smomfit3}
\begin{ruledtabular}
\begin{tabular}{cllllll}
{ } &  \multicolumn{1}{c}{$\beta=3.34$} &  \multicolumn{1}{c}{$\beta=3.40$}
    &  \multicolumn{1}{c}{$\beta=3.46$} &  \multicolumn{1}{c}{$\beta=3.55$}
    &  \multicolumn{1}{c}{$\beta=3.70$} &  \multicolumn{1}{c}{$\beta=3.85$} \\
\hline
$Z_{v_{3}}$          &     1.3792(36)(697) &     1.4228(16)(673) &     1.4593(13)(657) &     1.5092(20)(608) &     1.5826(49)(489) &     1.6493(68)(386) \\
$Z_{a_{2}}$          &     1.3729(37)(682) &     1.4168(17)(659) &     1.4536(13)(644) &     1.5039(19)(593) &     1.5776(48)(476) &     1.6445(66)(379) \\
$Z_{h_{2a}}$         &     1.4682(40)(725) &     1.5111(19)(707) &     1.5484(14)(690) &     1.5977(20)(653) &     1.6668(50)(580) &     1.7296(70)(514) \\
$Z_{h_{2b}}$         &     1.4829(41)(740) &     1.5251(19)(723) &     1.5616(14)(706) &     1.6099(20)(669) &     1.6776(51)(593) &     1.7389(71)(523) \\
$Z_{h_{2c}}$         &     1.4788(38)(739) &     1.5212(18)(721) &     1.5579(14)(703) &     1.6065(21)(665) &     1.6747(52)(589) &     1.7367(71)(521) \\
\end{tabular}
\end{ruledtabular}

\end{table*}
 
\begin{table*}[h]
\caption{Results for operators with less than two derivatives based on the
\RI-SMOM scheme, obtained by means of the fixed-scale method without the
perturbative subtraction of lattice artifacts.
The chiral extrapolation has been performed globally. The first
number in parentheses gives the statistical error, while the second number
is an estimate of the systematic uncertainty. All values refer to the
$\MS$ scheme at the scale $\mu_0^2 = 4 \, \mathrm {GeV}^2$. See
Table~\ref{tab.Nsmomextract1} of the Addendum for updated results.}
\label{tab.smomextract1}
\begin{ruledtabular}
\begin{tabular}{cllllll}
{ } &  \multicolumn{1}{c}{$\beta=3.34$} &  \multicolumn{1}{c}{$\beta=3.40$}
    &  \multicolumn{1}{c}{$\beta=3.46$} &  \multicolumn{1}{c}{$\beta=3.55$}
    &  \multicolumn{1}{c}{$\beta=3.70$} &  \multicolumn{1}{c}{$\beta=3.85$} \\
\hline
          $Z_q$                &     0.8308(3)(26) &     0.8302(2)(25) &     0.8307(1)(24) &     0.8335(1)(24) &     0.8400(3)(31) &     0.8484(4)(28) \\
          $Z_S$                &     0.6343(4)(28) &     0.6268(2)(28) &     0.6199(1)(28) &     0.6111(1)(27) &     0.5981(4)(26) &     0.5863(5)(25) \\
          $Z_P$                &     0.5769(5)(40) &     0.5635(2)(36) &     0.5537(1)(34) &     0.5439(1)(32) &     0.5344(4)(30) &     0.5283(5)(28) \\
          $Z_V$                &     0.7629(4)(24) &     0.7617(2)(22) &     0.7621(1)(22) &     0.7653(1)(21) &     0.7734(4)(23) &     0.7831(5)(22) \\
          $Z^\prime_V$         &     0.7611(5)(13) &     0.7601(3)(11) &     0.7606(1)(10) &     0.7636(2)(10) &     0.7716(5)(9) &     0.7807(7)(4) \\
          $Z_A$                &     0.7776(4)(23) &     0.7787(2)(22) &     0.7807(1)(22) &     0.7856(1)(22) &     0.7949(4)(25) &     0.8048(5)(25) \\
          $Z^\prime_A$         &     0.8171(3)(13) &     0.8123(2)(10) &     0.8092(1)(9) &     0.8074(1)(7) &     0.8082(4)(10) &     0.8120(5)(12) \\
          $Z_T$                &     0.8698(4)(35) &     0.8746(2)(35) &     0.8804(1)(35) &     0.8909(1)(36) &     0.9098(4)(41) &     0.9296(6)(41) \\
          $Z_{v_{2a}}$         &     1.0890(6)(107) &     1.1133(3)(111) &     1.1356(2)(114) &     1.1669(3)(118) &     1.2127(8)(134) &     1.2539(11)(130) \\
          $Z_{v_{2b}}$         &     1.0918(6)(107) &     1.1135(3)(110) &     1.1340(2)(113) &     1.1629(3)(117) &     1.2055(8)(133) &     1.2450(11)(130) \\
          $Z_{r_{2a}}$         &     1.0800(6)(107) &     1.1046(3)(110) &     1.1273(2)(113) &     1.1590(3)(117) &     1.2049(8)(135) &     1.2463(11)(131) \\
          $Z_{r_{2b}}$         &     1.1052(6)(109) &     1.1278(3)(112) &     1.1486(2)(115) &     1.1780(3)(119) &     1.2206(9)(136) &     1.2595(12)(133) \\
          $Z_{h_{1a}}$         &     1.1386(6)(197) &     1.1650(3)(202) &     1.1897(2)(207) &     1.2246(3)(214) &     1.2761(8)(231) &     1.3236(12)(234) \\
          $Z_{h_{1b}}$         &     1.1510(6)(199) &     1.1777(3)(204) &     1.2024(2)(209) &     1.2372(3)(216) &     1.2884(9)(233) &     1.3355(13)(236) \\
          $Z_S/Z_P$            &     1.0988(6)(32) &     1.1118(3)(26) &     1.1194(2)(23) &     1.1235(2)(20) &     1.1196(5)(25) &     1.1105(8)(24) \\
          $Z_A/Z_V$            &     1.0190(2)(5) &     1.0222(1)(4) &     1.0243(1)(4) &     1.0265(1)(4) &     1.0279(2)(4) &     1.0279(2)(5) \\
          $Z'_A/Z'_V$          &     1.0736(4)(4) &     1.0688(2)(4) &     1.0639(1)(7) &     1.0574(1)(10) &     1.0476(3)(21) &     1.0401(5)(20) \\
          $Z_A/Z_P$      &     1.3472(8)(94) &     1.3815(4)(94) &     1.4098(2)(95) &     1.4444(3)(96) &     1.4884(9)(103) &     1.5245(12)(102) \\
          $Z'_A/Z_P$     &     1.3977(10)(94) &     1.4231(5)(93) &     1.4430(3)(94) &     1.4660(4)(94) &     1.4946(10)(97) &     1.5194(14)(99) \\
          $Z_P/(Z_S Z_A)$      &     1.1706(7)(43) &     1.1552(3)(38) &     1.1443(2)(37) &     1.1329(3)(36) &     1.1231(7)(47) &     1.1183(11)(46) \\
          $Z_P/(Z_S Z'_A)$     &     1.1140(6)(20) &     1.1073(3)(11) &     1.1039(2)(10) &     1.1021(3)(10) &     1.1043(7)(26) &     1.1081(10)(31) \\
\end{tabular}
\end{ruledtabular}

\end{table*}
 
\begin{table*}[h]
\caption{Results for operators with less than two derivatives based on the
\RI-SMOM scheme, obtained by means of the fixed-scale method with the
perturbative subtraction of lattice artifacts.
The chiral extrapolation has been performed globally. The first
number in parentheses gives the statistical error, while the second number
is an estimate of the systematic uncertainty. All values refer to the
$\MS$ scheme at the scale $\mu_0^2 = 4 \, \mathrm {GeV}^2$. See
Table~\ref{tab.Nsmomextract2} of the Addendum for updated results.}
\label{tab.smomextract2}
\begin{ruledtabular}
\begin{tabular}{cllllll}
{ } &  \multicolumn{1}{c}{$\beta=3.34$} &  \multicolumn{1}{c}{$\beta=3.40$}
    &  \multicolumn{1}{c}{$\beta=3.46$} &  \multicolumn{1}{c}{$\beta=3.55$}
    &  \multicolumn{1}{c}{$\beta=3.70$} &  \multicolumn{1}{c}{$\beta=3.85$} \\
\hline
          $Z_q$                &     0.7970(3)(24) &     0.8028(2)(23) &     0.8082(1)(24) &     0.8166(1)(24) &     0.8292(3)(31) &     0.8414(4)(29) \\
          $Z_S$                &     0.6071(4)(26) &     0.6026(2)(25) &     0.5983(1)(25) &     0.5929(1)(25) &     0.5844(4)(24) &     0.5762(5)(24) \\
          $Z_P$                &     0.5368(5)(34) &     0.5300(2)(30) &     0.5256(1)(29) &     0.5221(1)(28) &     0.5200(4)(27) &     0.5187(5)(26) \\
          $Z_V$                &     0.7314(4)(21) &     0.7371(2)(20) &     0.7427(1)(21) &     0.7513(1)(21) &     0.7651(4)(22) &     0.7780(5)(22) \\
          $Z^\prime_V$         &     0.7268(5)(5) &     0.7328(3)(6) &     0.7386(1)(7) &     0.7473(2)(8) &     0.7614(5)(8) &     0.7742(7)(3) \\
          $Z_A$                &     0.7551(4)(22) &     0.7614(2)(21) &     0.7672(1)(21) &     0.7761(1)(22) &     0.7893(4)(25) &     0.8014(5)(25) \\
          $Z^\prime_A$         &     0.7760(3)(5) &     0.7793(2)(3) &     0.7823(1)(3) &     0.7873(1)(2) &     0.7955(4)(9) &     0.8039(5)(12) \\
          $Z_T$                &     0.8441(4)(35) &     0.8557(2)(35) &     0.8663(1)(36) &     0.8818(1)(37) &     0.9053(4)(42) &     0.9274(6)(42) \\
          $Z_{v_{2a}}$         &     1.0882(5)(108) &     1.1140(3)(111) &     1.1372(2)(114) &     1.1692(3)(118) &     1.2150(8)(134) &     1.2561(11)(130) \\
          $Z_{v_{2b}}$         &     1.0869(6)(108) &     1.1108(3)(111) &     1.1327(2)(114) &     1.1627(3)(117) &     1.2062(8)(133) &     1.2459(11)(130) \\
          $Z_{r_{2a}}$         &     1.0802(6)(107) &     1.1063(3)(111) &     1.1298(2)(114) &     1.1620(3)(118) &     1.2079(8)(135) &     1.2489(12)(131) \\
          $Z_{r_{2b}}$         &     1.1026(6)(110) &     1.1269(3)(113) &     1.1488(2)(116) &     1.1790(3)(120) &     1.2221(9)(137) &     1.2611(13)(133) \\
          $Z_{h_{1a}}$         &     1.1346(6)(197) &     1.1634(3)(202) &     1.1896(2)(207) &     1.2257(3)(214) &     1.2779(8)(232) &     1.3254(12)(235) \\
          $Z_{h_{1b}}$         &     1.1477(6)(199) &     1.1765(3)(205) &     1.2025(2)(210) &     1.2384(3)(216) &     1.2903(9)(234) &     1.3374(12)(236) \\
          $Z_S/Z_P$            &     1.1128(6)(28) &     1.1220(3)(22) &     1.1266(2)(20) &     1.1275(2)(18) &     1.1205(5)(24) &     1.1100(7)(24) \\
          $Z_A/Z_V$            &     1.0284(2)(3) &     1.0298(1)(3) &     1.0306(0)(3) &     1.0312(1)(3) &     1.0309(1)(4) &     1.0297(2)(5) \\
          $Z'_A/Z'_V$          &     1.0667(4)(5) &     1.0630(2)(4) &     1.0590(1)(7) &     1.0536(1)(10) &     1.0450(4)(21) &     1.0385(5)(20) \\
          $Z_A/Z_P$      &     1.3688(8)(94) &     1.4013(4)(93) &     1.4277(2)(94) &     1.4593(3)(95) &     1.4991(9)(102) &     1.5320(12)(101) \\
          $Z'_A/Z_P$     &     1.4008(10)(96) &     1.4271(5)(94) &     1.4474(3)(94) &     1.4703(4)(94) &     1.4981(10)(97) &     1.5221(14)(98) \\
          $Z_P/(Z_S Z_A)$      &     1.1805(7)(44) &     1.1633(3)(39) &     1.1514(2)(38) &     1.1390(3)(36) &     1.1281(8)(47) &     1.1223(11)(46) \\
          $Z_P/(Z_S Z'_A)$     &     1.1410(6)(24) &     1.1301(3)(15) &     1.1236(2)(13) &     1.1182(3)(12) &     1.1161(7)(26) &     1.1166(10)(31) \\
\end{tabular}
\end{ruledtabular}

\end{table*}
 
\begin{table*}[h]
\caption{Results for operators with two derivatives based on the
\RI-SMOM scheme, obtained by means of the fixed-scale method without the
perturbative subtraction of lattice artifacts.
The chiral extrapolation has been performed globally. The first
number in parentheses gives the statistical error, while the second number
is an estimate of the systematic uncertainty. All values refer to the
$\MS$ scheme at the scale $\mu_0^2 = 4 \, \mathrm {GeV}^2$.}
\label{tab.smomextract3}
\begin{ruledtabular}
\begin{tabular}{cllllll}
{ } &  \multicolumn{1}{c}{$\beta=3.34$} &  \multicolumn{1}{c}{$\beta=3.40$}
    &  \multicolumn{1}{c}{$\beta=3.46$} &  \multicolumn{1}{c}{$\beta=3.55$}
    &  \multicolumn{1}{c}{$\beta=3.70$} &  \multicolumn{1}{c}{$\beta=3.85$} \\
\hline
          $Z_{v_{3}}$          &     1.4250(16)(208) &     1.4633(9)(217) &     1.4976(5)(223) &     1.5451(8)(231) &     1.6112(22)(267) &     1.6714(30)(257) \\
          $Z_{a_{2}}$          &     1.4152(15)(207) &     1.4539(8)(215) &     1.4887(5)(222) &     1.5369(8)(231) &     1.6041(21)(268) &     1.6653(30)(258) \\
          $Z_{h_{2a}}$         &     1.4694(15)(366) &     1.5097(8)(378) &     1.5466(5)(388) &     1.5981(8)(402) &     1.6712(21)(438) &     1.7381(29)(442) \\
          $Z_{h_{2b}}$         &     1.4860(17)(371) &     1.5263(9)(382) &     1.5624(5)(392) &     1.6128(8)(406) &     1.6836(22)(444) &     1.7481(31)(446) \\
          $Z_{h_{2c}}$         &     1.4823(16)(369) &     1.5226(9)(381) &     1.5590(5)(391) &     1.6098(8)(405) &     1.6811(23)(442) &     1.7466(31)(446) \\
\end{tabular}
\end{ruledtabular}

\end{table*}

\end{appendix}

\FloatBarrier

\renewcommand{\theequation}{Z.\arabic{equation}}

\section*{Addendum}

After the publication of our article we noticed that some results from the
literature used were incorrect. This affects in particular the three-loop
conversion coefficient for the tensor density between the \RI-MOM and
the $\MS$ schemes and the subtraction of lattice artifacts at order $g^2$
in the coupling for our operators of dimension three and four.
The resulting changes are below 1\permil \ and smaller than the systematic
uncertainties given, in most cases even considerably smaller. This
will not lead to relevant differences in phenomenological results.
Nevertheless, we prefer to correct the affected tables accordingly.
We also include the four- and five-loop anomalous dimensions of the
tensor density and the quark field, respectively, in the determination.
The details are as follows.

In the perturbative subtraction of lattice artifacts we need the gluon
propagator for the L\"uscher-Weisz action with tree-level coefficients.
This has been taken from Eqs.~(11.26) -- (11.28) in
Ref.~\cite{Capitani:2002mp}. Unfortunately, there is an error in
Eq.~(11.28): The last line should read 
\begin{equation} 
-4 c_1^3 \sum_\rho \hat{k}^4_\rho \prod_{\tau \neq \rho} \hat{k}^2_\tau \,,
\end{equation}
as can be seen from the original literature~\cite{Weisz:1982zw,Weisz:1983bn}.

Moreover, the conversion factor (up to three loops) for the tensor density
$\mathcal T_{\mu \nu}$ in the \RI-MOM scheme was extracted from results
given in Ref.~\cite{Gracey:2003yr}. In Ref.~\cite{Gracey:2022vqr} it is
pointed out that the three-loop contribution is incorrect. The coefficient
$c_3$ should read

\begin{widetext}

\begin{equation} \label{eq.corrTconv}
c_3  = {} -\frac{9888899}{2916} + \frac{694633}{486} \zeta_3 
                   + \frac{464}{27} \zeta_4 - \frac{31720}{81} \zeta_5
         + \left( \frac{286262}{729} - \frac{2096}{27} \zeta_3 
                                     + \frac{80}{9} \zeta_4 \right) n_f
         - \left( \frac{13754}{2187} + \frac{32}{81} \zeta_3 \right) n_f^2 \,.
\end{equation}
For $n_f=3$ one finds $c_3 = - 1194.56$ to be compared with the value
$c_3 = - 1207.96$ given in Table~\ref{tab.confac} .

We also realized that the $\MS$ anomalous dimension of $\mathcal T_{\mu \nu}$
is actually known to four-loop accuracy~\cite{Baikov:2006ai}, while in
our paper we used only the three-loop anomalous dimension. The four-loop
coefficient reads
\begin{equation} 
\begin{split}
\gamma_3 = {} & \frac{2208517}{81} - \frac{247456}{243}\zeta_3 +
  \frac{10208}{9}\zeta_4 - \frac{1339520}{243}\zeta_5 +
  \left( - \frac{3074758}{729} - \frac{607328}{243}\zeta_3 +
   \frac{13984}{27}\zeta_4 + \frac{36800}{27}\zeta_5 \right) n_f \\ & {}
 + \left( \frac{39844}{729} + \frac{7360}{81}\zeta_3 -
    \frac{320}{9}\zeta_4 \right) n_f^2 +
  \left( \frac{56}{243} + \frac{128}{81}\zeta_3 \right) n_f^3 \,.
\end{split}
\end{equation}

Similarly, the five-loop $\MS$ anomalous dimension of the quark field in
the Landau gauge can be found from Ref.~\cite{Chetyrkin:2017bjc}.
In our paper we used only the four-loop result. The five-loop coefficent
reads
\begin{equation} 
\begin{split}
\gamma_4 = {} & \frac{4678028371}{7776} - \frac{161381953}{1728}\zeta_3 +
\frac{7530725}{108}\zeta_4 - \frac{447446437}{1944}\zeta_5
+ \frac{41883875}{1296}\zeta_6  + \frac{1914069143}{13824}\zeta_7
+ \frac{64558441}{5184}\zeta_3^2  \\ & {}
+ \left( - \frac{732321613}{5832} - \frac{157885211}{5832}\zeta_3 
  - \frac{355483}{162}\zeta_4 - \frac{27229511}{2916}\zeta_5
  + \frac{1894775}{1944}\zeta_6  + \frac{825587}{27}\zeta_7
+ \frac{4989469}{972}\zeta_3^2 \right) n_f \\ & {}
+ \left( \frac{4296115}{729} + \frac{1103240}{243}\zeta_3 
  - \frac{20422}{27}\zeta_4 - \frac{11936}{27}\zeta_5 - \frac{1600}{9}\zeta_6
  - 1323 \zeta_7 - \frac{256}{9}\zeta_3^2 \right) n_f^2 \\ & {}
+ \left( - \frac{42638}{2187} - \frac{2624}{81}\zeta_3 + \frac{320}{9}\zeta_4
   \right) n_f^3
+ \left( \frac{664}{729} - \frac{128}{81}\zeta_3 \right) n_f^4 \,.
\end{split}
\end{equation}

\end{widetext}

We have recalculated the lattice artifacts in one-loop lattice perturbation
theory for the quark-antiquark operators with less than two derivatives
using the correct gluon propagator and implemented the above conversion
factor for $\mathcal T_{\mu \nu}$ in the \RI-MOM scheme as well as the 
four-loop anomalous dimension of $\mathcal T_{\mu \nu}$ and the five-loop
anomalous dimension of the quark field.

With all these corrections in place, our analysis yields the
Tables~\ref{tab.Nmomfit1} -- \ref{tab.Nsmomextract2}, where we include
also the numbers that have not changed.

Tables~\ref{tab.Nmomfit1}, \ref{tab.Nmomfit2}, \ref{tab.Nmomextract1}
and \ref{tab.Nmomextract2} are the corrected versions
of Tables~\ref{tab.momfit1}, \ref{tab.momfit2}, \ref{tab.momextract1}
and~\ref{tab.momextract2}, respectively.
They differ from the latter tables due to the corrected subtraction,
the updated anomalous dimensions of $\mathcal T_{\mu \nu}$ and the
quark field as well as due to the corrected value of the three-loop
coefficient in the conversion factor for $\mathcal T_{\mu \nu}$.

Tables~\ref{tab.Nsmomfit1}, \ref{tab.Nsmomfit2}, \ref{tab.Nsmomextract1}
and \ref{tab.Nsmomextract2} are the corrected versions
of Tables~\ref{tab.smomfit1}, \ref{tab.smomfit2}, \ref{tab.smomextract1}
and~\ref{tab.smomextract2}, respectively.
They differ from the latter tables due to the corrected subtraction
as well as due to the updated anomalous dimensions of $\mathcal T_{\mu \nu}$
and the quark field.

\begin{table*}
\caption{Results from fits for operators with less than two derivatives
based on the \RI-MOM scheme without the perturbative subtraction of
lattice artifacts. The chiral extrapolation has been performed globally.
The first number in parentheses gives the statistical error, while the
second number is an estimate of the systematic uncertainty.
All values refer to the $\MS$ scheme at the scale
$\mu_0^2 = 4 \, \mathrm {GeV}^2$. For results with four-loop conversion
factors, see Table~\ref{tab.Nmomfit1a}.}
\label{tab.Nmomfit1}
\begin{ruledtabular}
\begin{tabular}{cllllll}
{ } &  \multicolumn{1}{c}{$\beta=3.34$} &  \multicolumn{1}{c}{$\beta=3.40$}
    &  \multicolumn{1}{c}{$\beta=3.46$} &  \multicolumn{1}{c}{$\beta=3.55$}
    &  \multicolumn{1}{c}{$\beta=3.70$} &  \multicolumn{1}{c}{$\beta=3.85$} \\
\hline
$Z_q$                &     0.7822(11)(67) &     0.7901(5)(61) &     0.7971(2)(54) &     0.8071(4)(48) &     0.8223(10)(34) &     0.8360(13)(30) \\
$Z_S$                &     0.6322(154)(300) &     0.6202(71)(255) &     0.6089(52)(261) &     0.5964(78)(267) &     0.5831(119)(233) &     0.5729(138)(183) \\
$Z_V$                &     0.6915(22)(78) &     0.7011(13)(61) &     0.7102(7)(53) &     0.7229(11)(52) &     0.7419(24)(48) &     0.7589(33)(54) \\
$Z^\prime_V$         &     0.6968(21)(63) &     0.7065(11)(47) &     0.7156(7)(39) &     0.7279(11)(38) &     0.7465(24)(37) &     0.7628(32)(45) \\
$Z_A$                &     0.7372(24)(74) &     0.7449(12)(56) &     0.7525(7)(55) &     0.7628(13)(59) &     0.7782(22)(58) &     0.7919(29)(57) \\
$Z^\prime_A$         &     0.7440(22)(42) &     0.7514(12)(30) &     0.7582(6)(27) &     0.7678(13)(29) &     0.7823(22)(30) &     0.7952(30)(36) \\
$Z_T$                &     0.8064(21)(125) &     0.8204(10)(98) &     0.8338(6)(97) &     0.8522(11)(102) &     0.8796(23)(98) &     0.9051(31)(98) \\
$Z_{v_{2a}}$         &     1.0527(18)(179) &     1.0786(9)(167) &     1.1020(6)(164) &     1.1335(10)(162) &     1.1798(21)(161) &     1.2228(27)(187) \\
$Z_{v_{2b}}$         &     1.0672(17)(136) &     1.0916(9)(128) &     1.1136(5)(122) &     1.1431(10)(118) &     1.1868(19)(116) &     1.2269(25)(134) \\
$Z_{r_{2a}}$         &     1.0514(20)(185) &     1.0776(10)(172) &     1.1012(6)(171) &     1.1329(11)(171) &     1.1791(23)(167) &     1.2217(30)(183) \\
$Z_{r_{2b}}$         &     1.0907(19)(142) &     1.1146(10)(135) &     1.1358(5)(129) &     1.1645(10)(120) &     1.2067(20)(107) &     1.2452(26)(113) \\
$Z_{h_{1a}}$         &     1.0794(19)(183) &     1.1083(8)(172) &     1.1344(5)(169) &     1.1696(11)(167) &     1.2218(22)(166) &     1.2702(30)(194) \\
$Z_{h_{1b}}$         &     1.0896(20)(186) &     1.1182(11)(175) &     1.1440(6)(170) &     1.1788(11)(163) &     1.2305(22)(158) &     1.2787(30)(186) \\
$Z_A/Z_V$            &     1.0643(32)(47) &     1.0613(15)(42) &     1.0585(11)(36) &     1.0546(16)(28) &     1.0486(30)(10) &     1.0433(41)(9) \\
$Z'_A/Z'_V$          &     1.0660(32)(40) &     1.0624(17)(29) &     1.0589(10)(24) &     1.0543(16)(21) &     1.0477(30)(19) &     1.0424(40)(24) \\
\end{tabular}
\end{ruledtabular}

\end{table*}

\begin{table*}
\caption{Results from fits for operators with less than two derivatives
based on the \RI-MOM scheme with the perturbative subtraction of lattice
artifacts. The chiral extrapolation has been performed globally.
The first number in parentheses gives the statistical error, while the
second number is an estimate of the systematic uncertainty.
All values refer to the $\MS$ scheme at the scale
$\mu_0^2 = 4 \, \mathrm {GeV}^2$. For results with four-loop conversion
factors, see Table~\ref{tab.Nmomfit2a}.}
\label{tab.Nmomfit2}
\begin{ruledtabular}
\begin{tabular}{cllllll}
{ } &  \multicolumn{1}{c}{$\beta=3.34$} &  \multicolumn{1}{c}{$\beta=3.40$}
    &  \multicolumn{1}{c}{$\beta=3.46$} &  \multicolumn{1}{c}{$\beta=3.55$}
    &  \multicolumn{1}{c}{$\beta=3.70$} &  \multicolumn{1}{c}{$\beta=3.85$} \\
\hline
$Z_q$                &     0.7809(11)(48) &     0.7894(5)(45) &     0.7971(2)(41) &     0.8077(4)(37) &     0.8234(10)(30) &     0.8373(13)(25) \\
$Z_S$                &     0.6044(154)(240) &     0.5942(71)(209) &     0.5850(49)(212) &     0.5748(78)(214) &     0.5643(119)(186) &     0.5564(136)(146) \\
$Z_V$                &     0.6996(22)(35) &     0.7099(13)(28) &     0.7191(7)(22) &     0.7318(11)(17) &     0.7507(24)(13) &     0.7670(32)(20) \\
$Z^\prime_V$         &     0.7028(21)(25) &     0.7131(11)(17) &     0.7224(7)(12) &     0.7349(11)(8) &     0.7534(24)(11) &     0.7691(32)(17) \\
$Z_A$                &     0.7433(24)(28) &     0.7517(12)(23) &     0.7594(7)(24) &     0.7699(13)(25) &     0.7853(23)(26) &     0.7984(29)(24) \\
$Z^\prime_A$         &     0.7476(22)(6) &     0.7556(12)(4) &     0.7627(6)(5) &     0.7725(13)(5) &     0.7871(22)(9) &     0.7996(30)(11) \\
$Z_T$                &     0.8185(22)(67) &     0.8333(10)(61) &     0.8468(6)(58) &     0.8652(11)(56) &     0.8925(23)(51) &     0.9170(31)(49) \\
$Z_{v_{2a}}$         &     1.0632(18)(167) &     1.0894(9)(159) &     1.1128(6)(153) &     1.1444(10)(145) &     1.1909(21)(137) &     1.2333(27)(154) \\
$Z_{v_{2b}}$         &     1.0726(17)(133) &     1.0973(9)(128) &     1.1193(5)(120) &     1.1489(10)(111) &     1.1927(19)(102) &     1.2323(25)(111) \\
$Z_{r_{2a}}$         &     1.0616(19)(168) &     1.0880(10)(161) &     1.1116(6)(156) &     1.1433(11)(148) &     1.1897(23)(139) &     1.2316(30)(148) \\
$Z_{r_{2b}}$         &     1.0969(19)(142) &     1.1210(10)(136) &     1.1423(5)(130) &     1.1711(10)(119) &     1.2135(20)(102) &     1.2516(26)(99) \\
$Z_{h_{1a}}$         &     1.0889(19)(174) &     1.1182(8)(166) &     1.1443(5)(160) &     1.1797(11)(151) &     1.2322(22)(141) &     1.2800(30)(157) \\
$Z_{h_{1b}}$         &     1.1002(21)(182) &     1.1291(11)(175) &     1.1549(6)(167) &     1.1900(11)(154) &     1.2419(22)(141) &     1.2896(30)(155) \\
$Z_A/Z_V$            &     1.0622(32)(52) &     1.0592(15)(46) &     1.0565(11)(41) &     1.0527(16)(33) &     1.0468(31)(17) &     1.0418(41)(10) \\
$Z'_A/Z'_V$          &     1.0636(31)(37) &     1.0600(17)(28) &     1.0566(10)(21) &     1.0521(16)(14) &     1.0456(30)(8) &     1.0406(39)(15) \\
\end{tabular}
\end{ruledtabular}

\end{table*}

\begin{table*}
\caption{Results for operators with less than two derivatives based on the
\RI-MOM scheme, obtained by means of the fixed-scale method without the
perturbative subtraction of lattice artifacts.
The chiral extrapolation has been performed globally.
The first number in parentheses gives the statistical error, while the
second number is an estimate of the systematic uncertainty.
All values refer to the $\MS$ scheme at the scale
$\mu_0^2 = 4 \, \mathrm {GeV}^2$. For results with four-loop conversion
factors, see Table~\ref{tab.Nmomextract1a}.}
\label{tab.Nmomextract1}
\begin{ruledtabular}
\begin{tabular}{cllllll}
{ } &  \multicolumn{1}{c}{$\beta=3.34$} &  \multicolumn{1}{c}{$\beta=3.40$}
    &  \multicolumn{1}{c}{$\beta=3.46$} &  \multicolumn{1}{c}{$\beta=3.55$}
    &  \multicolumn{1}{c}{$\beta=3.70$} &  \multicolumn{1}{c}{$\beta=3.85$} \\
\hline
$Z_q$                &     0.8092(4)(24) &     0.8114(2)(24) &     0.8147(1)(24) &     0.8207(1)(24) &     0.8317(4)(28) &     0.8430(5)(27) \\
$Z_S$                &     0.6978(54)(134) &     0.6713(31)(113) &     0.6473(23)(110) &     0.6220(30)(100) &     0.5880(62)(102) &     0.5648(85)(118) \\
$Z_V$                &     0.7005(8)(12) &     0.7074(5)(16) &     0.7156(2)(13) &     0.7275(3)(10) &     0.7472(9)(10) &     0.7639(13)(18) \\
$Z^\prime_V$         &     0.7070(8)(15) &     0.7137(4)(14) &     0.7213(3)(12) &     0.7322(3)(6) &     0.7501(9)(3) &     0.7648(13)(18) \\
$Z_A$                &     0.7456(8)(18) &     0.7509(4)(15) &     0.7565(3)(14) &     0.7660(4)(10) &     0.7812(9)(12) &     0.7947(12)(23) \\
$Z^\prime_A$         &     0.7561(9)(17) &     0.7604(5)(12) &     0.7648(3)(15) &     0.7725(4)(9) &     0.7851(10)(3) &     0.7965(14)(13) \\
$Z_T$                &     0.8085(7)(37) &     0.8216(4)(38) &     0.8347(2)(39) &     0.8543(3)(40) &     0.8855(8)(43) &     0.9135(11)(45) \\
$Z_{v_{2a}}$         &     1.0482(4)(111) &     1.0760(3)(114) &     1.1023(2)(118) &     1.1389(3)(124) &     1.1923(7)(140) &     1.2399(10)(137) \\
$Z_{v_{2b}}$         &     1.0666(4)(86) &     1.0920(3)(88) &     1.1155(2)(91) &     1.1478(3)(95) &     1.1944(7)(108) &     1.2364(9)(106) \\
$Z_{r_{2a}}$         &     1.0438(5)(111) &     1.0726(3)(115) &     1.0993(2)(119) &     1.1363(3)(124) &     1.1893(8)(142) &     1.2363(11)(138) \\
$Z_{r_{2b}}$         &     1.0898(5)(88) &     1.1147(3)(91) &     1.1374(2)(93) &     1.1685(3)(97) &     1.2135(7)(108) &     1.2537(10)(105) \\
$Z_{h_{1a}}$         &     1.0761(5)(116) &     1.1069(3)(121) &     1.1357(2)(125) &     1.1756(3)(131) &     1.2341(8)(149) &     1.2868(11)(147) \\
$Z_{h_{1b}}$         &     1.0887(5)(117) &     1.1185(3)(122) &     1.1465(2)(126) &     1.1857(3)(132) &     1.2433(8)(150) &     1.2956(11)(147) \\
$Z_A/Z_V$            &     1.0636(10)(5) &     1.0605(6)(7) &     1.0566(4)(7) &     1.0524(5)(10) &     1.0451(11)(7) &     1.0397(17)(13) \\
$Z'_A/Z'_V$          &     1.0688(11)(7) &     1.0648(6)(6) &     1.0597(4)(5) &     1.0541(6)(8) &     1.0456(13)(3) &     1.0399(19)(7) \\
\end{tabular}
\end{ruledtabular}

\end{table*}

\begin{table*}
\caption{Results for operators with less than two derivatives based on the
\RI-MOM scheme, obtained by means of the fixed-scale method with the
perturbative subtraction of lattice artifacts.
The chiral extrapolation has been performed globally.
The first number in parentheses gives the statistical error, while the
second number is an estimate of the systematic uncertainty.
All values refer to the $\MS$ scheme at the scale
$\mu_0^2 = 4 \, \mathrm {GeV}^2$. For results with four-loop conversion
factors, see Table~\ref{tab.Nmomextract2a}.}
\label{tab.Nmomextract2}
\begin{ruledtabular}
\begin{tabular}{cllllll}
{ } &  \multicolumn{1}{c}{$\beta=3.34$} &  \multicolumn{1}{c}{$\beta=3.40$}
    &  \multicolumn{1}{c}{$\beta=3.46$} &  \multicolumn{1}{c}{$\beta=3.55$}
    &  \multicolumn{1}{c}{$\beta=3.70$} &  \multicolumn{1}{c}{$\beta=3.85$} \\
\hline
$Z_q$                &     0.7897(4)(23) &     0.7965(2)(23) &     0.8031(1)(23) &     0.8126(1)(24) &     0.8270(4)(28) &     0.8403(5)(28) \\
$Z_S$                &     0.6509(54)(127) &     0.6300(31)(105) &     0.6107(23)(104) &     0.5916(30)(94) &     0.5657(64)(98) &     0.5485(85)(117) \\
$Z_V$                &     0.7039(8)(11) &     0.7129(5)(15) &     0.7221(2)(13) &     0.7342(3)(10) &     0.7530(9)(10) &     0.7684(13)(19) \\
$Z^\prime_V$         &     0.7081(8)(14) &     0.7169(4)(14) &     0.7256(3)(12) &     0.7369(3)(6) &     0.7542(9)(3) &     0.7681(13)(19) \\
$Z_A$                &     0.7447(8)(18) &     0.7525(4)(15) &     0.7595(3)(14) &     0.7698(4)(10) &     0.7849(9)(12) &     0.7978(12)(24) \\
$Z^\prime_A$         &     0.7523(9)(17) &     0.7593(5)(12) &     0.7651(3)(15) &     0.7738(4)(9) &     0.7869(10)(3) &     0.7982(14)(13) \\
$Z_T$                &     0.8189(7)(38) &     0.8338(4)(39) &     0.8474(2)(40) &     0.8664(3)(40) &     0.8955(8)(43) &     0.9211(11)(45) \\
$Z_{v_{2a}}$         &     1.0615(4)(112) &     1.0892(3)(115) &     1.1148(2)(119) &     1.1500(3)(124) &     1.2010(7)(140) &     1.2465(10)(137) \\
$Z_{v_{2b}}$         &     1.0730(4)(87) &     1.0986(3)(89) &     1.1219(2)(92) &     1.1535(3)(95) &     1.1989(7)(108) &     1.2397(9)(106) \\
$Z_{r_{2a}}$         &     1.0571(5)(112) &     1.0856(3)(116) &     1.1117(2)(119) &     1.1473(3)(124) &     1.1979(8)(142) &     1.2428(11)(138) \\
$Z_{r_{2b}}$         &     1.0961(5)(89) &     1.1214(3)(91) &     1.1441(2)(94) &     1.1748(3)(97) &     1.2187(7)(108) &     1.2578(10)(105) \\
$Z_{h_{1a}}$         &     1.0873(5)(118) &     1.1182(3)(122) &     1.1466(2)(126) &     1.1855(3)(131) &     1.2418(8)(149) &     1.2926(11)(147) \\
$Z_{h_{1b}}$         &     1.1003(5)(119) &     1.1305(3)(123) &     1.1582(2)(127) &     1.1964(3)(132) &     1.2520(8)(150) &     1.3023(11)(147) \\
$Z_A/Z_V$            &     1.0593(10)(6) &     1.0566(6)(7) &     1.0531(4)(8) &     1.0495(5)(10) &     1.0430(12)(7) &     1.0383(17)(13) \\
$Z'_A/Z'_V$          &     1.0639(11)(8) &     1.0604(6)(7) &     1.0557(5)(5) &     1.0508(6)(8) &     1.0433(14)(3) &     1.0383(21)(7) \\
\end{tabular}
\end{ruledtabular}

\end{table*}

\begin{table*}
\caption{Results from fits for operators with less than two derivatives
based on the \RI-SMOM scheme without the perturbative subtraction of
lattice artifacts. The chiral extrapolation has been performed globally. 
The first number in parentheses gives the statistical error, while the
second number is an estimate of the systematic uncertainty. All values
refer to the $\MS$ scheme at the scale $\mu_0^2 = 4 \, \mathrm {GeV}^2$.}
\label{tab.Nsmomfit1}
\begin{ruledtabular}
\begin{tabular}{cllllll}
{ } &  \multicolumn{1}{c}{$\beta=3.34$} &  \multicolumn{1}{c}{$\beta=3.40$}
    &  \multicolumn{1}{c}{$\beta=3.46$} &  \multicolumn{1}{c}{$\beta=3.55$}
    &  \multicolumn{1}{c}{$\beta=3.70$} &  \multicolumn{1}{c}{$\beta=3.85$} \\
\hline
$Z_q$                &     0.7858(9)(93) &     0.7938(4)(85) &     0.8006(2)(81) &     0.8106(3)(72) &     0.8255(9)(55) &     0.8391(12)(44) \\
$Z_S$                &     0.6233(14)(79) &     0.6179(7)(73) &     0.6128(4)(66) &     0.6062(5)(55) &     0.5962(11)(37) &     0.5864(16)(28) \\
$Z_P$                &     0.4958(16)(255) &     0.4968(6)(229) &     0.4978(3)(205) &     0.5006(5)(176) &     0.5056(13)(116) &     0.5092(19)(66) \\
$Z_V$                &     0.7151(11)(96) &     0.7236(5)(87) &     0.7310(3)(79) &     0.7417(4)(67) &     0.7581(10)(45) &     0.7730(15)(40) \\
$Z^\prime_V$         &     0.7104(13)(86) &     0.7192(6)(72) &     0.7268(3)(63) &     0.7378(5)(53) &     0.7550(13)(38) &     0.7702(18)(49) \\
$Z_A$                &     0.7464(10)(66) &     0.7543(4)(62) &     0.7612(3)(58) &     0.7711(4)(49) &     0.7857(10)(37) &     0.7988(14)(34) \\
$Z^\prime_A$         &     0.7625(9)(74) &     0.7679(4)(63) &     0.7724(2)(62) &     0.7795(4)(60) &     0.7909(10)(46) &     0.8017(14)(35) \\
$Z_T$                &     0.8312(12)(101) &     0.8446(5)(95) &     0.8565(3)(88) &     0.8731(4)(77) &     0.8982(12)(61) &     0.9215(16)(62) \\
$Z_{v_{2a}}$         &     1.0707(17)(304) &     1.0985(9)(290) &     1.1226(6)(278) &     1.1552(8)(251) &     1.2026(21)(203) &     1.2451(28)(177) \\
$Z_{v_{2b}}$         &     1.0655(16)(315) &     1.0920(8)(300) &     1.1150(6)(288) &     1.1465(8)(260) &     1.1927(20)(209) &     1.2347(28)(180) \\
$Z_{r_{2a}}$         &     1.0640(18)(298) &     1.0918(9)(286) &     1.1160(6)(273) &     1.1486(8)(248) &     1.1959(21)(202) &     1.2384(30)(172) \\
$Z_{r_{2b}}$         &     1.0842(18)(314) &     1.1106(9)(301) &     1.1334(6)(288) &     1.1644(9)(261) &     1.2096(22)(210) &     1.2505(30)(177) \\
$Z_{h_{1a}}$         &     1.1388(18)(374) &     1.1677(9)(361) &     1.1934(6)(351) &     1.2274(9)(334) &     1.2756(23)(299) &     1.3194(32)(270) \\
$Z_{h_{1b}}$         &     1.1529(18)(379) &     1.1817(10)(365) &     1.2071(6)(356) &     1.2408(10)(338) &     1.2884(24)(302) &     1.3315(32)(273) \\
$Z_S/Z_P$            &     1.2119(39)(635) &     1.2065(17)(586) &     1.2001(10)(546) &     1.1880(14)(477) &     1.1657(33)(318) &     1.1437(44)(160) \\
$Z_A/Z_V$            &     1.0412(7)(59) &     1.0407(4)(55) &     1.0400(2)(51) &     1.0387(2)(45) &     1.0360(5)(29) &     1.0333(8)(18) \\
$Z'_A/Z'_V$          &     1.0705(12)(60) &     1.0655(6)(46) &     1.0609(3)(35) &     1.0550(4)(25) &     1.0466(9)(22) &     1.0400(13)(29) \\
$Z_A/Z_P$            &     1.4529(39)(462) &     1.4732(16)(421) &     1.4908(8)(412) &     1.5102(12)(384) &     1.5335(35)(281) &     1.5543(51)(212) \\
$Z'_A/Z_P$           &     1.4582(47)(335) &     1.4753(15)(316) &     1.4891(9)(297) &     1.5045(15)(258) &     1.5239(41)(169) &     1.5422(59)(149) \\
$Z_P/(Z_S Z_A)$      &     1.0893(34)(439) &     1.0847(16)(418) &     1.0823(10)(399) &     1.0816(12)(339) &     1.0860(31)(204) &     1.0916(43)(126) \\
$Z_P/(Z_S Z'_A)$     &     1.0661(32)(377) &     1.0655(14)(351) &     1.0666(10)(330) &     1.0702(13)(265) &     1.0790(34)(137) &     1.0876(46)(94) \\
\end{tabular}
\end{ruledtabular}

\end{table*}
 
\begin{table*}
\caption{Results from fits for operators with less than two derivatives
based on the \RI-SMOM scheme with the perturbative subtraction of lattice
artifacts. The chiral extrapolation has been performed globally. The first
number in parentheses gives the statistical error, while the second number
is an estimate of the systematic uncertainty. All values refer to the
$\MS$ scheme at the scale $\mu_0^2 = 4 \, \mathrm {GeV}^2$.}
\label{tab.Nsmomfit2}
\begin{ruledtabular}
\begin{tabular}{cllllll}
{ } &  \multicolumn{1}{c}{$\beta=3.34$} &  \multicolumn{1}{c}{$\beta=3.40$}
    &  \multicolumn{1}{c}{$\beta=3.46$} &  \multicolumn{1}{c}{$\beta=3.55$}
    &  \multicolumn{1}{c}{$\beta=3.70$} &  \multicolumn{1}{c}{$\beta=3.85$} \\
\hline
$Z_q$                &     0.7822(9)(72) &     0.7908(4)(67) &     0.7983(2)(64) &     0.8088(3)(58) &     0.8243(9)(46) &     0.8383(12)(36) \\
$Z_S$                &     0.6115(14)(92) &     0.6068(7)(86) &     0.6025(4)(79) &     0.5968(5)(66) &     0.5880(11)(44) &     0.5793(16)(31) \\
$Z_P$                &     0.4881(15)(215) &     0.4901(6)(194) &     0.4921(3)(172) &     0.4959(5)(145) &     0.5020(13)(92) &     0.5064(19)(49) \\
$Z_V$                &     0.7137(11)(76) &     0.7228(5)(69) &     0.7308(3)(64) &     0.7420(4)(56) &     0.7588(10)(41) &     0.7738(15)(34) \\
$Z^\prime_V$         &     0.7072(13)(59) &     0.7168(6)(49) &     0.7250(3)(43) &     0.7367(5)(37) &     0.7544(13)(27) &     0.7699(19)(33) \\
$Z_A$                &     0.7457(10)(57) &     0.7541(4)(54) &     0.7614(3)(51) &     0.7716(4)(45) &     0.7864(10)(37) &     0.7996(14)(30) \\
$Z^\prime_A$         &     0.7579(9)(41) &     0.7641(4)(35) &     0.7695(2)(36) &     0.7774(4)(36) &     0.7895(10)(30) &     0.8006(14)(21) \\
$Z_T$                &     0.8321(12)(94) &     0.8462(5)(88) &     0.8585(3)(84) &     0.8756(4)(76) &     0.9010(12)(62) &     0.9243(16)(53) \\
$Z_{v_{2a}}$         &     1.0731(17)(306) &     1.1010(9)(293) &     1.1251(6)(282) &     1.1578(8)(256) &     1.2053(21)(207) &     1.2476(28)(176) \\
$Z_{v_{2b}}$         &     1.0672(16)(317) &     1.0938(8)(302) &     1.1169(6)(291) &     1.1484(8)(264) &     1.1948(20)(212) &     1.2367(28)(179) \\
$Z_{r_{2a}}$         &     1.0666(18)(303) &     1.0946(9)(290) &     1.1187(6)(280) &     1.1515(8)(254) &     1.1989(22)(206) &     1.2412(30)(173) \\
$Z_{r_{2b}}$         &     1.0861(18)(318) &     1.1127(9)(304) &     1.1355(6)(293) &     1.1666(9)(266) &     1.2119(22)(215) &     1.2527(30)(177) \\
$Z_{h_{1a}}$         &     1.1412(18)(372) &     1.1704(9)(360) &     1.1961(6)(351) &     1.2302(9)(333) &     1.2786(23)(297) &     1.3223(32)(268) \\
$Z_{h_{1b}}$         &     1.1553(18)(378) &     1.1843(10)(365) &     1.2098(6)(356) &     1.2436(10)(337) &     1.2913(24)(302) &     1.3343(32)(270) \\
$Z_S/Z_P$            &     1.2084(39)(615) &     1.2025(17)(566) &     1.1958(10)(528) &     1.1835(13)(461) &     1.1609(33)(308) &     1.1388(45)(155) \\
$Z_A/Z_V$            &     1.0423(7)(52) &     1.0415(4)(47) &     1.0406(2)(44) &     1.0390(2)(36) &     1.0362(6)(24) &     1.0333(9)(13) \\
$Z'_A/Z'_V$          &     1.0692(12)(64) &     1.0644(6)(50) &     1.0599(3)(39) &     1.0542(4)(26) &     1.0459(9)(16) &     1.0394(13)(27) \\
$Z_A/Z_P$            &     1.4583(39)(426) &     1.4787(16)(394) &     1.4960(8)(382) &     1.5150(12)(350) &     1.5380(35)(248) &     1.5583(50)(190) \\
$Z'_A/Z_P$           &     1.4600(46)(326) &     1.4772(15)(308) &     1.4910(9)(287) &     1.5064(14)(246) &     1.5260(41)(157) &     1.5442(59)(148) \\
$Z_P/(Z_S Z_A)$      &     1.0945(34)(445) &     1.0897(16)(423) &     1.0871(10)(404) &     1.0862(12)(342) &     1.0903(31)(205) &     1.0953(43)(128) \\
$Z_P/(Z_S Z'_A)$     &     1.0748(32)(404) &     1.0737(14)(373) &     1.0744(10)(351) &     1.0773(13)(285) &     1.0854(34)(153) &     1.0935(46)(102) \\
\end{tabular}
\end{ruledtabular}

\end{table*}
 
\begin{table*}
\caption{Results for operators with less than two derivatives based on the
\RI-SMOM scheme, obtained by means of the fixed-scale method without the
perturbative subtraction of lattice artifacts.
The chiral extrapolation has been performed globally. The first
number in parentheses gives the statistical error, while the second number
is an estimate of the systematic uncertainty. All values refer to the
$\MS$ scheme at the scale $\mu_0^2 = 4 \, \mathrm {GeV}^2$.}
\label{tab.Nsmomextract1}
\begin{ruledtabular}
\begin{tabular}{cllllll}
{ } &  \multicolumn{1}{c}{$\beta=3.34$} &  \multicolumn{1}{c}{$\beta=3.40$}
    &  \multicolumn{1}{c}{$\beta=3.46$} &  \multicolumn{1}{c}{$\beta=3.55$}
    &  \multicolumn{1}{c}{$\beta=3.70$} &  \multicolumn{1}{c}{$\beta=3.85$} \\
\hline
          $Z_q$                &     0.8311(3)(26) &     0.8305(2)(25) &     0.8310(1)(24) &     0.8338(1)(24) &     0.8403(3)(31) &     0.8487(4)(28) \\
          $Z_S$                &     0.6343(4)(28) &     0.6268(2)(28) &     0.6199(1)(28) &     0.6111(1)(27) &     0.5981(4)(26) &     0.5863(5)(25) \\
          $Z_P$                &     0.5769(5)(40) &     0.5635(2)(36) &     0.5537(1)(34) &     0.5439(1)(32) &     0.5344(4)(30) &     0.5283(5)(28) \\
          $Z_V$                &     0.7629(4)(24) &     0.7617(2)(22) &     0.7621(1)(22) &     0.7653(1)(21) &     0.7734(4)(23) &     0.7831(5)(22) \\
          $Z^\prime_V$         &     0.7611(5)(13) &     0.7601(3)(11) &     0.7606(1)(10) &     0.7636(2)(10) &     0.7716(5)(9) &     0.7807(7)(4) \\
          $Z_A$                &     0.7776(4)(23) &     0.7787(2)(22) &     0.7807(1)(22) &     0.7856(1)(22) &     0.7949(4)(25) &     0.8048(5)(25) \\
          $Z^\prime_A$         &     0.8171(3)(13) &     0.8123(2)(10) &     0.8092(1)(9) &     0.8074(1)(7) &     0.8082(4)(10) &     0.8120(5)(12) \\
          $Z_T$                &     0.8700(4)(35) &     0.8748(2)(35) &     0.8805(1)(35) &     0.8911(1)(36) &     0.9100(4)(41) &     0.9298(6)(41) \\
          $Z_{v_{2a}}$         &     1.0890(6)(107) &     1.1133(3)(111) &     1.1356(2)(114) &     1.1669(3)(118) &     1.2127(8)(134) &     1.2539(11)(130) \\
          $Z_{v_{2b}}$         &     1.0918(6)(107) &     1.1135(3)(110) &     1.1340(2)(113) &     1.1629(3)(117) &     1.2055(8)(133) &     1.2450(11)(130) \\
          $Z_{r_{2a}}$         &     1.0800(6)(107) &     1.1046(3)(110) &     1.1273(2)(113) &     1.1590(3)(117) &     1.2049(8)(135) &     1.2463(11)(131) \\
          $Z_{r_{2b}}$         &     1.1052(6)(109) &     1.1278(3)(112) &     1.1486(2)(115) &     1.1780(3)(119) &     1.2206(9)(136) &     1.2595(12)(133) \\
          $Z_{h_{1a}}$         &     1.1386(6)(197) &     1.1650(3)(202) &     1.1897(2)(207) &     1.2246(3)(214) &     1.2761(8)(231) &     1.3236(12)(234) \\
          $Z_{h_{1b}}$         &     1.1510(6)(199) &     1.1777(3)(204) &     1.2024(2)(209) &     1.2372(3)(216) &     1.2884(9)(233) &     1.3355(13)(236) \\
          $Z_S/Z_P$            &     1.0988(6)(32) &     1.1118(3)(26) &     1.1194(2)(23) &     1.1235(2)(20) &     1.1196(5)(25) &     1.1105(8)(24) \\
          $Z_A/Z_V$            &     1.0190(2)(5) &     1.0222(1)(4) &     1.0243(1)(4) &     1.0265(1)(4) &     1.0279(2)(4) &     1.0279(2)(5) \\
          $Z'_A/Z'_V$          &     1.0736(4)(4) &     1.0688(2)(4) &     1.0639(1)(7) &     1.0574(1)(10) &     1.0476(3)(21) &     1.0401(5)(20) \\
          $Z_A/Z_P$            &     1.3472(8)(94) &     1.3815(4)(94) &     1.4098(2)(95) &     1.4444(3)(96) &     1.4884(9)(103) &     1.5245(12)(102) \\
          $Z'_A/Z_P$           &     1.3977(10)(94) &     1.4231(5)(93) &     1.4430(3)(94) &     1.4660(4)(94) &     1.4946(10)(97) &     1.5194(14)(99) \\
          $Z_P/(Z_S Z_A)$      &     1.1706(7)(43) &     1.1552(3)(38) &     1.1443(2)(37) &     1.1329(3)(36) &     1.1231(7)(47) &     1.1183(11)(46) \\
          $Z_P/(Z_S Z'_A)$     &     1.1140(6)(20) &     1.1073(3)(11) &     1.1039(2)(10) &     1.1021(3)(10) &     1.1043(7)(26) &     1.1081(10)(31) \\
\end{tabular}
\end{ruledtabular}

\end{table*}
 
\begin{table*}
\caption{Results for operators with less than two derivatives based on the
\RI-SMOM scheme, obtained by means of the fixed-scale method with the
perturbative subtraction of lattice artifacts.
The chiral extrapolation has been performed globally. The first
number in parentheses gives the statistical error, while the second number
is an estimate of the systematic uncertainty. All values refer to the
$\MS$ scheme at the scale $\mu_0^2 = 4 \, \mathrm {GeV}^2$.}
\label{tab.Nsmomextract2}
\begin{ruledtabular}
\begin{tabular}{cllllll}
{ } &  \multicolumn{1}{c}{$\beta=3.34$} &  \multicolumn{1}{c}{$\beta=3.40$}
    &  \multicolumn{1}{c}{$\beta=3.46$} &  \multicolumn{1}{c}{$\beta=3.55$}
    &  \multicolumn{1}{c}{$\beta=3.70$} &  \multicolumn{1}{c}{$\beta=3.85$} \\
\hline
          $Z_q$                &     0.7974(3)(24) &     0.8032(2)(23) &     0.8086(1)(24) &     0.8170(1)(24) &     0.8297(3)(31) &     0.8419(4)(29) \\
          $Z_S$                &     0.6072(4)(26) &     0.6027(2)(25) &     0.5985(1)(25) &     0.5930(1)(25) &     0.5846(4)(24) &     0.5764(5)(24) \\
          $Z_P$                &     0.5369(5)(34) &     0.5302(2)(30) &     0.5257(1)(29) &     0.5222(1)(28) &     0.5202(4)(27) &     0.5189(5)(26) \\
          $Z_V$                &     0.7315(4)(21) &     0.7372(2)(20) &     0.7428(1)(21) &     0.7514(1)(21) &     0.7652(4)(22) &     0.7782(5)(22) \\
          $Z^\prime_V$         &     0.7269(5)(5) &     0.7329(3)(6) &     0.7387(1)(7) &     0.7474(2)(8) &     0.7615(5)(8) &     0.7743(7)(3) \\
          $Z_A$                &     0.7552(4)(22) &     0.7615(2)(21) &     0.7673(1)(21) &     0.7762(1)(22) &     0.7894(4)(25) &     0.8015(6)(25) \\
          $Z^\prime_A$         &     0.7761(3)(5) &     0.7794(2)(3) &     0.7824(1)(3) &     0.7874(1)(2) &     0.7956(4)(9) &     0.8040(5)(12) \\
          $Z_T$                &     0.8443(4)(35) &     0.8560(2)(35) &     0.8665(1)(36) &     0.8820(1)(37) &     0.9055(4)(42) &     0.9276(6)(42) \\
          $Z_{v_{2a}}$         &     1.0881(6)(108) &     1.1140(3)(111) &     1.1372(2)(114) &     1.1691(3)(118) &     1.2150(8)(134) &     1.2560(11)(130) \\
          $Z_{v_{2b}}$         &     1.0869(6)(108) &     1.1108(3)(111) &     1.1326(2)(114) &     1.1627(3)(117) &     1.2062(8)(133) &     1.2458(11)(130) \\
          $Z_{r_{2a}}$         &     1.0802(6)(107) &     1.1063(3)(111) &     1.1298(2)(114) &     1.1620(3)(118) &     1.2078(8)(135) &     1.2488(11)(131) \\
          $Z_{r_{2b}}$         &     1.1025(6)(110) &     1.1269(3)(113) &     1.1488(2)(116) &     1.1790(3)(119) &     1.2221(9)(136) &     1.2611(13)(132) \\
          $Z_{h_{1a}}$         &     1.1346(6)(197) &     1.1634(3)(202) &     1.1895(2)(207) &     1.2256(3)(214) &     1.2779(8)(232) &     1.3254(12)(235) \\
          $Z_{h_{1b}}$         &     1.1476(6)(199) &     1.1765(3)(205) &     1.2025(2)(210) &     1.2384(3)(216) &     1.2903(9)(234) &     1.3373(13)(236) \\
          $Z_S/Z_P$            &     1.1128(5)(28) &     1.1220(3)(22) &     1.1266(2)(20) &     1.1275(2)(18) &     1.1205(5)(24) &     1.1100(8)(24) \\
          $Z_A/Z_V$            &     1.0284(2)(3) &     1.0298(1)(3) &     1.0306(0)(3) &     1.0312(1)(3) &     1.0309(1)(4) &     1.0297(2)(5) \\
          $Z'_A/Z'_V$          &     1.0667(4)(5) &     1.0630(2)(4) &     1.0590(1)(7) &     1.0536(1)(10) &     1.0450(3)(21) &     1.0384(5)(20) \\
          $Z_A/Z_P$            &     1.3687(7)(94) &     1.4012(4)(93) &     1.4276(2)(94) &     1.4592(3)(95) &     1.4990(9)(102) &     1.5319(12)(101) \\
          $Z'_A/Z_P$           &     1.4007(10)(96) &     1.4271(5)(94) &     1.4473(3)(94) &     1.4702(4)(94) &     1.4980(10)(97) &     1.5220(14)(98) \\
          $Z_P/(Z_S Z_A)$      &     1.1804(6)(44) &     1.1632(3)(39) &     1.1513(2)(38) &     1.1389(3)(36) &     1.1280(8)(47) &     1.1222(11)(46) \\
          $Z_P/(Z_S Z'_A)$     &     1.1410(6)(24) &     1.1301(3)(15) &     1.1235(2)(13) &     1.1181(3)(12) &     1.1160(7)(26) &     1.1165(11)(31) \\
\end{tabular}
\end{ruledtabular}

\end{table*}

\FloatBarrier

The above-mentioned corrections and updates are contained in the
Erratum to our paper~\cite{RQCD:2020kuu}. We now present further
improvements resulting from the fact that for some operators the conversion
factors leading from the \RI-MOM scheme to the $\MS$ scheme are meanwhile
available to us in four-loop order, while in Ref.~\cite{RQCD:2020kuu}
only the three-loop conversion factors were used. We previously
overlooked that we could have extracted the four-loop expression for the
conversion factor of the quark wave function renormalization $C_q$ from
Ref.~\cite{Ruijl:2017eht}, and recently the four-loop coefficient
in the conversion factor for $\mathcal T_{\mu \nu}$ has been published
in Ref.~\cite{Gracey:2022vqr}, while for $\mathcal S$, $\mathcal V_\mu$ and
$\mathcal A_\mu$ it can be extracted from Ref.~\cite{Gracey:2022vjc}.

\begin{widetext}

One finds in the case of the quark wave function renormalization
\begin{equation} 
\begin{split}
c_4  = {} & {} - \frac{146722043}{864} + \frac{317781451}{2592} \zeta_3
  + \frac{565939}{864} \zeta_4 - \frac{356864009}{5184} \zeta_5
  + \frac{3807625}{10368} \zeta_6 + \frac{6747755}{288} \zeta_7 
  - \frac{29889697}{5184} \zeta_3^2 \\ & {}
  + \left( \frac{55476671}{1944} - \frac{1294381}{108} \zeta_3
     + \frac{2291}{72} \zeta_4 + \frac{1673051}{324} \zeta_5
     + \frac{100}{3} \zeta_6 - 1029 \zeta_7 - 24 \zeta_3^2 \right) n_f \\ & {}
  + \left( - \frac{1276817}{972} + \frac{5704}{27} \zeta_3
     - \frac{20}{3} \zeta_4 - \frac{440}{9} \zeta_5 \right) n_f^2
  + \left( \frac{21391}{1458} + \frac{8}{27} \zeta_3 \right) n_f^3
\end{split}
\end{equation}
in the Landau gauge. For $\mathcal S$ one gets
\begin{equation} 
\begin{split}
c_4  = {} & \frac{2876821079}{3888} - \frac{3967025}{12} \zeta_3
  + \frac{16960}{9} \zeta_4 - \frac{2618075}{144} \zeta_5
  - 5500\zeta_6 + \frac{3837631}{1728} \zeta_7 
  + \frac{981701}{54} \zeta_3^2 \\ & {}
  + \left( - \frac{288928891}{2916} + \frac{748015}{27} \zeta_3
     - \frac{14102}{9} \zeta_4 + \frac{359855}{81} \zeta_5
     + \frac{11500}{9} \zeta_6 + \frac{343}{2} \zeta_7
     + \frac{8776}{27} \zeta_3^2 \right) n_f \\ & {}
  + \left( \frac{55794733}{17496} - \frac{1220}{3} \zeta_3
     + \frac{100}{3} \zeta_4 - \frac{560}{3} \zeta_5 \right) n_f^2
  + \left( - \frac{96979}{4374} - \frac{40}{81} \zeta_3
     + \frac{8}{9} \zeta_4 \right) n_f^3 \,.
\end{split}
\end{equation}
For $\mathcal V_\mu$ and $\mathcal A_\mu$ one has
\begin{equation} 
c_4  = {} - \frac{4483139}{81} + \frac{15631129}{648} \zeta_3
  - \frac{15846715}{2592} \zeta_5 
  + \left( \frac{768940}{81} - \frac{60779}{27} \zeta_3
     + 415 \zeta_5 \right) n_f 
  + \left( - \frac{75355}{162} + 40 \zeta_3 \right) n_f^2
  + \frac{500}{81} n_f^3 
\end{equation}
and for $\mathcal T_{\mu \nu}$ 
\begin{equation} 
\begin{split}
c_4  = {} & {} - \frac{2280845783}{11664} + \frac{288171337}{2916} \zeta_3
  + \frac{15466}{81} \zeta_4 - \frac{106714573}{11664} \zeta_5
  + \frac{418600}{243} \zeta_6 - \frac{112538825}{15552} \zeta_7 
  - \frac{1243219}{486} \zeta_3^2 \\ & {}
  + \left( \frac{310694807}{8748} - \frac{2539997}{243} \zeta_3
     + \frac{37958}{81} \zeta_4 - \frac{59131}{243} \zeta_5
     - \frac{11500}{27} \zeta_6 + \frac{931}{2} \zeta_7
     - \frac{2096}{9} \zeta_3^2 \right) n_f \\ & {}
  + \left( - \frac{82363949}{52488} + \frac{14564}{81} \zeta_3
     - \frac{460}{27} \zeta_4 + \frac{2000}{27} \zeta_5 \right) n_f^2
  + \left( \frac{201739}{13122} + \frac{104}{243} \zeta_3
     - \frac{8}{27} \zeta_4 \right) n_f^3 \,.
\end{split}
\end{equation}

\end{widetext}

The numerical values are 188150.13 for $\mathcal S$, $- 14625.21$ for
$\mathcal V_\mu$ and $\mathcal A_\mu$, and $- 37788.57$ for
$\mathcal T_{\mu \nu}$.

In Tables~\ref{tab.Nmomfit1a} -- \ref{tab.Nmomextract2a} we present
the results obtained using the four-loop conversion factors for
$\mathcal S$, $\mathcal V_\mu$, $\mathcal A_\mu$, $\mathcal T_{\mu \nu}$
and for the quark wave function renormalization. Correspondingly, the
uncertainty due to the truncation of the expansion of the conversion
factors is estimated by comparing with the three-loop expressions. The
other elements of the evaluation are the same as in
Tables~\ref{tab.Nmomfit1} -- \ref{tab.Nmomextract2}.

The changes caused by the corrected subtraction, the updated anomalous
dimensions and the corrected value of the three-loop coefficient in the
conversion factor for $\mathcal T_{\mu \nu}$ look rather small.
However, replacing for $Z_q$, $Z_S$, $Z_V$, $Z_A$ and $Z_T$
the three-loop (two-loop) \RI-MOM conversion factors 
by the four-loop (three-loop) expressions leads to larger differences.

\begin{table*}
\caption{Results from fits for operators without derivatives
based on the \RI-MOM scheme without the perturbative subtraction of
lattice artifacts. The chiral extrapolation has been performed globally.
Four-loop conversion factors have been used.
The first number in parentheses gives the statistical error, while the
second number is an estimate of the systematic uncertainty.
All values refer to the $\MS$ scheme at the scale
$\mu_0^2 = 4 \, \mathrm {GeV}^2$.} 
\label{tab.Nmomfit1a}
\begin{ruledtabular}
\begin{tabular}{cllllll}
{ } &  \multicolumn{1}{c}{$\beta=3.34$} &  \multicolumn{1}{c}{$\beta=3.40$}
    &  \multicolumn{1}{c}{$\beta=3.46$} &  \multicolumn{1}{c}{$\beta=3.55$}
    &  \multicolumn{1}{c}{$\beta=3.70$} &  \multicolumn{1}{c}{$\beta=3.85$} \\
\hline
$Z_q$                &     0.7759(11)(86) &     0.7842(5)(76) &     0.7915(2)(68) &     0.8019(4)(62) &     0.8179(9)(43) &     0.8324(13)(32) \\
$Z_S$                &     0.6578(162)(230) &     0.6437(75)(191) &     0.6310(54)(189) &     0.6164(82)(192) &     0.5993(124)(169) &     0.5859(144)(136) \\
$Z_V$                &     0.6890(22)(78) &     0.6987(13)(62) &     0.7080(7)(53) &     0.7208(11)(49) &     0.7401(24)(42) &     0.7574(33)(50) \\
$Z_A$                &     0.7346(24)(70) &     0.7425(12)(53) &     0.7501(7)(50) &     0.7606(13)(53) &     0.7763(22)(51) &     0.7904(29)(52) \\
$Z_T$                &     0.7990(21)(106) &     0.8135(10)(83) &     0.8272(6)(78) &     0.8459(11)(79) &     0.8743(23)(76) &     0.9006(31)(83) \\
\end{tabular}
\end{ruledtabular}

\end{table*}

\begin{table*}
\caption{Results from fits for operators without derivatives
based on the \RI-MOM scheme with the perturbative subtraction of
lattice artifacts. The chiral extrapolation has been performed globally.
Four-loop conversion factors have been used.
The first number in parentheses gives the statistical error, while the
second number is an estimate of the systematic uncertainty.
All values refer to the $\MS$ scheme at the scale
$\mu_0^2 = 4 \, \mathrm {GeV}^2$.}
\label{tab.Nmomfit2a}
\begin{ruledtabular}
\begin{tabular}{cllllll}
{ } &  \multicolumn{1}{c}{$\beta=3.34$} &  \multicolumn{1}{c}{$\beta=3.40$}
    &  \multicolumn{1}{c}{$\beta=3.46$} &  \multicolumn{1}{c}{$\beta=3.55$}
    &  \multicolumn{1}{c}{$\beta=3.70$} &  \multicolumn{1}{c}{$\beta=3.85$} \\
\hline
$Z_q$                &     0.7747(10)(63) &     0.7837(5)(56) &     0.7916(2)(51) &     0.8025(4)(48) &     0.8190(9)(36) &     0.8337(13)(26) \\
$Z_S$                &     0.6286(162)(182) &     0.6165(75)(155) &     0.6060(51)(149) &     0.5938(82)(146) &     0.5798(124)(127) &     0.5689(142)(102) \\
$Z_V$                &     0.6971(22)(41) &     0.7075(13)(34) &     0.7169(7)(28) &     0.7297(11)(21) &     0.7490(24)(11) &     0.7655(32)(18) \\
$Z_A$                &     0.7406(24)(25) &     0.7492(12)(19) &     0.7570(7)(18) &     0.7677(13)(18) &     0.7834(22)(19) &     0.7969(29)(19) \\
$Z_T$                &     0.8110(21)(58) &     0.8263(10)(52) &     0.8400(6)(47) &     0.8588(10)(42) &     0.8871(23)(33) &     0.9124(31)(34) \\
\hline
\end{tabular}
\end{ruledtabular}

\end{table*}

\begin{table*}
\caption{Results for operators without derivatives based on the \RI-MOM
scheme, obtained by means of the fixed-scale method without the perturbative
subtraction of lattice artifacts. The chiral extrapolation has been
performed globally. Four-loop conversion factors have been used.
The first number in parentheses gives the statistical
error, while the second number is an estimate of the systematic uncertainty.
All values refer to the $\MS$ scheme at the scale
$\mu_0^2 = 4 \, \mathrm {GeV}^2$.}
\label{tab.Nmomextract1a}
\begin{ruledtabular}
\begin{tabular}{cllllll}
{ } &  \multicolumn{1}{c}{$\beta=3.34$} &  \multicolumn{1}{c}{$\beta=3.40$}
    &  \multicolumn{1}{c}{$\beta=3.46$} &  \multicolumn{1}{c}{$\beta=3.55$}
    &  \multicolumn{1}{c}{$\beta=3.70$} &  \multicolumn{1}{c}{$\beta=3.85$} \\
\hline
$Z_q$                &     0.8057(4)(19) &     0.8079(2)(18) &     0.8111(1)(18) &     0.8171(1)(18) &     0.8280(4)(23) &     0.8393(5)(22) \\
$Z_S$                &     0.7128(55)(116) &     0.6857(31)(91) &     0.6612(23)(89) &     0.6354(31)(78) &     0.6007(64)(83) &     0.5769(86)(105) \\
$Z_V$                &     0.6991(8)(11) &     0.7059(5)(15) &     0.7142(2)(12) &     0.7260(3)(9) &     0.7457(9)(9) &     0.7624(13)(18) \\
$Z_A$                &     0.7441(8)(17) &     0.7494(4)(14) &     0.7550(3)(13) &     0.7645(4)(9) &     0.7796(9)(11) &     0.7931(12)(23) \\
$Z_T$                &     0.8042(7)(24) &     0.8173(4)(25) &     0.8303(2)(26) &     0.8498(3)(27) &     0.8809(8)(30) &     0.9087(11)(33) \\
\end{tabular}
\end{ruledtabular}

\end{table*}

\begin{table*}
\caption{Results for operators without derivatives based on the \RI-MOM
scheme, obtained by means of the fixed-scale method with the perturbative
subtraction of lattice artifacts. The chiral extrapolation has been
performed globally. Four-loop conversion factors have been used.
The first number in parentheses gives the statistical
error, while the second number is an estimate of the systematic uncertainty.
All values refer to the $\MS$ scheme at the scale
$\mu_0^2 = 4 \, \mathrm {GeV}^2$.}
\label{tab.Nmomextract2a}
\begin{ruledtabular}
\begin{tabular}{cllllll}
{ } &  \multicolumn{1}{c}{$\beta=3.34$} &  \multicolumn{1}{c}{$\beta=3.40$}
    &  \multicolumn{1}{c}{$\beta=3.46$} &  \multicolumn{1}{c}{$\beta=3.55$}
    &  \multicolumn{1}{c}{$\beta=3.70$} &  \multicolumn{1}{c}{$\beta=3.85$} \\
\hline
$Z_q$                &     0.7863(4)(18) &     0.7930(2)(18) &     0.7996(1)(18) &     0.8090(1)(18) &     0.8234(4)(23) &     0.8366(5)(22) \\
$Z_S$                &     0.6649(55)(110) &     0.6435(31)(85) &     0.6238(23)(84) &     0.6043(31)(73) &     0.5779(65)(80) &     0.5603(86)(104) \\
$Z_V$                &     0.7025(8)(11) &     0.7115(5)(15) &     0.7206(2)(13) &     0.7327(3)(9) &     0.7515(9)(9) &     0.7669(13)(18) \\
$Z_A$                &     0.7432(8)(17) &     0.7510(4)(14) &     0.7580(3)(13) &     0.7683(4)(9) &     0.7833(9)(11) &     0.7962(12)(23) \\
$Z_T$                &     0.8146(7)(26) &     0.8294(4)(26) &     0.8429(2)(27) &     0.8619(3)(27) &     0.8908(8)(29) &     0.9163(11)(32) \\
\hline
\end{tabular}
\end{ruledtabular}

\end{table*}

\FloatBarrier

\end{document}